\documentclass[preprint,12pt,authoryear]{elsarticle}

\usepackage{amssymb}
\usepackage{amsmath}
\usepackage[top=2.5cm,bottom=2.5cm,left=2.5cm,right=2.5cm]{geometry}
\usepackage[table,xcdraw,svgnames]{xcolor}
\usepackage{natbib}
\usepackage{tikz}
\usepackage[hyphens,spaces,obeyspaces]{url}
\usepackage{mathrsfs}
\usepackage{caption}
\usepackage{subcaption}
\usepackage{bm}
\usepackage{import}
\usepackage[most]{tcolorbox}
\usepackage{calc}
\usepackage[utf8]{inputenc} 
\usepackage[T1]{fontenc}
\usepackage{textcomp}

\usepackage{verbatim}
\usepackage[colorlinks]{hyperref}
\usepackage{algorithm} 
\usepackage{algpseudocode} 
\usepackage{xcolor}

\usepackage{scalerel}
\usepackage{tikz}
\usetikzlibrary{shapes}
\usepackage{caption,graphicx,newfloat}
\DeclareCaptionType{InfoBox}
\usepackage{amsthm}

\usepackage{svg}
\newcommand{\orcid}[1]{\href{https://orcid.org/#1}{\includegraphics[width=10pt]{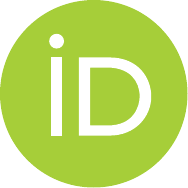}}}

\DeclareRobustCommand{\moatomnew}{\raisebox{-2pt}{\tikz{
\shade[ball color = {rgb,255:red,0;green,0;blue,255}] (0,0) circle(0.2cm);}}}

\DeclareRobustCommand{\satomnew}{\raisebox{-2pt}{\tikz{
\shade[ball color = {rgb,255:red,255;green,0;blue,0}] (0,0) circle(0.15cm);}}}

\DeclareUnicodeCharacter{2212}{-}

\AtBeginDocument{%
  \hypersetup{
    citecolor=blue,
    linkcolor=blue,   
    urlcolor=Magenta}}
\journal{Journal of the Mechanics and Physics of Solids}

\begin{document}
\begin{frontmatter}
\title{An Atomistic-based Finite Deformation Continuum Membrane Model for Monolayer Transition Metal Dichalcogenides}
\author[inst1]{Upendra Yadav \orcid{0000-0002-4863-3929}}
\affiliation[inst1]{organization={Mechanical Engineering - Engineering Mechanics, Michigan  Technological University},
            city={Houghton},
            state={MI},
            country={USA}}
\author[inst1]{Susanta Ghosh \orcid{0000-0002-6262-4121} \corref{sg}}

\cortext[sg]{Corresponding Author}
\ead{susantag@mtu.edu}
\begin{abstract}
A finite-deformation \emph{crystal-elasticity} membrane model for Transition Metal Dichalcogenide (TMD) monolayers is presented. Monolayer TMDs are multi-atom-thick two-dimensional (2D) crystalline membranes having atoms arranged in three parallel surfaces. In the present formulation, the deformed configuration of a TMD-membrane is represented through the deformation map of its middle surface and two stretches normal to the middle surface. \emph{Crystal-elasticity} based kinematic rules are employed to express the deformed bond lengths and bond angles of TMDs in terms of the continuum strains. The continuum hyper-elastic strain energy of the TMD membrane is formulated from its inter-atomic potential. The relative shifts between two simple lattices of TMDs are also considered in the constitutive relation. A smooth finite element framework using B-splines is developed to numerically implement the present  continuum membrane model. The proposed model generalizes the \emph{crystal-elasticity}-based membrane theory of purely 2D membranes, such as graphene, to the multi-atom-thick TMD crystalline membranes. The significance of relative shifts and two normal stretches are  demonstrated through numerical results. 
The proposed atomistic-based continuum model accurately matches the material moduli, complex post-buckling deformations, and the equilibrium energies predicted by the purely atomistic simulations. It also accurately reproduces the experimental results for large-area TMD samples containing tens of millions of atoms.

\end{abstract}
%
%
%
\begin{keyword}
 Crystal-Elasticity \sep  Molybdenum Disulfide \sep Transition Metal Dichalcogenide (TMD) \sep Cauchy-Born rule  \sep 2D materials 
\end{keyword}
\end{frontmatter}




\section{Introduction}
\label{sec:intro}
\begin{figure}[t]
  \centering

  \includegraphics[width=\linewidth]{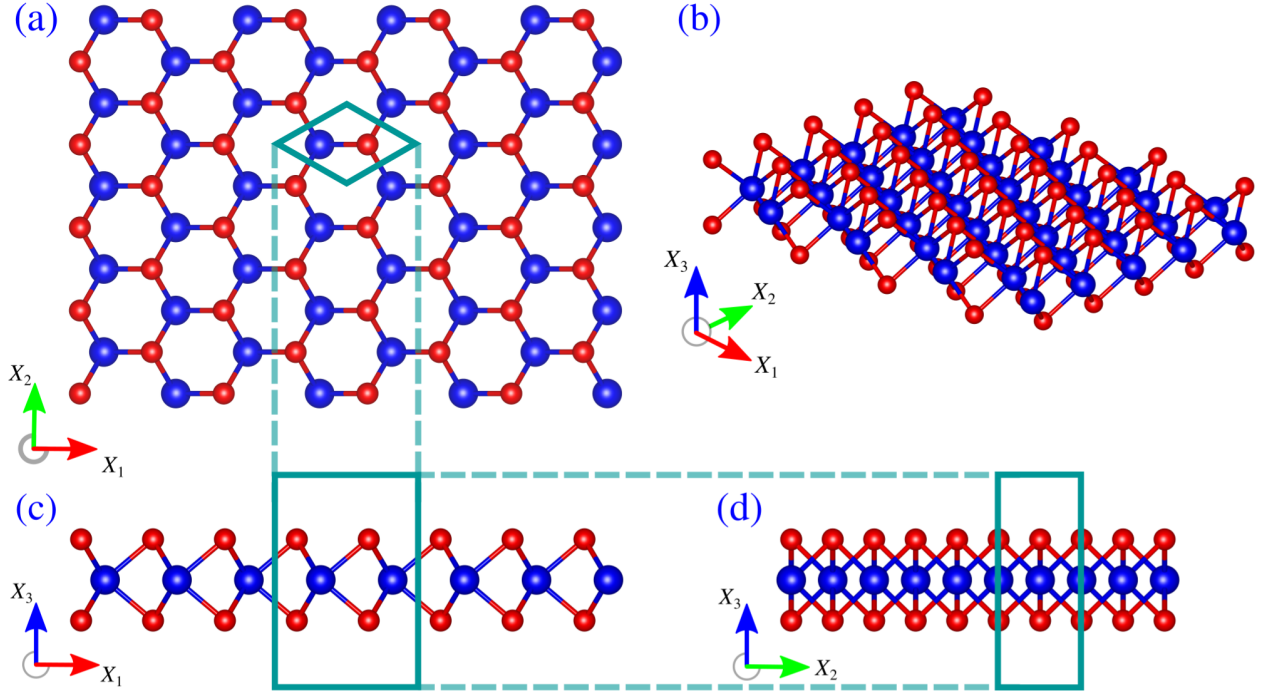}
  \caption{Different views showing the atomic arrangement of monolayer Transition Metal Dichalcogenides. (a) Top view where the rhombus represents the unit cell containing one transition metal atom (\moatomnew) and two chalcogen atoms (\satomnew) (one superimposed on the other and straight line denotes the bond)  (b) Isometric view, (c) Front view and (d) Side view.}
  \label{fig:mos2_monolayer}
\end{figure}

Transition metal dichalcogenides (TMDs) are emerging two-dimensional (2D) materials that exhibit exceptional electrical, optical, and chemical properties \citep{kolobov2016two,Bhimanapati2015acsnano,Akinwande2017EML}. 
TMDs are made of  transition metal   (Mo, W, etc.) and   chalcogen  (S, Se, Te, etc.)  atoms covalently bonded with each other, yielding a range of compositions   such as $\mathrm{MoS_{2}}$, $\mathrm{WS_{2}}$, $\mathrm{MoSe_{2}}$, $\mathrm{WSe_{2}}$, and $\mathrm{MoTe_{2}}$. The atomic arrangements of TMD monolayers are shown in figure~\ref{fig:mos2_monolayer}.  Monolayer TMDs  have a direct bandgap, and hence they can be used in electronics as transistors and in optics as emitters and detectors \citep{Splendiani2010NanoLetters,Sundaram2013NanoLetter}.  
Similar to other 2D materials, TMDs  behave like  nonlinear-elastic membranes having stiff in-plane and very-weak bending rigidity, thus often showing distributed buckles like wrinkles or folds. 
The electronic band-gap in TMDs can be reversibly tuned via mechanical strain \citep{He2013NanoLett,Conley2013NanoLett,Zhu2013PRB}.
Wrinkles or folds in TMDs also can reversibly alter their electronic, opto-electronic, and surface properties \citep{Castellanos-Gomez2013NanoLett}, which is promising for various high-impact applications. 
Therefore, predicting the mechanical deformations, such as wrinkles and folds, \citep{colas2019strength,zhao2019geometry} are important as it can allow us to alter their properties in very controlled manner. 

Despite the tremendous growth both in experiments \citep{dai2020radial,yang2018edge} as well as in small-scale simulations for TMDs, still there is no effective predictive modeling framework at the length-scale of experiments or  devices. 
The modeling techniques to simulate 2D materials like TMDs can be divided in two categories: atomistic models and continuum models.
The pure atomistic models, such as Molecular Dynamics (MD), and Density Functional Theories (DFT)  
are highly reliable but computationally prohibitive for experimental length-scales
\citep{zhao2018study,Jiang2013JAP,Li2012PRB,Jiang2013Nanotech,ansari2016dft,gupta2018understanding}.
On the contrary, the phenomenological continuum models 
are efficient and easy to use, but they ignore the underlying atomistic physics \citep{Castellanos-Gomez2011AdvMat,Bertolazzi2011ACSNano,cooper2013nonlinear}. 

To fill the gap between atomistic and continuum models, \emph{crystal--elasticity}-based models 
 have been developed for three-dimensional materials,  which encode the inter-atomic interactions in the continuum theory instead of using phenomenological constitutive models \citep{xin2012atomistic,martin1975many,cousins1978inner,hill1975elasticity,milstein}.  
In \emph{crystal-elasticity} theories the \emph{Cauchy-Born} rule is the key kinematic assumption that represents the lattice deformation through continuum strain measures for the bulk crystalline solids.
It is shown in \cite{Arroyo2002JMPS}, that the standard \emph{Cauchy-Born} rule is not applicable for curved 2D crystalline membranes.
This is due to the fact that the standard \emph{Cauchy-Born} rule incorrectly maps the deformed bonds to the tangent of the surface representing the membrane.   
To overcome this limitation of the standard \emph{Cauchy-Born} rule, a new kinematic rule, namely the \emph{exponential Cauchy--Born} rule is derived in \citep{Arroyo2002JMPS,Arroyo2003MOM,Arroyo2004IJNME}.
The \emph{exponential Cauchy-Born} rule extends the standard \emph{Cauchy-Born} rule for single-atom thick curved crystalline membranes, like graphene.

Several extensions to the \emph{Cauchy-Born} rule are reported in various works; for instance,
a modified \emph{Born} rule is proposed to develop a finite-deformation shell theory for the single-wall carbon nanotubes in \cite{Huang-JMPS-2008}.
A higher-order \emph{Cauchy-Born} rule is proposed for the curved crystalline membranes to obtain the deformation of single-wall carbon nanotubes in \cite{Guo-IJSS}.
Based on the higher-order \emph{Cauchy-Born} rule, a mesh-free computational framework is developed to simulate single-wall carbon nanotube under various loading conditions in \cite{liew-IJNME}.
Derivation of continuum theories for sheets, plates and rods from the atomistic models is summarized in \cite{weinan-prb}.
While these aforementioned models provide continuum representation for the bonded interactions, the van der Waals interactions between the multi-layers is calculated in a discrete manner.
A three-dimensional continuum model for multi-layered cyrstalline membranes that provides a continuum representation for the non-bonded interaction is developed in \cite{GhoshJMPS2013}.
{Quasi-continuum (QC) modeling is yet another approach developed to achieve the atomistic accuracy in a continuum setting  
 \citep{Tadmor1996PhilMagA,Tadmor1999PRB,Shenoy1999JMPS,quasi4}.
These QC models provide simultaneous resolution of  atomistic and continuum length-scales. In contrast to QC models for bulk crystalline solids,
a  QC formulation is proposed for curved crystalline membranes such as carbon nanotubes in \cite{quasi-CNT}.} 

At present, finite deformation \emph{crystal-elasticity} models are available only for single-atom-thick crystalline membranes, e.g. graphene. 
These models cannot be applied to multi-atom-thick 2D materials such as TMDs since they have covalent bonds located out of the middle surface of the membrane. 
In the present work, a  finite deformation \emph{crystal--elasticity} membrane model for monolayer TMDs is presented. 
{The deformed configuration of a TMD-membrane is represented through the deformation map of its middle surface and two stretches normal to the middle surface. Herein, the  middle surface represents the layer made of transition metal atoms and the two stretches provide  the location of the top and bottom  layers made of chalcogen atoms.
Based on this deformation map, the deformation of bond lengths and angles is obtained as a function of continuum  strains.
Finally, the continuum  constitutive relation for TMD membrane is derived  from its inter-atomic  potential.}
{This continuum membrane model is numerically implemented through a smooth finite element framework that uses the B-spline-based approximation.}
The relative shifts between the simple lattices constituting the complex lattice of TMDs is also considered in the formulation.
To demonstrate the efficiency and accuracy of the proposed model, the results obtained are compared against atomistic simulations and experimental results. 

The present paper is organized as follows.
The kinematics of the proposed membrane formulation is provided in section~\ref{sec:formulation}.
Section~\ref{sec:formulation} also includes the calculations of  deformed bond lengths and bond angles as a function of continuum strains.
Section~\ref{sec:inner} deals with the computation of strain energy density from the inter-atomic potentials for TMDs. This section explains the atomic arrangement of monolayer TMDs and the concept of relative shifts between two simple lattices.
The section also includes the non-bonded interactions.
{The description of the boundary value problem is provided in section~\ref{sec:bvp}.}
The numerical implementation to solve the boundary value problem is explained in section~\ref{sec:numerical_implementation}.
The validation of the numerical results is provided in section~\ref{sec:results}.
All the results presented in this section are for MoS$_2$ monolayer.
Finally, the conclusion and discussions are provided in the section~\ref{sec:conclusion}.

\section{Present Membrane Formulation}
\label{sec:formulation}
This section presents the kinematics used in the present formulation for multi-atom-thick 2D membranes.
Monolayer TMDs are multi-atom thick where the atoms lie on three different surfaces, as shown in figure~\ref{fig:mos2_monolayer}.
However, their thickness is very small compared to its other two dimensions and hence can be modeled as 2D membranes. 
In the present formulation, the deformation of each surface is represented in terms of the deformation of the middle surface and the thickness variation. 
Covalent bonds of TMDs are inclined to the surfaces, connecting atoms on the middle surface to the atoms on the other surfaces. 
These covalent bonds between atoms from different surfaces restrict them from deforming individually.
The interactions between the atoms lying on the same surface can also be represented through the deformation of the middle surface and normal stretches.
Therefore, we divide the bonds into two components i) tangential to the middle layer and ii) normal to the middle surface.
The deformation of the tangential component of the covalent bonds is obtained in terms of the deformation of the middle surface.
The deformation of the normal component of covalent bonds is obtained by using the stretch variables.
The details of the differential geometric concepts used to represent the deformation of the middle surface can be found in \cite{diff_geom,do2016differential}.
The notations used here follow  \citet{marsden1994mathematical,Arroyo2002JMPS} except that the points in the Euclidean space are represented through their position vectors. 

\subsection{Kinematics}
The kinematics used in the present formulation is presented in this section. 
This section includes the continuum deformation maps and strain measures. The schematics of the kinematics is shown in figure~\ref{fig:def_map}. 
The atoms in TMDs are arranged in three parallel layers in the undeformed configuration. The atoms in the deformed TMDs are also arranged in three layers, which are approximated as three locally parallel surfaces in the present formulation. 
In the present formulation, the middle surface corresponds to the transition metal atoms and other two surfaces (top and bottom) corresponds to chalcogen atoms.


\subsubsection{Deformation map for the TMD membranes}
\begin{figure}[h]
    \centering
    \includegraphics[width=\linewidth]{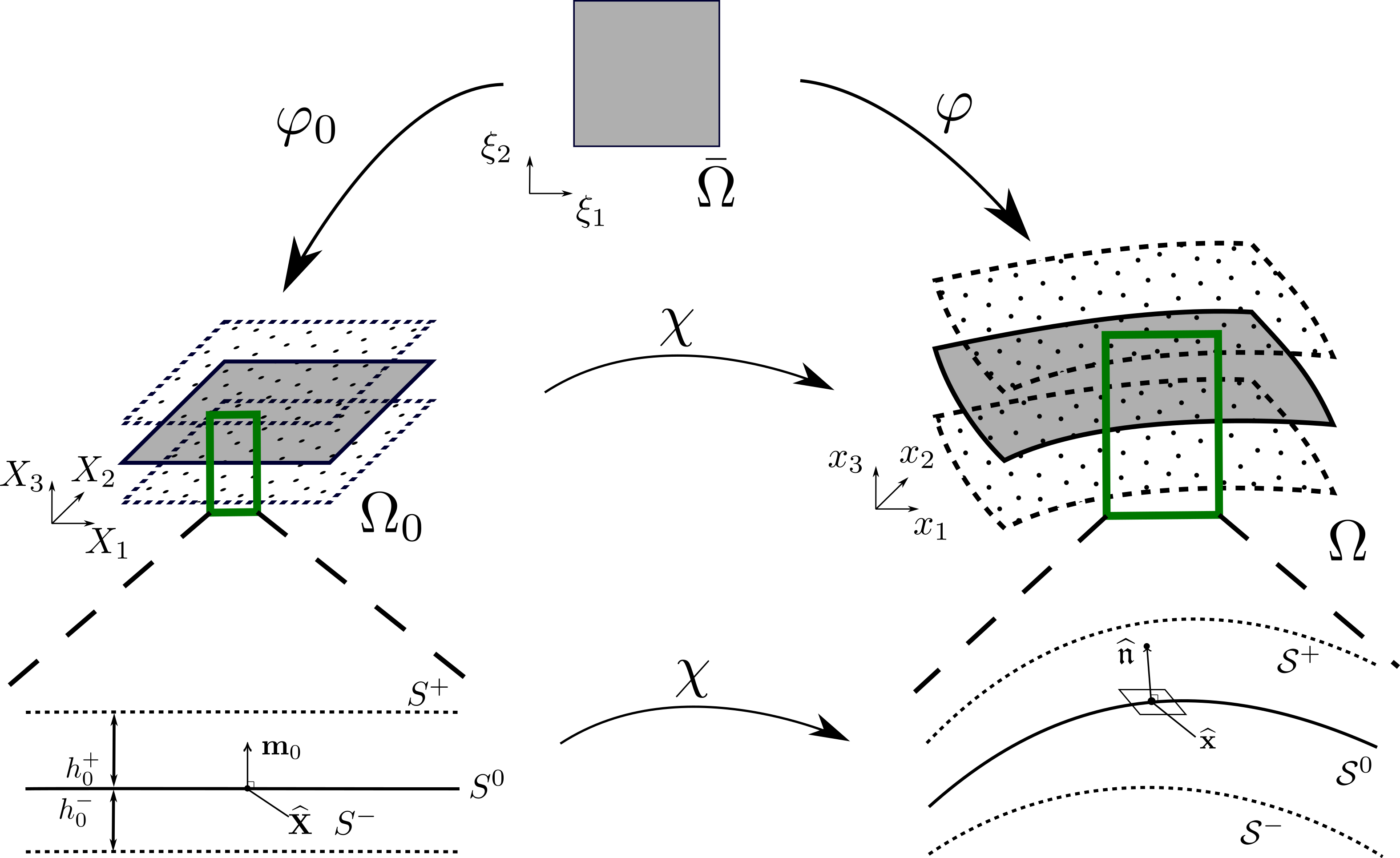}
    \caption{Kinematics showing the deformation map for monolayer TMDs. 
    }
    \label{fig:def_map}
\end{figure}
Let $\Omega_0 \subset \mathbb{R}^3$ be the undeformed configuration.
The undeformed configuration is considered as a collection of three parallel 2D surfaces $S^0$, $S^+$ and $S^-$ as,
\begin{equation}
    \Omega_0 = S^+ \cup S^0 \cup S^-
\end{equation}
where $S^0$, $S^+$ and $S^-$ be open sets in $\mathbb{R}^3$ representing the middle, top and bottom surface of TMDs, as shown in figure~\ref{fig:def_map}. The quantities associated with top and bottom surfaces are denoted with superscript ``$+$'' and ``$-$'', respectively.
The co-ordinate of any point in $\Omega_0$ can be denoted as $\mathbf{X} = \{X_1,X_2,X_3\}$, and the corresponding basis is given by $\mathscr{B}_0 = \{ \mathbf{I}_1, \mathbf{I}_2, \mathbf{I}_3\}$, such that any point on the undeformed configuration is given by
\begin{equation}
    \Omega_0 = \{ \mathbf{X} = (\widehat{\mathbf{X}} \times X_3) \in S^0 \times \{-h_0,0,h_0\} \}
    \label{eq:undef_config}
\end{equation}
The three surfaces $S^0$, $S^+$ and $S^-$ are parallel to the $X_1\,X_2$ plane and have values $0$, $+h_0$ and $-h_0$ for the $X_3$ coordinate. Here, $2h_0$ represents the distance between the top and bottom layers of atoms. 
Points on $S^0$ can be defined as $\{\widehat{\bf{X}},0\}$ where $\widehat{\bf{X}} = \{X_1,X_2\}$.
On the surface $S^0$ at each $\widehat{\bf{X}}$ a tangent plane $T_{\widehat{\mathbf{X}}}S^0$ can be defined, where $\mathbf{m}$ is the unit normal defined on the tangent plane at that point. For the chosen planar undeformed configuration, $\bf{m}$ is aligned with $\mathbf{I}_3$. 
The basis for the tangent space $T_{\widehat{\mathbf{X}}}S^0$ is $\mathscr{C}_0 = \{\mathbf{I}_1,\mathbf{I}_2\}$. 
For the chosen planar undeformed configuration, the tangent space $T_{\widehat{\mathbf{X}}}S^0$ and its convected bases $\mathscr{C}_0$ coincide with the surface $S^0$ and the first two components of the bases $\mathscr{B}_0$, respectively.
Following the undeformed configuration defined in  equation~\ref{eq:undef_config}, 
the co-ordinate of any point on the $S^+$ and $S^-$ surfaces are represented as $\mathbf{X}^+ = \{\widehat{\mathbf{X}},h_0\}$ and $\mathbf{X}^- = \{\widehat{\mathbf{X}},-h_0\}$, respectively. {Here, $\mathbf{X}^+$ and $\mathbf{X}^-$ represents points on surfaces $S^+$ and $S^-$, respectively.} \\


\textit{Deformation map of the middle surface}

Let $\varphi_0$ be the map from the parametric domain $\bar{\Omega} \subset \mathbb{R}^2$ to the middle surface $S^0$ of undeformed configuration $\Omega_0$. Let the co-ordinate of the parametric domain be expressed as $\{\xi_1, \xi_2\}$ and the corresponding basis set is $\bar{\mathscr{B}} = \{\bar{\mathbf{I}}_1, \bar{\mathbf{I}}_2\}$.
 Another map $\varphi$ takes the parametric domain $\bar{\Omega}$ to the deformed surface $\mathcal{S}^0$ lying in the deformed configuration $\Omega$. 
 Let $\mathcal{S}^0 \subset \Omega$ be a smooth, open and orientable surface representing the deformed middle surface, $\widehat{\mathbf{n}}$ be the unit normal and $T_{\widehat{\mathbf{x}}}\mathcal{S}^0$ be the tangent space defined at $\widehat{\mathbf{x}}$ on $\mathcal{S}^0$.
 The map from the $S^0$ to the $\mathcal{S}^0$ can be expressed as  
 \begin{equation}
    \Phi =  \varphi \circ \varphi_0^{-1} : \widehat{\mathbf{X}} \to \widehat{\mathbf{x}}\: , \; \widehat{\mathbf{X}} \in S^0 \: \text{and} \: \widehat{\mathbf{x}} \in \mathcal{S}^0
\label{eq:mid_map}
\end{equation} 
{Note that the point mappings (e.g. $\varphi, \varphi_0, \Phi, \chi$) are denoted by lightface symbols. Whereas the boldface symbols are used for vectors and tensors. However, to represent the points in the Euclidean space, their position vectors are used (e.g.  $\mathbf{X}$, ${\widehat{\mathbf{X}}}$, $\mathbf{x}$).}

\textit{Normal stretches} 

The scalar field $\lambda(\widehat{\mathbf{X}},X_3)$ is used to represent the thickness variation in the deformed configuration and can be expressed as
\begin{equation}
    \lambda(\widehat{\mathbf{X}},X_3) = \begin{cases} \lambda^+(\widehat{\mathbf{X}}) \in \mathbb{R}^+ \: & \text{for} \: X_3 = h_0 \\
                            0         \: &  \text{for} \: X_3 = 0 \\
                            \lambda^-(\widehat{\mathbf{X}}) \in \mathbb{R}^+ \: & \text{for} \: X_3 = -h_0
    \end{cases}  
    \label{eq:lambda_map}
\end{equation}
Here, $\lambda^{+}(\widehat{\mathbf{X}}) \, h_0 = h^{+}(\widehat{\mathbf{X}})$ and $\lambda^{-}(\widehat{\mathbf{X}}) \, h_0 = h^{-}(\widehat{\mathbf{X}})$ are the deformed thicknesses at $\widehat{\mathbf{X}}$ in the middle surface $\mathcal{S}^0$ along $\widehat{\mathbf{n}}$ and $-\widehat{\mathbf{n}}$ respectively. \\

\textit{Total deformation map} 

Let $\chi$ be the deformation map that takes $\Omega_0$ to the deformed configuration $\Omega = \chi(\Omega_0)$. 
The map $\chi$ can be expressed as,
\begin{equation}
    \Omega = \chi(\Omega_0) = \{ \chi : \mathbf{X} \to \mathbf{x} = \Phi(\widehat{\mathbf{X}}) + h_0\, \lambda(\widehat{\mathbf{X}},X_3)\, \widehat{\mathbf{n}}(\Phi(\widehat{\textbf{X}})) \},\quad \mathbf{X} \in \Omega_0 \: , \: \mathbf{x} \in \Omega
    \label{eq:def_config}
\end{equation}
and the deformed configuration can also be expressed as the combination of three surfaces as,
\begin{equation}
    \Omega = \mathcal{S}^+ \cup \mathcal{S}^0 \cup \mathcal{S}^-
\end{equation}
where $\mathcal{S}^+$ and $\mathcal{S}^-$ represents the deformed top and bottom surfaces. Thus, the map $\chi$ represents any point in the deformed configuration $\Omega$ only through the deformation of the middle surface, $\Phi$, and two normal stretches, $\lambda^+$ and $\lambda^-$.
The coordinate of any point in the deformed configuration can be represented as $\mathbf{x} = \{x_1,x_2,x_3\} \in \mathbb{R}^3$ in the standard basis set $\mathscr{B} = \{\mathbf{i}_1,\mathbf{i}_2,\mathbf{i}_3\}$. 
To obtain the deformed middle, top and bottom surfaces, the map $\chi$ is restricted to $X_3=0$, $X_3=h_0$, and $X_3=-h_0$ respectively, such that,
\begin{align}
      \chi(\widehat{\mathbf{X}} \in S^0 \subset \Omega_0) & = \widehat{\mathbf{x}} \in \mathcal{S}^0 \subset \Omega, \nonumber \\
      \chi(\mathbf{X}^+ \in S^+ \subset \Omega_0) & = \mathbf{x}^+ \in \mathcal{S}^ +\subset \Omega, \nonumber \\ 
      \chi(\mathbf{X}^- \in S^- \subset \Omega_0) & = \mathbf{x}^- \in \mathcal{S}^- \subset \Omega. 
\end{align}
The next subsection follows \cite{Arroyo2002JMPS} to represent the strains for the middle surface.  

\subsubsection{Strains for the middle surface}

Following the deformation map for the middle surface, $\Phi$
, $T_{\widehat{\mathbf{X}}}S^0$ defines the tangent plane in the undeformed and $T_{\widehat{\mathbf{x}}}\mathcal{S}^0$ defines the tangent plane in the deformed configurations. 
For the current case where the undeformed configuration is planar, the tangent space $T_{\widehat{\mathbf{X}}}S^0$ and the surface $S^0$ are coincident. The convected basis $\mathscr{C} = \{\bf{g}_1,\bf{g}_2\}$, for the tangent space $T_{\widehat{\mathbf{x}}}\mathcal{S}^0$ of the deformed configuration can be defined as,
\begin{equation}
    \mathbf{g}_{\alpha} = \frac{\partial \varphi^a}{\partial \xi^{\alpha}} \mathbf{i}_a
    \label{eq:conv_basis}
\end{equation}
The deformation gradient of the middle surface defined in the basis $\mathscr{C}_0 - \mathscr{C}$ can be expressed as,
\begin{eqnarray}
\widehat{\mathbf{F}} & = T\Phi = T\varphi \circ T\varphi_0^{-1} \nonumber \\
                     & = [T\varphi]_{\mathscr{C}\bar{\mathscr{B}}} [T\varphi_0^{-1}]_{\bar{\mathscr{B}}\mathscr{C}_0} \nonumber \\
                     & = [T\varphi_0^{-1}]_{\bar{\mathscr{B}}\mathscr{C}_0}
                     \label{eq:def_grad_mid}
\end{eqnarray}
The first part of the deformation gradient, $[T\varphi]_{\mathscr{C}\bar{\mathscr{B}}}$, becomes identity as the information about the deformation is contained in the convected basis vectors defining the tangent plane.\\

\textit{Metric Tensor and Green Strain Tensor} 

The metric tensor containing the information of the deformation can be expressed as
\begin{equation}
    [\mathbf{g}] = \bigg[ \begin{array}{cc} g_{11} & g_{12} \\
                                       g_{21} & g_{22} \end{array} \bigg]
                                       \label{eq:metric_tensor}
\end{equation}
where $g_{\alpha \beta} = \langle \mathbf{g}_{\alpha} | \mathbf{g}_{\beta} \rangle$ ($\langle \cdot | \cdot \rangle$ denotes the Euclidean norm). 
The right Cauchy-Green strain tensor for the middle surface ($\widehat{\mathbf{C}}$) in the undeformed configuration can be obtained as the pull-back of the metric tensor, $[\widehat{\mathbf{C}}]_{\mathscr{C}_0} = \Phi^* \mathbf{g}$, and expressed as, 
\begin{align}
[\widehat{\mathbf{C}}]_{\mathscr{C}_0} & = \Phi^* [\mathbf{g}] \nonumber \\
                                       & = [\widehat{\mathbf{F}}]_{\mathscr{C} \mathscr{C}_0}^T[\mathbf{g}]_{\mathscr{C}} [\widehat{\mathbf{F}}]_{\mathscr{C} \mathscr{C}_0} 
    \label{eq:cauchy_green}
\end{align} \\

\textit{Curvature Tensor} 

The unit normal at each point on the middle surface can be expressed as,
\begin{equation}
\widehat{\mathbf{n}} = \frac{\mathbf{g}_1 \times \mathbf{g}_2}{||\mathbf{g}_1 \times \mathbf{g}_2||}
\end{equation}
The matrix elements of curvature tensor in the convected basis can be expressed as
\begin{equation}
    k_{\alpha \beta} = \langle \widehat{\mathbf{n}} | \mathbf{g}_{\alpha,\beta} \rangle
    \label{eq:curv_tensor_def}
\end{equation}
where $\mathbf{g_{\alpha,\beta}} = \frac{\partial \mathbf{g}_{\alpha}}{{\partial \xi_{\beta}}}$.
Similar to metric tensor, the pull-back of the curvature tensor in the undeformed configuration can be expressed as
\begin{align}
[\widehat{\mathscr{K}}]_{\mathscr{C}_0}  & =  \Phi^* [\mathbf{k}] \nonumber \\
                                         & =  [\widehat{\mathbf{F}}]_{\mathscr{C} \mathscr{C}_0}^T[\mathbf{k}]_{\mathscr{C}} [\widehat{\mathbf{F}}]_{\mathscr{C} \mathscr{C}_0} 
\end{align}
Corresponding to the curvature tensor, $\widehat{\mathscr{K}}$, and the Cauchy-Green tensor, $\widehat{\mathbf{C}}$, the principal curvatures, $k_1$ and $k_2$, and the principal curvature vectors, $\mathbf{v}_1$ and $\mathbf{v}_2$, are the eigenvalues and eigenvectors of the Weingarten map.
Further details of the Weingarten map are given in \ref{app:surf}.

The principal curvatures can be obtained by solving the generalized eigenvalue problem 
\begin{equation}
    [\mathbf{k}]_{\mathscr{C}} [\mathbf{v}]_{\mathscr{C}} = k [\mathbf{g}]_{\mathscr{C}} [\mathbf{v}]_{\mathscr{C}}
\end{equation}
Here, $k$ represents a principal curvature and $\mathbf{v}$ represents the corresponding principal direction in the convected basis. The principal curvatures and the corresponding principal directions can be obtained in the undeformed configuration as, 
\begin{equation}
    [\widehat{\mathscr{K}}]_{\mathscr{C}_0} [\mathbf{V}]_{\mathscr{C}_0} = k [\widehat{\mathbf{C}}]_{\mathscr{C}_0} [\mathbf{V}]_{\mathscr{C}_0} 
    \label{eq:eig}
\end{equation} 

Further details on the eigenvalue problem and the derivative calculation of the principal curvatures and the principal directions with respect to the $\widehat{\mathbf{C}}$ and $\widehat{\mathscr{K}}$ are provided in~\ref{ap:princi_curv_ders}.
\subsection{Calculation of lattice deformation for TMDs}
In the present formulation, we use the \emph{crystal-elasticity} theory to represent the energy of the deformed lattice of TMDs in terms of the continuum strain measures. \emph{Crystal-elasticity} uses  the \emph{Cauchy-Born} rule for this purpose that links the deformed and undeformed lattice vectors through the continuum deformations. Following the  \emph{Cauchy-Born} rule, any deformed lattice vector $\mathbf{a}$ can be obtained as $\mathbf{a} = \mathbf{F}\,\mathbf{A}$, where $\mathbf{F}$ represents the deformation gradient and $\mathbf{A}$ is the undeformed bond vector.
Since, The deformation gradient maps a vector to the tangent plane of a curved membrane whereas the deformed bonds of 2D crystalline membranes are the chords to the surface and not the tangents.
Therefore, the \emph{Cauchy-Born} rule cannot be directly applied to the purely two-dimensional (2D) crystalline membranes (surfaces).  
The \emph{exponential Cauchy--Born} (ECB) rule resolves this issue by projecting the deformed bond vector on to the chord \citep{Arroyo2002JMPS}. 
Since graphene is a single atom thick, it is a purely 2D membrane, hence it is amenable to the ECB rule. This fact has been validated in  \cite{Arroyo2002JMPS,Arroyo2004IJNME}. 
The ECB rule can not be directly used for TMDs since they are not purely 2D membranes but have finite thickness.  
On the contrary,  due to its membrane characteristics, the use of the \emph{Cauchy-Born} rule will be inefficient. 
Therefore, to efficiently model TMDs a new \emph{crystal-elasticity} membrane model is needed that considers the  effect of thickness in its deformation. 
In the present formulation, the deformed configuration is represented through the  deformation of three surfaces and their relative distances (see figures~\ref{fig:mos2_monolayer}\textcolor{blue}{(c,d)} and~\ref{fig:def_map}).
To compute the deformation of a bond inclined to the middle surface, it is first decomposed into a tangential and a normal component to the middle surface.
The component of the bond tangential to the middle surface is denoted by $\mathbf{A}_t$, and the component of the bond along the normal to the middle surface is denoted by $\mathbf{A}_n$.
The deformation of the tangential component of the bond is obtained by the \emph{exponential Cauchy-Born} rule applied to the deformation of the middle surface ($\Phi$) 
whereas the deformation of the normal component of the bond is obtained by using the stretches ($\lambda^+$ and $\lambda^-$) normal to the middle surface.

\subsubsection{Deformation of component of bonds tangential to the middle surface} 
In the present work, the deformation of the tangential component of a bond is obtained following the ECB rule.
The tangential component of an undeformed bond vector is obtained as $\mathbf{A}_t = \mathbb{P}_ {\mathbf{m}_0} \mathbf{A}$. 
Here, $\mathbf{A}$ is the bond (inclined to the middle surface $S^0$) in the undeformed configuration and  $\mathbb{P}_{{\mathbf{m}}_0}$ is the perpendicular projection operator that projects any vector in $\mathbb{R}^3$ on the tangent plane of the middle surface  having the $\mathbf{m}_0$ as unit normal. 
The perpendicular projection operator can be expressed as
\begin{equation}
    \mathbb{P}_{\mathbf{m}_0} = \mathcal{I} - \mathbf{m}_0 \otimes \mathbf{m}_0
    \label{eq:app_projection}
\end{equation}
where $\mathcal{I}$ represents the identity operator and can be expressed as, $\mathcal{I} = \mathbf{I}_i \otimes \mathbf{I}_i$.
Further details on the perpendicular projection operator are provided in~\ref{app:perpendicular_proj}.
Using the \emph{exponential Cauchy--Born} rule, the deformed lattice vector ($\mathbf{a}$) corresponding to the undeformed lattice vector ($\mathbf{A}$) can be obtained as $\mathbf{a} = \text{exp}_{\mathbf{X}}\widehat{\mathbf{F}}\mathbf{A}$. Here, the $\text{exp}_{\mathbf{X}}$ denotes the \emph{exponential map} to a nonlinear surface at   $\mathbf{X}$ 
\footnote{The exponential map as defined by \cite{morgan1993}:
``The exponential map $\text{exp}_{\mathbf{p}}$ at a point $\mathbf{p}$ in $M$ maps the tangent space $T_{\mathbf{p}}M$ into $M$ by sending a vector $\mathbf{v}$ in the $T_{\mathbf{p}}M$ to a point in $M$ a distance $|\mathbf{v}|$ along the geodesic from $\mathbf{p}$ in the direction of $\mathbf{v}$.''
For our case $M$ is the middle surface $S_0$ and $\mathbf{v}$ is the vector that is obtained after applying the Cauchy--Born rule on the undeformed lattice vector. \\
In a simplistic way a geodesic is the shortest curve between two points lying on a non-linear surface. A more precise definition of geodesic is given in \cite{diff_geom} as: 
``A curve $\gamma$ on a surface is called geodesic if $\ddot{\gamma}$ is zero or perpendicular to the tangent plane of the surface at the point $\gamma(t)$, i.e., parallel to its unit normal, for all values of the parameter $t$.'' 
}. 
The ECB rule is applied on the tangential component of the undeformed bond $\mathbf{A}_t$ to obtain the deformed bond $\mathbf{a}_t$ following \cite{Arroyo2002JMPS,Arroyo2004IJNME}  as,
\begin{equation}[\mathbf{a}_t]_{\tilde{\mathscr{B}}}=\left\{\begin{array}{l}
a^{1} \\
a^{2} \\
a^{3}
\end{array}\right\}=\left\{\begin{array}{c}
w^{1} \mathscr{Q}\left(k_{1} w^{1}\right) \\
w^{2} \mathscr{Q}\left(k_{2} w^{2}\right) \\
\frac{k_{1}\left(w^{1}\right)^{2}}{2} \mathscr{Q}^{2}\left(k_{1} w^{1} / 2\right)+\frac{k_{2}\left(w^{2}\right)^{2}}{2} \mathscr{Q}^{2}\left(k_{2} w^{2} / 2\right)
\end{array}\right\}
\label{eq:proj_def_bond}
\end{equation}
Here, $\mathbf{A}_t$ lies on the undeformed middle surface $S^0$ and $\mathbf{a}_t$ lies as the chord of the middle surface 
$\mathcal{S}^0$ in the deformed configuration. 

\subsubsection{Deformation of the component of bonds normal to the middle surface} 

The deformed normal component of the bonds are computed by using 
two scalar stretch fields $\lambda^{+}(\widehat{\mathbf{X}})$ and $\lambda^{-}(\widehat{\mathbf{X}})$, which define the stretches above and below the middle surface.
Following the map defined in equation~\ref{eq:def_config}, the deformed normal components of bonds can be obtained as
\begin{align}
\mathbf{a}_n^+ = & 
\lambda^+(\widehat{\mathbf{X}})\mathbf{A}_n^+ \nonumber \\ 
\mathbf{a}_n^- = & 
\lambda^-(\widehat{\mathbf{X}})\mathbf{A}_n^- 
\label{eq:perp_def_bond}
\end{align}
where $\mathbf{a}_n^+$ and $\mathbf{a}_n^-$ denote the deformed   normal components of the undeformed bonds ($\mathbf{A}_n^+$ and $\mathbf{A}_n^-$) above and below the middle surface.

The sum of the two components $\mathbf{a}_t$ and $\mathbf{a}_n^{+}$ is the deformed Mo--S bond vector for an S--atom on the top surface. Similarly, the deformed Mo--S bond for an S--atom on the bottom surface is given by the sum of $\mathbf{a}_t$ and $\mathbf{a}_n^-$.
The $\mathbf{a}_t$ is obtained in the basis $\tilde{\mathscr{B}} = \{\mathbf{v}_1, \mathbf{v}_2, \mathbf{v}_1 \times \mathbf{v}_2 \}$ (following equation~\ref{eq:proj_def_bond}), where $\mathbf{v}_i$ are the principal curvature vectors.
The basis for $\mathbf{a}_n$ is $\mathcal{B}$. 
Hence, the final deformed lattice vectors ($\mathbf{a}$) corresponding to the undeformed lattice vector ($\mathbf{A}$) can be obtained as a function of strain measures of the middle surface and two scalar fields defining the thickness variation such as,
\begin{equation}
    \mathbf{a} = f(\widehat{\mathbf{C}},\widehat{\mathscr{K}},\lambda^+,\lambda^-,\mathbf{A})
\end{equation}
Similarly, the deformed bond angles can be obtained in terms of the strain measures as
\begin{equation}
    \mathbf{\theta} = f(\widehat{\mathbf{C}},\widehat{\mathscr{K}},\lambda^+,\lambda^-,\mathbf{A},\mathbf{B})
\end{equation}
where $\mathbf{B}$ is another bond vector in the undeformed configuration.
The energy of the deformed crystal-lattice are computed through an inter-atomic potential in terms of the deformed bond lengths and deformed bond angles.
Therefore, the continuum energy of the deformed body can be represented in terms of the strains.
For TMDs, the most widely used inter-atomic potential is the Stillinger-Weber (SW) potential \citep{PhysRevB.31.5262}. 
The details of the computation of the strain energy using the SW potential is provided in the next section.
Henceforth, for the inter-atomic potential, the Molybdenum Disulfide (MoS$_2$) is   considered as a representative TMD. Nevertheless, the present approach can be used for any TMD.

\section{Constitutive Model}
\label{sec:inner}
In this section, the lattice structure of MoS$_2$ is described.
The lattice structure of MoS$_2$ is a complex lattice structure as it constitutes two simple Bravais lattices. 
These two simple lattices rigidly shift relative to each other during deformation. This relative shift is presented in this section. 
The formulation of the hyperelastic continuum constitutive model based on the Stillinger-Weber inter-atomic potential is also described.  
The steps including the non-bonded interaction are also presented in this section.

\subsection{Lattice structure of MoS$_2$}
\begin{figure}[h!]
    \centering
    
    \includegraphics[width=1.0\linewidth]{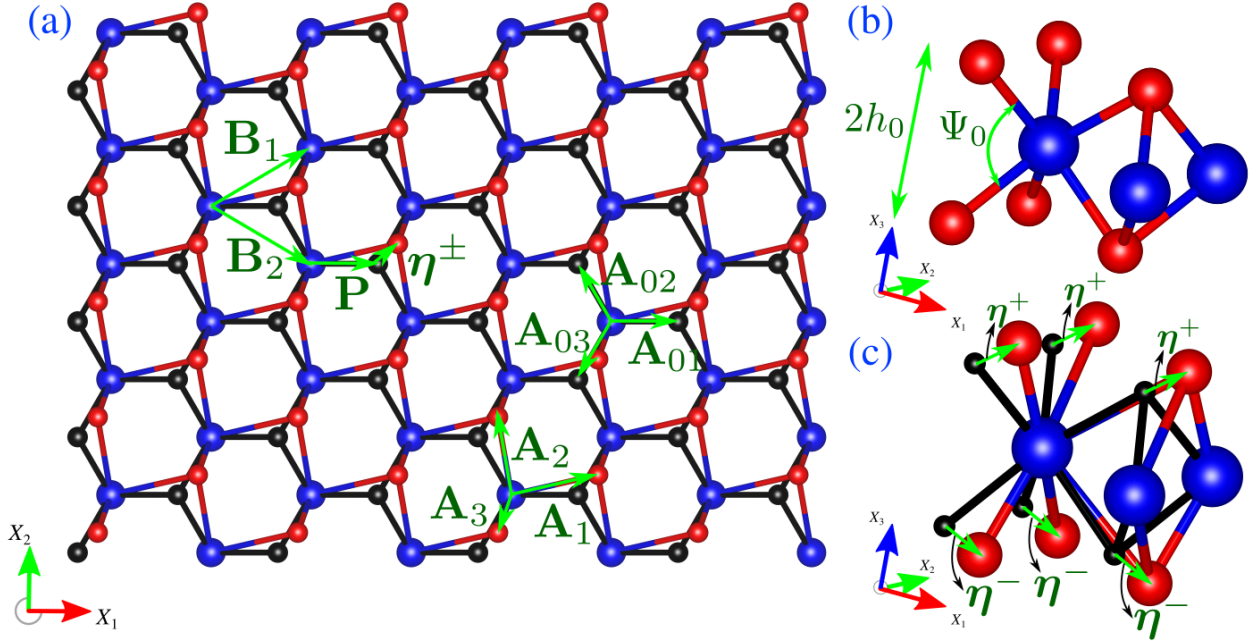}

    \caption{(a) Top view of the MoS$_2$ monolayer. $\mathbf{B}_1$ and $\mathbf{B}_2$ represents the two basis vectors to define the MoS$_2$ crystal lattice projected on the $X_1-X_2$ plane. $\mathbf{P}$ represents the  shift vector and $\bm{\eta}^{+}$ and $\bm{\eta}^-$ represents the \emph{inner displacement} vectors for the S--atoms located above and below the middle Mo-atom. Three Mo-S bonds from the central Mo-atom before \emph{inner displacements} are defined by $\mathbf{A}_{01}$,$\mathbf{A}_{02}$ and $\mathbf{A}_{03}$, whereas bonds after \emph{inner displacements} are defined as $\mathbf{A}_{1}$, $\mathbf{A}_2$ and $\mathbf{A}_3$.
    A magnified view of the unit cell along with the nearest neighborhood atoms (b) without \emph{inner displacements} and 
    (c) with \emph{inner displacements}.}
    \label{fig:lattice_inner}
\end{figure}
Monolayer MoS$_2$ exhibits a trigonal prismatic crystal structure. 
Figure~\ref{fig:lattice_inner}{\color{blue}a} shows the top view of a monolayer MoS$_2$ lattice where the green trapezoid represents the unit cell.
The atoms lying in the unit cell and their first nearest neighboring atoms  are given in figure~\ref{fig:lattice_inner}{\color{blue}b}.
The bond vectors and bond angles in the undeformed configuration of the unit cell can be defined through two variables, which are the height of the unit cell $2h_0$ and the angle $\Psi_0 = \angle (S2-Mo1-S5)$, as shown in figure~\ref{fig:lattice_inner}.
The length of the Mo-S bonds can be expressed as a function of $h_0$ and $\Psi_0$ as,
\begin{equation}\label{eq:Latticeparameters}
b_0 = h_0 \text{sin}^{-1}\bigg(\frac{\Psi_0}{2}\bigg)
\end{equation}
Considering Mo1 atom as the central atom positioned at $(0,0,0)$, the bond vectors between Mo1 and S-atoms of top surface in the undeformed configuration can be written as,


\begin{equation*}
[\mathbf{A}]_{\text{Mo1-S1}} = 
\left\{ \begin{array}{c} 
-b_0 \text{cos}(\frac{\Psi_0}{2})\text{sin}(\frac{\pi}{3}) \\
\\
b_0 \text{cos}(\frac{\Psi_0}{2})\text{cos}(\frac{\pi}{3}) \\
\\
h_0
\end{array}  \right\}, \:
[\mathbf{A}]_{\text{Mo1-S2}} = 
\left\{ \begin{array}{c} 
-b_0 \text{cos}(\frac{\Psi_0}{2})\text{sin}(\frac{\pi}{3}) \\ 
\\
-b_0 \text{cos}(\frac{\Psi_0}{2})\text{cos}(\frac{\pi}{3}) \\
\\
h_0
\end{array}  \right\}
\end{equation*}

\begin{align}
\text{and,} \: \: \: 
[\mathbf{A}]_{\text{Mo1-S3}} & = 
\left\{ \begin{array}{c} 
b_0 \text{cos}(\frac{\Psi_0}{2}) \\
\\
0 \\
\\
h_0
\end{array} \right\}
\label{eq:mo_s_top}
\end{align}
Similarly, the bond vectors between the central Mo--atom and S--atoms of bottom surfaces can be expressed as

\begin{equation*}
[\mathbf{A}]_{\text{Mo1-S4}} = 
\left\{ \begin{array}{c} 
-b_0 \text{cos}(\frac{\Psi_0}{2})\text{sin}(\frac{\pi}{3}) \\
\\
b_0 \text{cos}(\frac{\Psi_0}{2})\text{cos}(\frac{\pi}{3}) \\
\\
-h_0
\end{array}  \right\} , \:
[\mathbf{A}]_{\text{Mo1-S5}} = 
\left\{ \begin{array}{c} 
-b_0 \text{cos}(\frac{\Psi_0}{2})\text{sin}(\frac{\pi}{3}) \\
\\
-b_0 \cos(\frac{\Psi_0}{2})\cos(\frac{\pi}{3}) \\
\\
-h_0
\end{array}  \right\}
\end{equation*}
\begin{align}
\vspace{5pt}
\text{and,} \: \: \:
[\mathbf{A}]_{\text{Mo1-S6}} & = 
\left\{ \begin{array}{c} 
b_0 \cos(\frac{\Psi_0}{2}) \\
\\
0 \\
\\
-h_0
\end{array}  \right\}
\label{eq:mo_s_bottom}
\end{align}

The Mo-Mo bond vectors can be expressed as
\begin{equation*}
[\mathbf{A}]_{\text{Mo1-Mo2}} = 
\left\{ \begin{array}{c} 
\bigg[1+ b_0 \text{cos}(\frac{\Psi_0}{2})\bigg]\text{cos}(\frac{\pi}{3})  \\
\\
b_0\text{cos}(\frac{\Psi_0}{2})\text{sin}(\frac{\pi}{3})\\
\\
0
\end{array}  \right\} \: \: \text{and} \: \:
[\mathbf{A}]_{\text{Mo1-Mo3}} = 
\left\{ \begin{array}{c} 
\bigg[1+ b_0 \text{cos}(\frac{\Psi_0}{2})\bigg]\text{cos}(\frac{\pi}{3})  \\
\\
-b_0\text{cos}(\frac{\Psi_0}{2})\text{sin}(\frac{\pi}{3})\\
\\
0
\end{array}  \right\} 
\label{eq:mo-mo}
\end{equation*}


\subsection{Deformation of lattice}
The lattice structure of MoS$_2$ comprises of two simple interpenetrating lattices, hence it is a complex lattice.
In complex lattice structures, there are more than one basis nuclei, and hence the crystal structure can be considered as combinations of inter-penetrating lattices. 
In the case of MoS$_2$ all the Mo-atoms can be defined by basis vectors $\mathbf{B}_1$ and $\mathbf{B}_2$. Whereas, the positions of S-atoms cannot be defined only by the positions of the  Mo--atoms and the basis vectors. Therefore,  an additional shift vector $\mathbf{P}$ is used to obtain the positions of the S-atoms.
The bond vectors can be expressed as $\mathbf{A} = n^i \mathbf{B}_i + m\mathbf{P}$. Here, $n^i$ is an integer and a summation is implied over the repeated index $i$, where $i = \{1,2\}$ and $m$ takes the value of 0 and 1 for Mo--atoms and S--atoms respectively. 

Kinematic variables are commonly used to define the relative shifts between the simple lattices \citep{Tadmor1999PRB,Arroyo2002JMPS}. These relative shifts are also referred to as  \emph{inner displacements}. Hereafter, relative shifts and \emph{inner displacements} will be used interchangeably.  In the case of MoS$_2$, two sets of \emph{inner displacements} are considered for S--atoms, one affects the lattice arrangement of top S--atoms and the other affects the lattice arrangement of bottom S--atoms. 
Hence, the \emph{inner displacements} ($\bm{\eta}$) is a set consisting the \emph{inner displacements} for top ($\bm{\eta}^+$) and bottom ($\bm{\eta}^-$) S--atoms as $\bm{\eta} = \{\bm{\eta}^+,\bm{\eta}^-\}$.
In figure \ref{fig:lattice_inner}, the lattice structure is shown where all the black and red dots are the S-atoms before and after the relative shift, respectively.
The shift vector $\mathbf{P}$ and the \emph{inner displacements} $\bm{\eta}$ are shown in figure~\ref{fig:lattice_inner}. 
The equilibrium lattice is obtained by minimizing the energy with respect to the \emph{inner displacements} for a given continuum deformation. 

Note that $\bm{\eta}$ is defined in the reference configuration. 
Since the Mo-S bond vectors don't lie on the Mo-surface, the \emph{inner displacements} $\bm{\eta}$
are 3D vectors. 
Therefore,  $\bm{\eta}$ includes the relative shift along the thickness as well, as shown in figure~\ref{fig:lattice_inner}{\color{blue}c}.
Upon incorporation of the \emph{inner displacements},  the bond vectors are given by
\begin{equation}
    \mathbf{A}^{\pm} = n_i\mathbf{B}_i + m(\mathbf{P} + \bm{\eta}^{\pm}),
    \label{eq:bonds_after_shift}
\end{equation}
here $\mathbf{A}^{+}$ and $\mathbf{A}^-$ corresponds to the S--atoms at top and bottom surfaces, respectively. Here
$\bm{\eta}^{+}$ and $\bm{\eta}^-$ are the \emph{inner displacements} corresponding to $\mathbf{A}^{+}$ and $\mathbf{A}^-$.

The deformed tangential components of $\mathbf{A}^{+}$ and $\mathbf{A}^-$ are obtained following equation~\ref{eq:proj_def_bond} as,
\begin{equation}
           \mathbf{a}_t^{\pm}   =  \mathrm{exp}_{\chi(\widehat{\mathbf{X}})}\circ \widehat{\mathbf{F}} \circ \mathbb{P}_{\mathbf{m}}\big((n_i\mathbf{B}_i + m(\mathbf{P} + \bm{\eta}^{\pm}))\big)
           \label{eq:tangential_def_bond}
\end{equation}
and the deformed normal components of $\mathbf{A}^{+}$ and $\mathbf{A}^-$ are obtained following equation~\ref{eq:perp_def_bond} as,
\begin{equation}
               \mathbf{a}_n^{\pm}   =  \lambda^{\pm}(\widehat{\mathbf{X}})[\mathcal{I}-\mathbb{P}_m]\big((n_i\mathbf{B}_i + m(\mathbf{P} + \bm{\eta}^{\pm}))\big)
               \label{eq:perpendicular_def_bond}
\end{equation}
From the above equations, the lengths of the deformed lattice vectors can be represented in terms of the strains, undeformed lattice vectors, and the \emph{inner displacements} as
\begin{equation}
a  = f(\widehat{\mathbf{C}},\widehat{\mathscr{K}},\lambda^+,\lambda^-,\bm{\eta}^+,\bm{\eta}^-,\mathbf{A}) 
\label{eq:bond_length}
\end{equation}
Similarly, the bond angles can also be represented in terms of the strains, undeformed lattice vectors, and the \emph{inner displacements} as
\begin{equation}
\theta = f(\widehat{\mathbf{C}},\widehat{\mathscr{K}},\lambda^+,\lambda^-,\bm{\eta}^+,\bm{\eta}^-,\mathbf{A},\mathbf{B}) 
\label{eq:bond_angle}
\end{equation}
Here, $\mathbf{A}$ and $\mathbf{B}$ are any undeformed lattice vectors given in equations~\ref{eq:mo_s_top},~\ref{eq:mo_s_bottom} and~\ref{eq:mo-mo}.
Therefore, the bond lengths and the bond angles of the deformed TMD lattice can be obtained in terms of the strains, undeformed lattice vectors, and the \emph{inner displacements} through equations~\ref{eq:bond_length} and~\ref{eq:bond_angle}. \\

\subsection{Inter-atomic Potential for monolayer TMDs}
Inter-atomic potentials describe the interaction between an atom with its neighboring atoms and express the potential energy in terms of the bond lengths and bond angles.
Therefore, from the above formulation, the potential energy can be computed using appropriate inter-atomic potentials in terms of the continuum strains. 
In the case of TMDs, Stillinger--Weber (SW) potential \citep{PhysRevB.31.5262} is the commonly used inter-atomic potential. 
Several efforts have been made to obtain the parameters of SW potential for MoS$_2$ from different experiments and purely atomistic simulations.
The parameters of SW potential are found by fitting to the experimentally obtained phonon spectrums \citep{Jiang2013JAP}, the energies obtained from molecular dynamics simulations based on valence force-field \citep{Jiang2015a}, the lattice geometry, elastic constants and phonon frequencies obtained from First Principal calculations \citep{Kandemir2016}, and  the lattice geometry and atomic forces obtained from \emph{ab-initio} molecular dynamics simulations \citep{Wen2017}. 

The SW potential energy $\mathscr{E}$ of a system consisting of $N$ atoms is
\begin{equation}
\mathscr{E} = \sum_{i=j}^N \sum_{j>i}^N V_2(r_{ij}) + \sum_{i=1}^N\sum_{j \neq i}^N \sum_{\begin{array}{c} k > j \\ k\neq i \end{array}}^N V_3(r_{ij},r_{ik},\theta_{ijk})
\end{equation}
where the two-body interaction takes the form
\begin{equation}
    V_2(r_{ij}) = A \times \text{exp}\bigg({\frac{\rho}{r_{ij}-r_{max}}}\bigg)\bigg(\frac{B}{r_{ij}^4}-1\bigg)
\end{equation}
and the three-body term can be expressed as
\begin{equation}
    V_3(r_{ij},r_{ik},\theta_{ijk}) = K \times \text{exp}\bigg(\frac{\rho_1}{r_{ij}-r_{maxij}} + \frac{\rho_2}{r_{ik}-r_{maxik}}\bigg)(\text{cos}(\theta_{ijk}) - \text{cos}(\theta_{0ijk}))^2
\end{equation}
Here, $r_{ij}$ is the bond length between atoms $i$ and $j$ and $\theta_{ijk}$ is the angle formed by atoms $i$, $j$ and $k$, where the $i$-th atom is the vertex of the angle.
Here $A$, $\rho$, $r_{max}$ and $B$ are the parameters for the two-body interactions and $K$, $\rho_1$, $\rho_2$, $r_{maxij}$, $r_{maxik}$ and $\theta_{0ijk}$ are the parameters for the three-body interactions.

The continuum strain energy density i.e. the energy per unit area of the middle surface in the undeformed configuration, can be expressed as 
\begin{equation}
  \mathscr{W} = \mathscr{W}(\widehat{\mathbf{C}},\widehat{\mathscr{K}},\lambda^{+},\lambda^{-},\bm{\eta}^{+},\bm{\eta}^{-}) = \frac{1}{A_{rc}}\bigg[\sum_{i=j}^{N_a} \sum_{j>i}^{N_a} V_2(r_{ij}) + \sum_{i=1}^{N_a}\sum_{j \neq i}^{N_a} \sum_{\begin{array}{c} k > j \\ k\neq i \end{array}}^{N_a} V_3(r_{ij},r_{ik},\theta_{ijk})\bigg]
\end{equation}
where $A_{rc}$ is the area of the representative cell in the undeformed configuration and $N_a$ are the number of atoms in the representative cell.
For the MoS$_2$ system, there are three types of two body interactions (i.e. $IJ \in \{Mo-Mo, Mo-S, S-S\}$) and two types of three body interactions (i.e. $IJK \in \{Mo-S-S,S-Mo-Mo\}$).

The present work uses the parameters of SW potential for MoS$_2$ obtained in \cite{Jiang2015a}, since it is the most widely used and can produce elastic properties that matches with the  experiments. However, we found that it yields negative energy under small  compressive strain. We investigated this anomaly by exploring the energy in the parameter space. We found that the values of the lattice parameters ($h_0$ and $\Psi_{0}$ of  equation~\ref{eq:Latticeparameters}) reported in \cite{Jiang2015a} do not  correspond to the equilibrium. We have made the necessary corrections to the parameters and used them in our model. Further detail on the correction of the parameters is provided in \ref{app:opt}.

\subsubsection{Inner relaxation: Optimal relative shifts between different simple lattices}

The energy is minimized to obtain the optimal \emph{inner displacements} (relative shifts) while all strains are held constant. This minimization step is termed as \emph{inner relaxation}. 
To obtain the optimal \emph{inner displacements}, the strain energy density of the representative cell is minimized with respect to the \emph{inner displacements} ($\bm{\eta}^{+}$ and $\bm{\eta}^{-}$) as,
\begin{equation}
    \bar{\bm{\eta}}^+(\widehat{\mathbf{C}},\widehat{\mathscr{K}},\lambda^{+},\lambda^-) = \text{arg}\Big(\underset{\bm{\eta}^+}{\text{min}} \mathscr{W}(\widehat{\mathbf{C}},\widehat{\mathscr{K}},\lambda^{+},\lambda^-,\bm{\eta}^{+},\bm{\eta}^-)\Big) \Rightarrow \frac{\partial \mathscr{W}}{\partial{\bm{\eta}}^+}\Bigg|_{\bar{\bm{\eta}}^+} = \bf{0}
\end{equation}
and 
\begin{equation}
    \bar{\bm{\eta}}^-(\widehat{\mathbf{C}},\widehat{\mathscr{K}},\lambda^{+},\lambda^-) = \text{arg}\Big(\underset{\bm{\eta}^-}{\text{min}} \mathscr{W}(\widehat{\mathbf{C}},\widehat{\mathscr{K}},\lambda^{+},\lambda^-,\bm{\eta}^{+},\bm{\eta}^-)\Big) \Rightarrow \frac{\partial \mathscr{W}}{\partial{\bm{\eta}}^-}\Bigg|_{\bar{\bm{\eta}}^-} = \bf{0}
\end{equation}
Here, $\bar{\bm{\eta}}^{+}$ and $\bar{\bm{\eta}}^-$ are the optimum values of the \emph{inner displacements} for given the strains to define the optimal positions of S-atoms on the top and bottom surfaces, respectively.
The detailed steps to obtain the optimal shifts are provided below. 

 To begin with, two 3D vectors $\bm{\eta}^+$ and $\bm{\eta}^-$ are chosen to define the \emph{inner displacements} corresponding to top and bottom of S-atoms relative to Mo-atoms. 
Energy density is minimized with respect to these two vectors. The minimization is done using Newton's method. For a fixed $\widehat{\mathbf{C}}$, $\widehat{\mathscr{K}}$, $\lambda^+$ and $\lambda^{-}$, energy density is minimized with respect to the relative shift between the two lattices. 

The \emph{inner displacements} do not affect the distance between two Mo-atoms as all the three Mo-atoms in the unit cell are a part of one lattice in the complex lattice. Similarly, \emph{inner displacements} do not affect the distance between two S-atoms in the same plane as all the S-atoms are displaced equally by either $\bm{\eta}^+$ or $\bm{\eta}^-$ corresponding to S--atoms lying on the top or bottom surface respectively. The optimal values of \emph{inner displacements} are compactly written as $\bar{\bm{\eta}} = \{\bar{\bm{\eta}}^+,\bar{\bm{\eta}}^-\}$.

The continuum strain energy density can be expressed in terms of the optimum shifts between two simple lattices as,
\begin{eqnarray}
       \bar{\mathscr{W}}(\widehat{\mathbf{C}},\widehat{\mathscr{K}},\lambda^+,\lambda^-) = \mathscr{W}(\widehat{\mathbf{C}},\widehat{\mathscr{K}},\lambda^+,\lambda^-,\bar{\bm{\eta}}^+(\widehat{\mathbf{C}},\widehat{\mathscr{K}},\lambda^+,\lambda^-),\bar{\bm{\eta}}^-(\widehat{\mathbf{C}},\widehat{\mathscr{K}},\lambda^+,\lambda^-)).
       \label{eq:energy}
\end{eqnarray}
Based on the optimum shift, the second Piola-Kirchhoff stress tensor can be obtained as
\begin{equation}
    \textbf{S} = 2 \frac{\partial \bar{\mathscr{W}}}{\partial \widehat{\textbf{C}}} = 2 \frac{\partial \mathscr{W}}{\partial \widehat{\textbf{C}}}\bigg|_{\bm{\eta} = \bar{\bm{\eta}}} 
\end{equation}
and the Lagrangian bending tensor can be defined as
\begin{equation}
    \textbf{M} =  \frac{\partial \bar{\mathscr{W}}}{\partial \widehat{\mathscr{K}}} =  \frac{\partial \mathscr{W}}{\partial \widehat{\mathscr{K}}}\bigg|_{\bm{\eta} = \bar{\bm{\eta}}} 
\end{equation}
Two other stresses corresponding to two stretches denoting the separation of top and bottom surfaces from the middle surface can be expressed as,
\begin{equation}
L^+ = \frac{\partial \bar{\mathscr{W}}}{\partial \lambda^+}\bigg|_{\bm{\eta} = \bar{\bm{\eta}}} \; \; \text{and} \; \; L^- = \frac{\partial \bar{\mathscr{W}}}{\partial \lambda^-}\bigg|_{\bm{\eta} = \bar{\bm{\eta}}} 
\end{equation}

\subsubsection{Inner forces and inner elastic constants}
In order to minimize the energy density with respect to $\bm{\eta}$ using Newton's method, the derivative (residual, $\mathbf{r} = \mathscr{W},_{\bm{\eta}}$) and double derivative (jacobian, $\mathbf{J} = \mathscr{W},_{\bm{\eta}\bm{\eta}}$) of energy with respect to $\bm{\eta}$ must be computed. 
The residual $\mathbf{r}$ and the jacobian $\mathbf{J}$ can be be interpreted as inner out-of-balance forces and inner elastic constants, respectively~\citep{cousinsPhD}.
Consider $\mathbf{p}$ as a set containing all the bond lengths and bond angles of the representative cell, $\mathbf{p} = \{r_{ij},r_{ik},\theta_{ijk}\}$.
The inner forces of the system can be obtained by applying the chain rule as,

\begin{equation}
    \mathscr{W},_{\bm{\eta}} = \frac{1}{A_{rc}}\bigg[\sum_{u=1}^{N_{2b}} \frac{\partial V_2}{\partial p_u}\frac{\partial p_u}{\partial \bm{\eta}} \ + \sum_{v=1}^{N_{3b}} \frac{\partial V_3}{\partial p_v}\frac{\partial p_u}{\partial \bm{\eta}}\bigg]
    \label{eq:dW_by_deta}
\end{equation}
where $N_{2b}$ is the number of two-body interactions and $N_{3b}$ is the number of three-body interactions present in the representative cell. ${\partial p_u}/{\partial \bm{\eta}}$ and ${\partial p_v}/{\partial \bm{\eta}}$ represent the derivative of $u$-th and $v$-th components of the set $\mathbf{p}$.
To obtain the derivatives of the bond lengths and bond angles with respect to the inner displacements, the derivatives of the deformed bond vectors must be defined. 
Since the deformed bond vectors are obtained in two components, their derivative can be obtained separately.
The derivative of deformed tangential component of the bond can be obtained as
\begin{equation}
    \frac{\partial [\bm{a}_t]}{\partial \bm{\eta}} = \left\{\begin{array}{c}
\left(\mathscr{Q}_I + k_{1} w^{1}\mathscr{Q}'_I\right)w_{,\bm{\eta}}^1 \\
\left(\mathscr{Q}_{II} + k_{2} w^{2}\mathscr{Q}'_{II}\right)w_{,\bm{\eta}}^2\\
k_2 w^2 \mathscr{Q}_{22}\left(\mathscr{Q}_{I/2} + \frac{k_2 w^2}{2} \mathscr{Q}'_{I/2}\right) w^2_{,\bm{\eta}} + k_2 w^2 \mathscr{Q}_{II/2}\left(\mathscr{Q}_{II/2} + \frac{k_2 w^2}{2} \mathscr{Q}'_{II/2}\right) w^2_{,\bm{\eta}}
\end{array}\right\} 
\label{eq:der}
\end{equation}
where 
\begin{align*}
    \mathscr{Q}_{I} = \mathscr{Q}(k_1 w^1),  \; \; \;     \mathscr{Q}_{II} = \mathscr{Q}(k_2 w^2) , \; \; \;
    \mathscr{Q}_{I/2} = \mathscr{Q}(k_2 w^2/2), \; \; \; 
    \mathscr{Q}_{II/2} = \mathscr{Q}(k_2 w^2/2) \\
    \mathscr{Q}_{I}' = \mathscr{Q}'(k_1 w^1),  \; \; \;     \mathscr{Q}_{II}' = \mathscr{Q}'(k_2 w^2) , \; \; \;
    \mathscr{Q}_{I/2}' = \mathscr{Q}'(k_2 w^2/2), \; \; \; 
    \mathscr{Q}_{II/2}' = \mathscr{Q}'(k_2 w^2/2) \\
       \mathscr{Q}_{I}'' = \mathscr{Q}''(k_1 w^1),  \; \; \;     \mathscr{Q}_{II}'' = \mathscr{Q}''(k_2 w^2) , \; \; \;
    \mathscr{Q}_{I/2}'' = \mathscr{Q}''(k_2 w^2/2), \; \; \; 
    \mathscr{Q}_{II/2}'' = \mathscr{Q}''(k_2 w^2/2)
\end{align*}
The derivative of the deformed normal component of the bonds can be obtained as
\begin{equation}
    \frac{\partial [\bm{a}_n^{+}]}{\partial \bm{\eta}} = \lambda^{+}\: \: ;     \frac{\partial [\bm{a}_n^{-}]}{\partial \bm{\eta}} = \lambda^{-}
\end{equation}
Similarly, the double derivative of the deformed tangential component of the bonds can be expressed as
\begin{equation}
    \frac{\partial^2 [\bm{a}_t]}{\partial \bm{\eta}^2} = \left\{\begin{array}{c}
k_1 \left(2\mathscr{Q}'_I + k_1 w^1 \mathscr{Q}''_I\right) w^1_{,\bm{\eta}} \otimes w^1_{,\bm{\eta}} \\
k_2 \left(2\mathscr{Q}'_{II} + k_2 w^2 \mathscr{Q}''_{II}\right) w^2_{,\bm{\eta}} \otimes w^2_{,\bm{\eta}}\\
k_1 \left[ \left( \mathscr{Q}_{I/2} + \frac{k_1 w^1}{2}\mathscr{Q}'_{I/2}\right)^2 + k_1 w^1 \mathscr{Q}_{I/2} \left( \mathscr{Q}'_{I/2} + \frac{k_1 w^1}{4} \mathscr{Q}''_{I/2}\right) \right] w^1_{,\bm{\eta}} \otimes w^1_{,\bm{\eta}} + \cdots  \\
k_2 \left[ \left( \mathscr{Q}_{II/2} + \frac{k_2 w^2}{2}\mathscr{Q}'_{II/2}\right)^2 + k_2 w^2 \mathscr{Q}_{II/2} \left( \mathscr{Q}'_{II/2} + \frac{k_2 w^2}{4} \mathscr{Q}''_{II/2}\right) \right] w^2_{,\bm{\eta}} \otimes w^2_{,\bm{\eta}}
\end{array}\right\} 
\label{eq:dder}
\end{equation} 
and the double derivative of the deformed normal component of the bond can be expressed as, 
\begin{equation}
    \frac{\partial^2 [\bm{a}_n^{\pm}]}{\partial \bm{\eta}^2} = \bm{0}
\end{equation}
In Eq. \ref{eq:der} and \ref{eq:dder} $w^1_{,\bm{\eta}}$ and $w^2_{,\bm{\eta}}$ can be expressed as
\begin{equation}
    w^n_{,\bm{\eta}} = \widehat{\mathbf{C}} \mathbf{V}_n 
\end{equation}
The double derivative of energy with respect to $\bm{\eta}$ can be expressed as

\begin{align}
    \mathscr{W},_{\bm{\eta} \bm{\eta}} & = \frac{1}{A_{rc}}\sum_{u=1}^{N_{2b}} \bigg[ \frac{\partial V_2}{\partial p_u}\frac{\partial^2 p_u}{\partial \bm{\eta}^2} + \frac{\partial^2 V_2}{\partial p^2_u}\frac{\partial p_u}{\partial \bm{\eta}}\otimes \frac{\partial p_u}{\partial \bm{\eta}} \bigg] \\ 
    & + \frac{1}{A_{rc}} \sum_{v=1}^{N_{3b}} \bigg[  \frac{\partial V_3}{\partial p_v}\frac{\partial^2 p_v}{\partial \bm{\eta}^2} + \frac{\partial^2 V_3}{\partial p^2_v}\frac{\partial p_v}{\partial \bm{\eta}}\otimes \frac{\partial p_v}{\partial \bm{\eta}} + \sum_{v <  w\le N_{3b}} 2 \frac{\partial^2 V_3}{\partial p_v \partial p_w} \frac{\partial p_v}{\partial \bm{\eta}} \otimes_{symm} \frac{\partial p_w}{\partial \bm{\eta}} \bigg]
    \label{eq:ddW_by_ddeta}
\end{align}
where the operation $\otimes_{symm}$ is defined as,
\begin{equation}
    \frac{\partial p_v}{\partial \bm{\eta}} \otimes_{symm} \frac{\partial p_w}{\partial \bm{\eta}} = \frac{1}{2}\bigg[\frac{\partial p_v}{\partial \bm{\eta}} \otimes \frac{\partial p_w}{\partial \bm{\eta}} + \frac{\partial p_w}{\partial \bm{\eta}} \otimes \frac{\partial p_v}{\partial \bm{\eta}} \bigg]
\end{equation}



The steps for the inner relaxation are summarized in Box \textcolor{blue}{1}. The calculations for strains, strain energy density and stresses are summarized in Box \textcolor{blue}{2}.

\begin{center}
\begin{tcolorbox}[colback=gray!40,
                  colframe=black,
                  width=\linewidth,
                  arc=3mm, auto outer arc,
                 ]
\begin{center} \textbf{Box 1: Algorithm to obtain the optimum relative shifts} 
\end{center}
\begin{itemize}
    \item Initiate $\bm{\eta} = \{\bm{\eta}^+,\bm{\eta}^-\}$, set $k = 0$, $r^k = 1$ and $\Delta\bm{\eta}^k = 1$
    \item WHILE $||\mathbf{r}^k|| >$ Tolerance1 .OR. $||\Delta\bm{\eta}^k|| > $ Tolerance2
    \begin{itemize}
        \item Compute residual $\bf{r}^k = \frac{\partial \mathscr{W}}{\partial \bm{\eta}}\big|_{\bm{\eta} = \bm{\eta^k}}$ (Equation~\ref{eq:dW_by_deta})
        \item Compute Jacobian $\bf{J}^k = \frac{\partial^2 \mathscr{W}}{\partial^2 \bm{\eta}}\big|_{\bm{\eta} = \bm{\eta^k}}$ (Equation~\ref{eq:ddW_by_ddeta})
        \item $\Delta\bm{\eta}^k = -[\bf{J}^k]^{-1}\bf{r}^k$
        \item $\bm{\eta}^{k+1} = \bm{\eta}^k + \Delta\bm{\eta}^k$
        \item $k = k+1$
    \end{itemize}
    \item Check if det $\bf{J}^k >0$, i.e. energy is minimum. If det $\bf{J}^k < 0$ the minimization is performed through quasi-newton method. 
    \item $\widehat{\mathscr{W}} = \mathscr{W}(\bm{\eta^k})$ and $\bar{\bm{\eta}}= \bm{\eta}^k$.
\end{itemize}
\end{tcolorbox}
\end{center}

\subsection{Non-bonded interactions}
To obtain the total energy of the system, interactions between atoms which are not bonded also need to be considered in this subsection following~\cite{Arroyo2004IJNME}.
Such interactions are the result of (a) electrostatic interactions between two permanently charged atoms, (b) attractive interaction between a mono-pole and an induced mono-pole and (c) attractive interaction between two induced mono-poles.
Van der Waals interaction between two atoms is the combination of all these interactions and hence is considered as the non-bonded interaction.
The total non-bonded interaction between all the non-bonded atoms is given by
\begin{equation}
\mathscr{E}_{nb} = \sum_i \sum_{j>i,j \notin B_i} V_{nb}(r_{ij})
\end{equation}
where $V_{nb}$ represents the non-bonded interaction between atom $i$ and atom $j$, and $r_{ij}$ is the distance between those two atoms. Here, $B_i$ is the set containing all the atoms bonded to atom $i$.
In the present model, a 6-12 Lennard-Jones potential \citep{lennard1931cohesion} is used to represent the non-bonded interaction which can be expressed as
\begin{equation}
V_{nb}(r_{ij}) = 4\epsilon \bigg[ \bigg(\frac{\sigma}{r_{ij}}\bigg)^{12} - \bigg(\frac{\sigma}{r_{ij}}\bigg)^6\bigg]
\label{eq:lj}
\end{equation}
where $\epsilon$ and $\sigma$ denotes the non-bonded energy at equilibrium and spacing between atoms at equilibrium, respectively.

\begin{center}
\begin{tcolorbox}[colback=gray!40,
                  colframe=black,
                  width=\linewidth,
                  arc=3mm, auto outer arc,
                 ]
   \begin{center} \textbf{Box 2: Calculation of continuum strains, stresses and energy density}.  \end{center} 
   \begin{enumerate}
    \item \textit{Deformation map}: 
    The deformation map consists of two parts, i) the deformation map of the middle surface and ii) the stretches above and below the middle surface.
    \begin{equation}
    \Omega = \chi(\Omega_0) = \{ \chi : \mathbf{X} \to \mathbf{x} = \Phi(\widehat{\mathbf{X}}) + h_0\, \lambda(\widehat{\mathbf{X}},X_3)\, \widehat{\mathbf{n}}(\widehat{\textbf{X}}) \},\quad \mathbf{X} \in \Omega_0 \: , \: \mathbf{x} \in \Omega \nonumber
    \end{equation}
 
    \item \textit{Strain measures of the middle surface}: Obtain the Cauchy green tensor $\widehat{\mathbf{C}}$ and the curvature tensor $\widehat{\mathscr{K}}$ following equations~\ref{eq:cauchy_green} and~\ref{eq:eig}.


   \item \textit{Principal curvatures for middle surface}: Obtain the principal directions and principal values for the curvature at each point on the middle surface by solving the eigenvalue problem in equation~\ref{eq:eig}. Following equations~\ref{eq:dk_dCK}-\ref{eq:dV_dCK}, the derivatives of principal curvature and principal directions can be obtained
   \begin{equation*}
       k_n, \: \: \frac{\partial k_n}{\partial \widehat{\textbf{C}}}, \: \:  \frac{\partial k_n}{\partial \widehat{\mathscr{K}}}, \: \: \textbf{V}_n, \: \: \frac{\partial \textbf{V}_n}{\partial \widehat{\textbf{C}}}, \: \frac{\partial \textbf{V}_n}{\partial \widehat{\mathscr{K}}}, \: \: \text{for} n=1,2
   \end{equation*}
   
   \item \textit{Optimal shift}: Obtain the optimal value of $\bm{\eta}$, denoted by $\bar{\bm{\eta}}$, by performing the energy minimization, as shown in Box \textcolor{blue}{1}. 
   Update the undeformed bonds following equation~\ref{eq:bonds_after_shift}
   as, $
   \mathbf{A} = n_i\mathbf{B}_i + m(\mathbf{P} + \bar{\bm{\eta}})
  $

   \item \textit{Deformed lattice parameters and their derivatives}: Obtain the deformed lattice parameters following equations~\ref{eq:tangential_def_bond} and~\ref{eq:perpendicular_def_bond}. Compute the derivatives of the deformed lattice parameters with respect to continuum variables, following \ref{ap:bond_ders}. 
   \begin{equation*}
        a_i, \: \frac{\partial a_i}{\partial \widehat{\textbf{C}}}, \: \frac{\partial a_i}{\partial \widehat{\mathscr{K}}}, \: \frac{\partial a_i}{\partial \lambda^{+}}, \frac{\partial a_i}{\partial \lambda^{-}}, \: \theta_i, \: \frac{\partial \theta_i}{\partial \widehat{\textbf{C}}}, \: \frac{\partial \theta_i}{\partial \widehat{\mathscr{K}}}, \: \frac{\partial \theta_i}{\partial \lambda^{+}}, \: \frac{\partial \theta_i}{\partial \lambda^{-}}
   \end{equation*}
   
   \item \textit{Energy density and stresses}: Calculate energy densities for unit cell using equation~\ref{eq:energy} and its derivatives with respect to continuum strains to obtain stress tensors 
   \begin{equation*}
       \textbf{S} = 2\sum_i \bigg(\frac{\partial \bar{\mathscr{W}}}{\partial a_i} \frac{\partial a_i}{\partial \widehat{\textbf{C}}} + \frac{\partial \bar{\mathscr{W}}}{\partial \theta_i} \frac{\partial \theta_i}{\partial \widehat{\textbf{C}}}\bigg); \: \:
      \textbf{M} = \sum_i \bigg(\frac{\partial \bar{\mathscr{W}}}{\partial a_i} \frac{\partial a_i}{\partial \widehat{\mathscr{K}}} + \frac{\partial \bar{\mathscr{W}}}{\partial \theta_i} \frac{\partial \theta_i}{\partial \widehat{\mathscr{K}}}\bigg)
   \end{equation*}
   \begin{equation*}
   L^+ = \sum_i \bigg( \frac{\partial \bar{\mathscr{W}}}{\partial a_i}\frac{\partial a_i}{\partial \lambda^+} + \frac{\partial \bar{\mathscr{W}}}{\partial \theta_i}\frac{\partial \theta_i}{\partial \lambda^+}\bigg) \; \; \text{and} \; \;       
   L^- = \sum_i \bigg( \frac{\partial \bar{\mathscr{W}}}{\partial a_i}\frac{\partial a_i}{\partial \lambda^-} + \frac{\partial \bar{\mathscr{W}}}{\partial \theta_i}\frac{\partial \theta_i}{\partial \lambda^-}\bigg) 
   \end{equation*}
   \end{enumerate}
\end{tcolorbox}
\end{center}

To represent the non-bonded energy density in the continuum form, the interaction between two representative cells is considered as
\begin{equation}
\mathscr{V}_{nd}(d) = \bigg(\frac{n}{A_{rc}}\bigg)^2V_{nb}(d)
\end{equation}
where $n$ is the number of atoms in the representative cell, which is 3 for the case of MoS$_2$. $A_{rc}$ represents the area of the representative cell and $d$ is the distance between the centroid of the two representative cells.

The total non-bonded energy for the system can then be expressed as
\begin{equation}
\Pi_{nb}(\chi) = \frac{1}{2} \int_{S_0} \int_{S_0-B_{\mathbf{X}}} \mathscr{V}_{nb}(||\mathbf{x} - \mathbf{y}||)\,dS_{0\mathbf{Y}}\,dS_{0\mathbf{X}}
\label{eq:non_bonded}
\end{equation}
where $S_0$ represents the undeformed surface, $B_{\mathbf{X}}$ represents the set containing the representative cell within the cut-off distance to account for the bonds that are not a part of non-bonded interactions. 
Here, $\mathbf{x} = \chi(\mathbf{X})$ and $\mathbf{y} = \chi(\mathbf{Y})$.  


\section{Boundary value problem}
\label{sec:bvp}
The total energy of the continuum membrane is obtained by integrating the energy densities over the entire domain. 
In the presence of any external force whose energy can be described through the potential $\Pi_{ext}(\chi)$, the total energy can be expressed as 
\begin{equation}
    \Pi(\chi) = \Pi_{int}(\chi) + \Pi_{nb}(\chi) - \Pi_{ext}(\chi)
\end{equation}
The total internal energy of the system due to the deformation $\chi$ can be obtained by integrating the energy density given in equation~\ref{eq:energy} as
\begin{equation}
    \Pi_{int}(\chi) = \int_{S_0} \bar{\mathscr{W}}(\widehat{\mathbf{C}}(\chi),\widehat{\mathscr{K}}(\chi),\lambda^{+}(\chi),\lambda^{-}(\chi)) \: \text{d}S_0
\end{equation}
The external potential due to the  external body force per unit area, $\mathbf{B}$, is given by 
\begin{equation}
\Pi_{ext}(\chi) = \int_{S_0} \mathbf{B}\cdot {\chi} dS_0
\end{equation}
and $\Pi_{nb}(\chi)$ is the total energy due to the non-bonded interaction as computed in equation~\ref{eq:non_bonded}.

The equilibrium configuration can be obtained by minimizing the total energy as,
\begin{equation}
    \bar{\chi} = \text{arg}\bigg(\underset{\chi}{\text{min}} \; \; \; \Pi(\chi)\bigg)
\end{equation}
The equilibrium configuration $\bar{\chi}$ is one of the stationary points of the potential energy functional, and hence its first variation must vanish as, 
\begin{equation}
\int_{S_0} \bigg(\frac{1}{2}\mathbf{S}:\delta \widehat{\mathbf{C}} + \mathbf{M}:\delta \widehat{\mathscr{K}} + L^{+} \: \delta\lambda^{-} + L^{-} \: \delta\lambda^{+}\bigg)dS_0 + \delta\Pi_{nb}[\bar{\chi};\delta\bar{\chi}]
- \delta\Pi_{ext}[\bar{\chi};\delta\bar{\chi}] = 0
\end{equation}
where $\delta(\cdot)$ represents the variation of the quantity $(\cdot)$.
The variation of non-bonded interaction can be expressed as
\begin{equation}
\delta\Pi_{nb}[\bar{\chi};\delta\bar{\chi}] = \frac{1}{2} \int_{S_0} \int_{S_0-B_{\mathbf{X}}} \mathscr{V}'_{nb}(d) \delta d[\bar{\chi};\delta\bar{\chi}] dS_{0\mathbf{X}} dS_{0\mathbf{Y}} 
\end{equation}
where $d = (||\bar{{\chi}}(\mathbf{X}) - \bar{{\chi}}(\mathbf{Y})||)$ is the distance between two points $\bf{X}$ and $\bf{Y}$ on the two representative cells
The variation of the external body force potential can be expressed as
\begin{equation}
\delta \Pi[\bar{\chi};\delta\bar{\chi}] = \int_{S_0} \mathbf{B} \cdot \delta\bar{{\chi}} dS_0
\end{equation}

\section{Numerical implementation }
\label{sec:numerical_implementation}
This section describes the finite element discretization to solve the boundary value problem. 
The membrane is discretized using a Ritz-Galerkin finite element scheme  through B-spline basis functions. 
Since the potential energy is a function of curvature, it requires the second-order derivatives to be  square-integrable. B-splines provide a smoother approximation than the standard finite element approximations and ensure the  square integrability of the second derivatives.
A brief explanation of B-splines is provided in \ref{app:b_splines}. 
For a more detailed account of the B-splines and the finite element methods using B-splines, the reader is referred to \cite{PiegTill96} and \cite{fe_bsplines}, respectively.
A quasi-Newton method is used to perform the energy minimization to obtain the equilibrium configuration. 
The numerical implementation to minimize the total energy using a Ritz-Galerkin formulation is described in this section.
\subsection{Finite element discretization using B-splines}
%
\begin{figure}[htbp]
\centering
\includegraphics[width=0.75\linewidth]{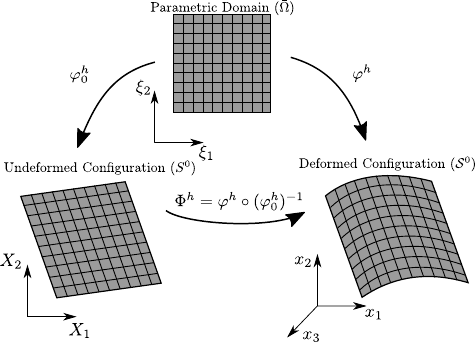}
\caption{Schematic representing the finite element discretization used for the numerical implementation of the proposed continuum formulation.}
\label{fig:num_fe}
\end{figure}
 We discretize the parametric domain, $\bar{\Omega}$, by partitioning it into a regular grid of knots using the B-spline basis functions. We have used second-order B-splines in both directions, $\xi_1$ and $\xi_2$. The schematic for the present finite element discretization is shown in figure~\ref{fig:num_fe}. 
 The superscript $\left(\cdot\right)^h$ denotes the discretized fields. The superscript $\left(\cdot\right)^e$ denotes the quantities restricted to an element $e$.
 
The map ${\varphi}^h_0$ is  homogeneous. The map  ${\varphi}^h_0$, its inverse, and its derivatives are obtained analytically. 
The B-spline approximation of the map $\varphi^{h}$ is given by
\begin{equation}
{\varphi}^{h}(\xi_1,\xi_2) = \sum_I N_I(\xi_1,\xi_2)\,\mathbf{Q}^s_{I} \:, \qquad \mathbf{Q}^s_I \in \mathbb{R}^3
\label{eq:map}
\end{equation}
where 
$N_I$ represents the $I$-th B-spline basis function and $\mathbf{Q}^s_I$ represents the associated control point in the global node numbering. The control points associated with the  surface  $\mathcal{S}^0$ are distinguished with a superscript $\left(\cdot\right)^s$. 
The deformation map for the middle surface is approximated as $\Phi^{h} = \varphi^{h} \circ  {\varphi^{h}_0}^{\,-1}$.

The two normal stretches are  approximated through B-spline as,
\begin{equation}
    ({\lambda^{\pm}})^{h} = \sum_I N_I(\xi_1,\xi_2)\,Q^{\pm}_{I}
    \label{eq:lambda}
\end{equation} 
Here, $Q^+_{I}$ and $Q^-_{I}$ are the $I$-th control points defining the stretches above and below the middle surface, respectively.

\subsubsection{Calculation of Strains}
The convected basis at each point in the domain can be computed as,
\begin{equation}
\mathbf{g}_{\alpha} = \sum_I N_{I,\alpha} \mathbf{Q}^s_I
\label{eq:num_convected_basis}
\end{equation}
and the individual components can be expressed as,
\begin{equation}
g^a_{\alpha} = \sum_I N_{I,\alpha}\, (Q^s_I)^a \qquad \alpha = \{1,2\},\; and\; a=\{1,2,3\}
\end{equation}
Here, $N_{I,\alpha}$ represents the derivative of the basis function $N_{I}$ with respect to $\xi_{\alpha}$.
The calculation of the metric tensor and the right Cauchy-Green strain tensor follows equations~\ref{eq:metric_tensor} and~\ref{eq:cauchy_green}. 
Following equation~\ref{eq:curv_tensor_def}, the calculation of the curvature tensor  requires the derivative of the convected basis which can be obtained as
\begin{equation}
\mathbf{g}_{\alpha,\beta} = \sum_I N_{I,\alpha \beta}\, \mathbf{Q}^s_I
\end{equation}
where,
\begin{equation}
N_{I,\alpha \beta} = \frac{\partial^2 N_I}{\partial \xi_{\alpha}\, \partial \xi_{\beta}}
\end{equation}
The deformed bond lengths and bond angles can be computed from the strains of the middle surface and the two normal  stretches computed in the aforementioned steps.

\subsection{Energies and out--of--balance forces}
\label{sec:energy_derivative}
The total internal energy is obtained by integrating the energy per unit area of the reference configuration. The numerical integration is performed by defining the Gauss quadrature over the parametric domain as,
\begin{align}
\Pi_{int}[\chi^h] & = \int_{S_0} \bar{\mathscr{W}}(\widehat{\mathbf{C}},\widehat{\mathscr{K}},\lambda^{+},\lambda^{-})) \text{d}S_0 \nonumber \\ 
                & = \sum_{e=1}^{nel} \int_{\bar{\Omega}^e} \bar{\mathscr{W}}(\widehat{\mathbf{C}},\widehat{\mathscr{K}},\lambda^{+},\lambda^-) \text{det}(T\varphi^e_0)d\bar{\Omega}^e \nonumber \\
& = \sum_{e=1}^{nel} \sum_{i=1}^{nint} \bar{\mathscr{W}}(\widehat{\mathbf{C}},\widehat{\mathscr{K}},\lambda^{+},\lambda^{-})|_{\bm{\xi}_i} \text{det}[(T\varphi_0^e)]\omega_i
\end{align}
where $nel$ denotes the  number of elements and $nint$ denotes the number of integration points in each element. 
$\bm{\xi}_i = ({\xi}_{i1},{\xi}_{i2})$ defines the $i$-th integration point in the parametric domain. Here, 
$\omega_i$ represents the Gauss weight corresponding to the $i$-th Gauss point. The $T\varphi_0$ is constant and the same for all elements.  
$\bar{\Omega}^e$ represents an element in the parametric domain. 

Internal forces for the bonded potential are obtained by computing the derivative of internal energy with respect to the control points $\mathbf{Q}_I \in \mathbb{R}^5$. The control points $\mathbf{Q}_I$ are obtained by combining the control points for $\Phi$, $\lambda^+$ and $\lambda^-$ as $\mathbf{Q}_I = \{\mathbf{Q}^s_I, Q^+_I,Q^-_I\}$. {Here, $I$ corresponds to global node numbering.}
The elementwise out--of--balance forces can be expressed as,
\begin{align}
    (\mathbf{f}^e_{int})_J & = \frac{\partial \Pi_{int}^e}{\partial \mathbf{Q}^e_J} \nonumber \\
    & = \int_{\bar{\Omega}^e} \bigg( \frac{\partial \bar{\mathscr{W}}}{\partial \widehat{\mathbf{C}}} \frac{\partial \widehat{\mathbf{C}}}{\partial \mathbf{Q}_J} + \frac{\partial \bar{\mathscr{W}}}{\partial \widehat{\mathscr{K}}} \frac{\partial \widehat{\mathscr{K}}}{\partial \mathbf{Q}_J} + \frac{\partial \bar{\mathscr{W}}}{\partial \lambda^{+}} \frac{\partial \lambda^{+}}{\partial \mathbf{Q}_J} +  \frac{\partial \bar{\mathscr{W}}}{\partial \lambda^{-}} \frac{\partial \lambda^{-}}{\partial \mathbf{Q}_J}\bigg) \: \text{det}[(T\varphi^e_0)] \: \text{d}\bar{\Omega}^e \nonumber \\
    & = \sum_{i=1}^{nint} \bigg( \frac{\partial \bar{\mathscr{W}}}{\partial \widehat{\mathbf{C}}} \frac{\partial \widehat{\mathbf{C}}}{\partial \mathbf{Q}_J} + \frac{\partial \bar{\mathscr{W}}}{\partial \widehat{\mathscr{K}}} \frac{\partial \widehat{\mathscr{K}}}{\partial \mathbf{Q}_J} + \frac{\partial \bar{\mathscr{W}}}{\partial \lambda^{+}} \frac{\partial \lambda^{+}}{\partial \mathbf{Q}_J} + \frac{\partial \bar{\mathscr{W}}}{\partial \lambda^{-}} \frac{\partial \lambda^{-}}{\partial \mathbf{Q}_J}\bigg)\bigg|_{\bm{\xi}_i} \:  \text{det}[(T\varphi^e_0)] \:  \omega_i 
\end{align}
{Here, $(\mathbf{f}^e_{int})_J$ represents the elemental forces corresponding to local node numbering.}

Similarly, following equation~\ref{eq:non_bonded}, the non-bonded energy is computed by integrating the non-bonded energy density over the interacting surfaces as,
\begin{align}
\Pi_{nb}[\chi^h] & = \frac{1}{2} \int_{S_0} \int_{S_0-B_{\mathbf{X}}} \mathscr{V}_{nb}(||\mathbf{x}-\mathbf{y}||)dS_{0\mathbf{Y}}dS_{0\mathbf{X}}  \nonumber \\
& = \sum_{e=1}^{nel} \int_{\bar{\Omega}^e} \sum_{f = e +1 }^{nel} \int_{\bar{\Omega}^f} \mathscr{V}_{nb}(||\mathbf{x}-\mathbf{y}||) \: \text{det}(T\varphi_0^e) \: \text{det}(T\varphi_0^f)\: d\bar{\Omega}^f \: d\bar{\Omega}^e \nonumber \\
&  = \sum_{e=1}^{nel} \sum_{i=1}^{ngpt} \sum_{f = e +1 }^{nel} \sum_{j=1}^{ngpt} \mathscr{V}_{nb}(||\mathbf{r}_{i-j}^{e-f}||)\:\text{det}(T\varphi_0^e)\:\text{det}(T\varphi_0^f)\:\omega_i \: \omega_j
\label{eq:num_nb_ener}
\end{align}
where $\mathbf{x}$ and $\mathbf{y}$ are two points on the deformed configuration mapped from the points $\mathbf{X}$ and $\mathbf{Y}$ located in the undeformed configuration.
Here, $e$-th element interacts with all other elements $f > e$ out of the bonded region $B_{\mathbf{X}}$, and $\mathbf{r}_{i-j}^{e-f}$ represents the vector from the $i$-th integration point in element $e$ to the $j$-th integration point on element $f$.  
The out--of--balance forces due to the non-bonded interaction for element $e$ with element $f$ can be computed as the derivative of the non-bonded energy with respect to the control points as,
\begin{align}
(\mathbf{f}^{e-f}_{nb})_J & = \frac{\partial \Pi^{e-f}_{nb}}{\partial \mathbf{Q}^e_J} \nonumber \\
                  & = \sum_{i=1}^{ngpt} \sum_{j=1}^{ngpt} \frac{1}{||\mathbf{r}_{i-j}^{e-f}||}\mathscr{V}'_{nb}(||\mathbf{r}_{i-j}^{e-f}||)\:\mathbf{r}_{i-j}^{e-f} \:N_J(\bm{\xi}_i)\:\text{det}(T\varphi_0^e)\:\text{det}(T\varphi_0^f)\:\omega_i \: \omega_j 
\label{eq:force_e-f}
\end{align}
Here, $\mathscr{V}'_{nb}(||\mathbf{r}_{i-j}^{e-f}||)$ represents the derivative of $\mathscr{V}_{nb}(||\mathbf{r}_{i-j}^{e-f}||)$ with respect to the distance between two integration points  ($||\mathbf{r}_{i-j}^{e-f}||$).
Corresponding to this, an elemental force in element $f$ can also be obtained as
\begin{align}
(\mathbf{f}^{f-e}_{nb})_J & = \frac{\partial \Pi^{f-e}_{nb}}{\partial \mathbf{Q}_J^e} \nonumber \\
                  & = -\sum_{i=1}^{ngpt} \sum_{j=1}^{ngpt} \frac{1}{||\mathbf{r}_{i-j}^{e-f}||}\mathscr{V}'_{nb}(||\mathbf{r}_{i-j}^{e-f}||)\:\mathbf{r}_{i-j}^{e-f}\: N_J(\bm{\xi}_i)\:\text{det}(T\varphi_0^e)\:\text{det}(T\varphi_0^f)\:\omega_i \: \omega_j 
\label{eq:force_f-e}
\end{align}
{The local elemental forces $\mathbf{f}_J$ are assembled to obtain the global forces $\mathbf{f}_I$.}

In equations~\ref{eq:num_nb_ener}, \ref{eq:force_e-f} and \ref{eq:force_f-e}, $ngpt$ denotes  the number of Gauss-quadrature points for the calculation of non-bonded interaction.
These Gauss-quadrature points can be different than that of the bonded energy calculations.
The implementation of non-bonded interaction is required in the present formulation to compute the interaction due to self contact when the TMD folds to touch itself. This formulation can be used for non-bonded interaction between multiple TMDs as well.
The parameters for 6-12 Lennard-Jones potential are considered for two neighboring S-atoms \citep{liang2009parametrization}.  
The parameters for this potential are taken from \cite{Jiang2013JAP} and~\cite{Jiang2015d}.

   \begin{algorithm}
   \caption{Algorithm for energy minimization for a given loading condition}
   \label{alg:num_implement}
   \begin{algorithmic}
   \State \textbullet~ Initiate the control points $\mathbf{Q}$ or take from last iteration.
   \State \textbullet~ iter = 1
   \While{||$\mathbf{f}$|| $>tolerance$} 
   \For{$ele=1$ to $ele=nel$}
     \For{$gpt=1$ to $gpt=nint$}
        \State \textbullet~ Compute the convected basis $\mathbf{g}$ and metric tensor $[\mathbf{g}]$.
        \State \textbullet~ Compute the deformation gradient for the middle surface $\widehat{\mathbf{F}}$
        \State \textbullet~ Compute the strains of the middle surface and their derivatives $\frac{\partial \widehat{\mathbf{C}}}{\partial \mathbf{Q}_I}$, $\frac{\partial \widehat{\mathscr{K}}}{\partial \mathbf{Q}_I}$
        \State \textbullet~ Compute the stretches $\lambda^+$ and $\lambda^-$ their derivatives 
        $\frac{\partial \lambda^+}{\partial \mathbf{Q}_I}$ and $\frac{\partial \lambda^-}{\partial \mathbf{Q}_I}$
        \State \textbullet~ Perform the inner relaxation using the algorithm provided in Box.1
        \State \textbullet~ Compute bond lengths $a_i$ and their derivatives as $\frac{\partial a_i}{\partial \widehat{\mathbf{C}}}$, $\frac{\partial a_i}{\partial \widehat{\mathscr{K}}}$, $\frac{\partial a_i}{\partial \lambda^{+}}$ and $\frac{\partial a_i}{\partial \lambda^{-}}$.
        \State \textbullet~ Compute bond angles $\theta_i$ and their derivatives as $\frac{\partial \theta_i}{\partial \widehat{\mathbf{C}}}$, $\frac{\partial \theta_i}{\partial \widehat{\mathscr{K}}}$, $\frac{\partial \theta_i}{\partial \lambda^{+}}$ and $\frac{\partial \theta_i}{\partial \lambda^{-}}$.
        \State \textbullet~ Compute the internal energy $\Pi_{int}$ and the stresses $\mathbf{S}$, $\mathbf{M}$, $L^+$ and $L^-$.
        \State \textbullet~ Compute non-bonded interaction and other external interactions.
       \State \textbullet~ Compute the forces as the derivative of energy with respect to  $\mathbf{Q}_I$ as
       \begin{equation}
         (\mathbf{f}_{gpt}^{ele})_J = \frac{\partial \Pi}{\partial \mathbf{Q}_J} = \frac{\partial \Pi_{int}}{\partial \widehat{\mathbf{C}}} \frac{\partial \widehat{\mathbf{C}}}{\partial \mathbf{Q}_J} + \frac{\partial \Pi_{int}}{\partial \widehat{\mathscr{K}}} \frac{\partial \widehat{\mathscr{K}}}{\partial \mathbf{Q}_J} + \frac{\partial \Pi_{int}}{\partial \lambda^+} \frac{\partial \lambda^+}{\partial \mathbf{Q}_J} + \frac{\partial \Pi_{int}}{\partial \lambda^-} \frac{\partial \lambda^-}{\partial \mathbf{Q}_J} + {\frac{\partial \Pi_{nb}}{\partial \mathbf{Q}_J}} \nonumber
       \end{equation}
       \begin{equation*}
           (\mathbf{f}^{ele})_J \leftarrow (\mathbf{f}^{ele})_J + (\mathbf{f}_{gpt}^{ele})_J
       \end{equation*}
      \EndFor
   \EndFor
   \State \textbullet~ Assemble the local forces $\mathbf{f}_J$ to the global forces $\mathbf{f}_I$.
   \State \textbullet~ {Compute norm of the total global force vector as $||\mathbf{f}_I||$.}
   \State \textbullet~ Supply the total energy and force to L-BGFS which in turn will provide the direction toward lower energy i.e. control points for next iteration as $\mathbf{Q}^{new} \leftarrow \text{L-BFGS}(\Pi,\mathbf{f}_I)$ 
   \State \textbullet~ iter = iter + 1
   \EndWhile
   \end{algorithmic}
   \end{algorithm}

\subsection{Energy minimization to obtain the equilibrium configuration}
The equilibrium configuration $\bar{\chi}$ for a given applied boundary condition is obtained by minimizing the total energy $\Pi(\bar{\chi})$. To minimize the energy, its derivatives with respect to the control points $\mathbf{Q}_I$ are obtained as explained in section~\ref{sec:energy_derivative}.
However, the derivation of the Hessian is difficult due to the complexity of the inter-atomic potential.
Therefore, Newton's method can not be used as it requires the calculation of Hessian. 
Quasi-Newton methods  (\cite{gilbert1992global},\cite{liu1989limited}) provide an attractive alternative to Newton's method as they do not require the calculation of Hessian but still attain a super-linear convergence rate  \citep{wright1999numerical}.
In the present work, L-BFGS, a quasi-Newton optimization technique is used~\citep{nocedal1980updating}. An algorithm to perform the minimization of the total energy is provided in Algorithm~\ref{alg:num_implement}.

\section{Numerical Validations for the present model}
\label{sec:results}
This section describes the numerical experiments that are performed to validate the continuum model and its numerical implementation
The membrane model derived in the proposed work is valid for all TMDs.
However, for numerical implementation, we choose the inter-atomic potential for MoS$_2$.
Therefore, to simulate other TMDs (such as WSe$_2$, MoSe$_2$ etc.) using the proposed formulation, only the parameters for inter-atomic potentials need to be changed.
For MoS$_2$ the Stillinger-Weber (SW) inter-atomic potential whose parameters are reported in \cite{Jiang2015a} is widely used. However, we found that the values for the  lattice parameters for the undeformed MoS$_2$ reported in \cite{Jiang2015a} do not correspond to the minimum of the potential. Thus, it shows an anomolous decrease in strain energy from the equilibrium under uniaxial compression. We found the correct lattice parameters corresponding to the minimum of the energy while keeping the other parameters provided in~\cite{Jiang2015a} unchanged. The corrections required in the lattice parameters are less than 1.5\%. A brief comparison between new and reported values is provided in~\ref{app:opt}. 
The SW potential (\cite{Jiang2015a}) with the updated parameters are used in all the numerical examples presented here except in one case where the results by \cite{Jiang2015a} are compared.

To validate the present continuum model,  it is compared with the molecular mechanics simulations, using the same inter-atomic potential under a wide variety of boundary conditions that create complex post-buckling responses. The material modulus, deformed shapes and energies (with and without inner relaxation) are compared. While comparing the results with the atomistic simulations, only small scale samples of a few nanometers are considered. In addition, the continuum model is also validated by comparing against the nano-indentation experiment performed on a micron-scale sample. 
To impose the boundary condition, the displacements for the control points at the boundary are prescribed while the rest of the control points are obtained through energy minimization.
A large displacement boundary condition is reached in successive increments.

The sub-sections~\ref{sec:tension}-\ref{sec:shear_compression} contains the validation against molecular mechanics simulations for various loading conditions.
In the sub-sections~\ref{sec:tension}-\ref{sec:bending_rigidity}, the elastic constants are computed from the continuum simulation and compared against first principal calculations and experiments.
The effect of inner relaxation between two simple lattices is also discussed in these sub-sections. 
Sub-sections~\ref{sec:uniaxial_compression} and~\ref{sec:shear_compression} deals with various loading conditions that leads to complex post-buckling deformations. 
The total energies and the deformed configurations are compared with the atomistic calculations.
The results obtained from the proposed continuum model are presented in the form of stresses, energy and the deformation patterns and compared against purely atomistic simulations.
The sub-section~\ref{sec:indent} deals with the nano-indentation simulation for a large scale sample and experimental validation.
\subsection{Uniaxial and Biaxial Tension Test}
\label{sec:tension}
\begin{figure}[h!]
\centering
\includegraphics[width=\textwidth]{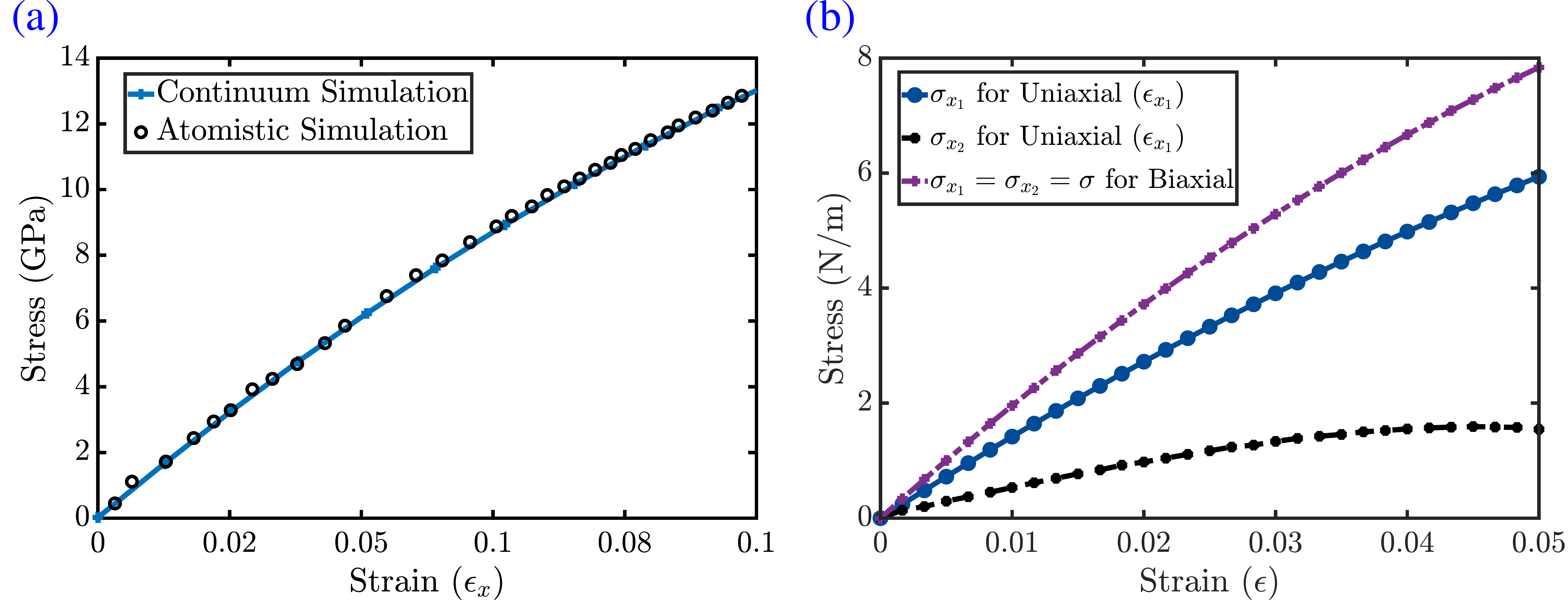}
\caption{{Stress-strain curves under tension.} (a) Uniaxial tension test for monolayer MoS$_2$ of dimension $27.0 \times 28.1$ \AA. The result obtained from the proposed continuum formulation is compared with atomistic simulation given in \cite{Jiang2015a}. {Only for this numerical uniaxial tension  experiment the lattice parameters provided in \cite{Jiang2015a} are used in the continuum model without updating it since their results are based on the the same lattice parameters.} (b) Stress-strain curve for pure uniaxial tension and biaxial tension test plotted with equilibrium bond lengths and angles.}
\label{fig:uniaxial_tension}
\end{figure}

In this section, the present continuum model is compared against the atomistic models for MoS$_2$ subjected to uniaxial and biaxial tension.
The stress-strain curves for uniaxial tension obtained by the present model are compared against the molecular dynamics simulation result reported in \cite{Jiang2015a}, as shown in figure~\ref{fig:uniaxial_tension}\textcolor{blue}{a}. 
The curve is fitted to $\sigma = E\epsilon + \frac{1}{2}D\epsilon^2$, where $E$ is the Young's modulus and $D$ is the Third order elastic constant. 
By considering {the strain range} $\epsilon \in [0,0.01]$, the value of Young's modulus obtained is 167.0 GPa. 
The experimental results reported in \cite{cooper2013nonlinear} measures Young's modulus as $120 \pm 30$ Nm$^{-1}$, which corresponds to $195.12 \pm 49.7$ GPa, by considering an inter-layer distance of 6.15 \AA. In another experiment \citep{Bertolazzi2011ACSNano}, the reported value of Young's modulus is $180 \pm 60$ Nm$^{-1}$, which corresponds to $297.9 \pm 99.3$ GPa, by considering an inter-layer distance of 6.092 \AA.

While performing uniaxial tension, the deformation in the other direction is not allowed to mimic the boundary condition applied in~\cite{cooper2013nonlinear}.
Figure~\ref{fig:uniaxial_tension}\textcolor{blue}{b}, shows the stress-strain results obtained from uniaxial and biaxial tension tests. 
{The elastic properties, such as Young's modulus ($E$) and the Poisson's ratio ($\nu$), are obtained by  fitting to the results for the small strain regime,  $\epsilon \in [0, 0.01]$.}
The values of the material constants thus obtained are compared against the other methods in  Table~\ref{tab:elastic_constants}.
The elastic constants obtained through the present formulation matches well with various atomistic calculations and experiments reported in the literature. 
 
\begin{table}[h!]
    \centering
    \begin{tabular}{ccccccccc} \hline \hline
   \small{Property} & \small{Continuum}$^a$ & \small{ReaxFF}$^b$ & \small{GGA}$^c$ & \small{LDA}$^d$ & \small{GGA}$^d$ & \small{TM}$^e$ & \small{AFM}$^f$ & \small{HSE06-D2}$^g$ \\ \hline
$C_{11}$ (N/m) & 142.35 & 205.08 & 128.4 & 130 & 140 & & & 145.0 \\
$C_{12}$ (N/m) & 53.3 & 81.59 & 32.6 & 40 & 40 & & & 82.90 \\
$E$ (N/m) & 125.41 & 176.32 & 120.1 & 118 & 129 & 123 & 180$\pm$60 & 134.60 \\
$\nu$ & 0.37 & 0.39 & 0.254 & 0.31 & 0.29 & 0.25 & 0.27 (bulk) & 0.57\\
$G$ (N/m) & 46.01 & 61.745 & 47.9 & 45 & 50 & & & 31.05 \\
\hline \hline
\multicolumn{9}{l}{$^a$Present Atomistic-based continuum Model.} \\ \multicolumn{9}{l}{$^b$\cite{ostadhossein2017reaxff}.  $^c$\cite{peng2013theoretical}. $^d$\cite{cooper2013nonlinear}.} \\
\multicolumn{9}{l}{$^e$\cite{Li2012PRB}. $^f$\cite{Bertolazzi2011ACSNano}. $^g$\cite{peelaers2014elastic}.} \\
\hline
\multicolumn{9}{l}{ReaxFF: Reactive Force Field} \\
\multicolumn{9}{l}{GGA: Generalized Gradient Approximation (first principal calculation)} \\
\multicolumn{9}{l}{LDA: Local Density Approximation (first principal calculations)} \\
\multicolumn{9}{l}{TM: Trouiller-Martins (first principal calculations)} \\
\multicolumn{9}{l}{AFM: Atomic Force Microscopy} \\
\multicolumn{9}{l}{HSE: Heyd, Scuseria, and Ernzerhof (first principal calculations)} \\
\hline

\end{tabular}
    \caption{Comparison of elastic constants.}
    \label{tab:elastic_constants}
\end{table}

\subsection{Shear Test}
\label{sec:shear}
\begin{figure}[h!]
\centering
\includegraphics[width=0.5\textwidth]{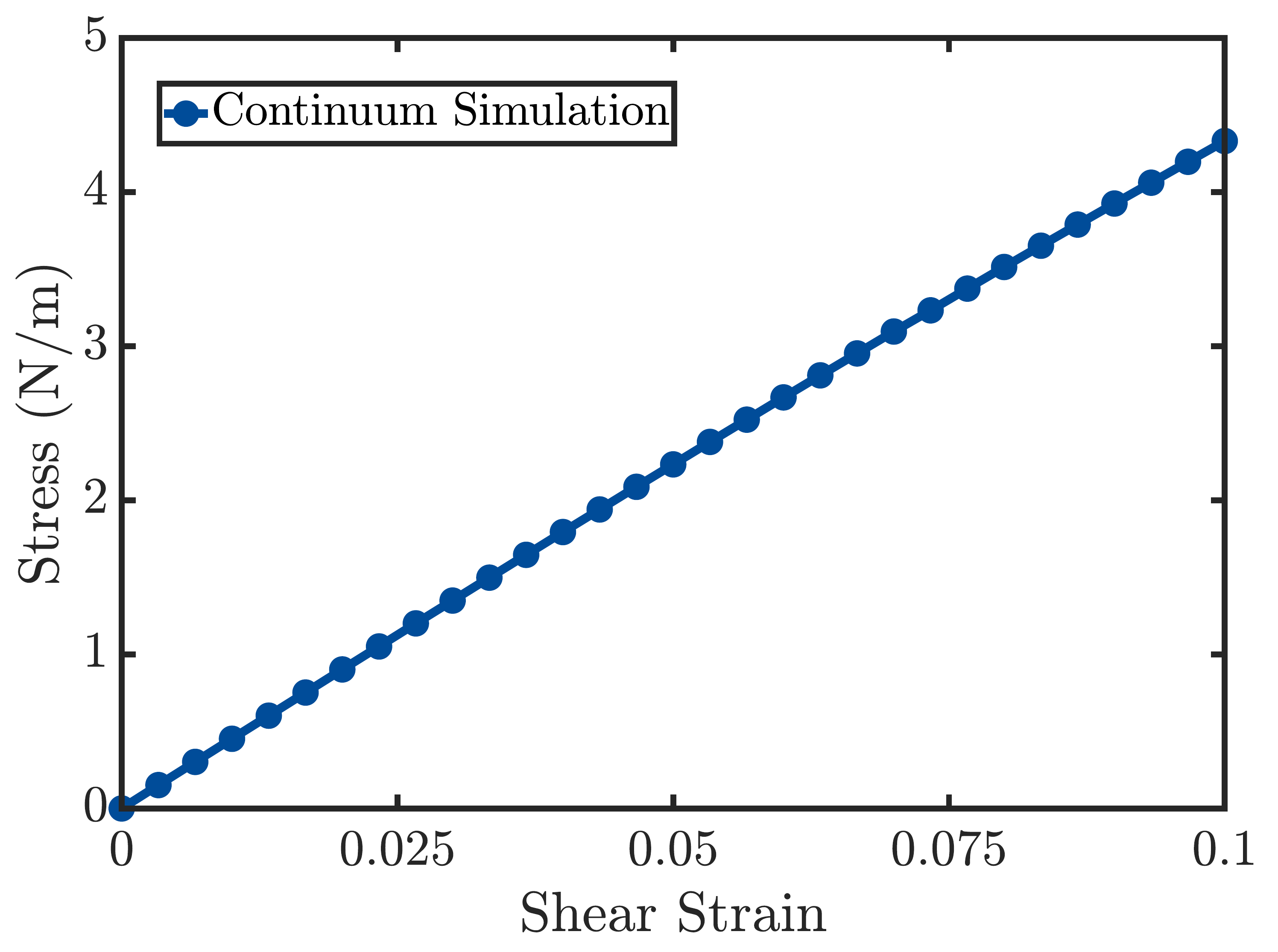}
\caption{Shear stress vs shear strain obtained using the present continuum model.}
\label{fig:shear}
\end{figure}

To perform the shear test, pure shear is applied on the same sample used in the tension test, and the corresponding stress-strain curve is obtained, as shown in figure~\ref{fig:shear}. 
Considering only the the linear regime, the shear modulus obtained from the shear test is 45.2176 N/m. The shear modulus obtained from the biaxial tension test is 46.01 N/m following $G=(C_{11}-C_{12})/2$.
Both the values lie well within the range provided in various atomistic simulations and experiments, given in Table~\ref{tab:elastic_constants}.

\subsection{Bending Modulus}
\label{sec:bending_rigidity}
\begin{figure}[htbp]
\centering 
\includegraphics[width=\textwidth]{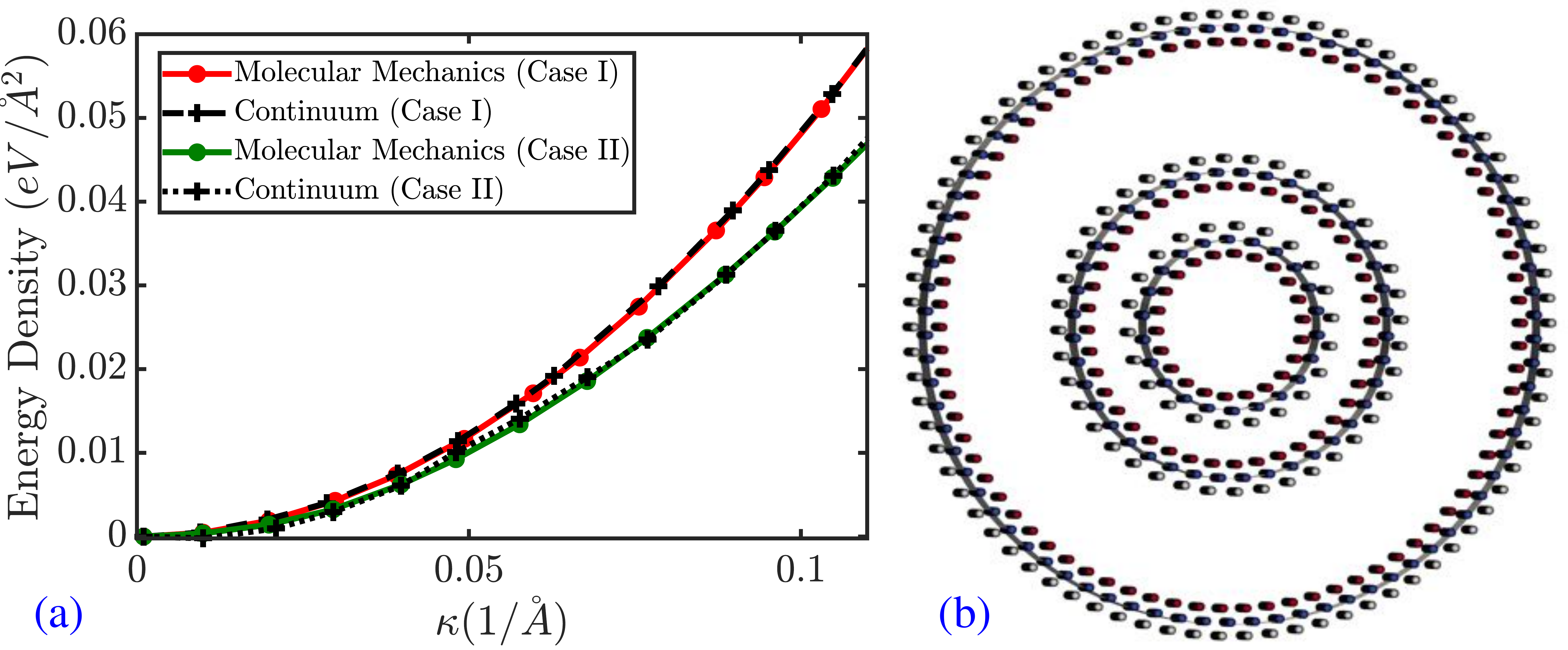} 
\caption{(a) Energy density of MoS$_2$ cylinders for various curvatures obtained by the continuum model and the molecular mechanics simulations.
{Case I: none of the atoms are allowed to move freely from the ideal cylindrical shape. The corresponding bending modulus is $D = 9.61eV$. Case II: all atoms are allowed to move freely and the energy is minimized. The corresponding bending modulus is $D = 7.565eV$.} 
(b) The deformed shapes (for Case II) obtained by both continuum and molecular mechanics simulations for various curvatures ( $\kappa = 0.0295, 0.0576 \: \text{and} \: 0.1048\, \text{\AA}^{-1}$). The continuum membrane and the atomic positions are shown by the shaded surface and the colored spheres respectively. The three tubes are obtained from three different simulations.}
\label{fig:bending_curvature}
\end{figure}

In this section, the bending modulus is computed and compared with the values reported in the literature obtained from the simulations and experiments. 
To obtain the bending modulus of an MoS$_2$ sheet, cylinders of different radii are considered, and their energy is minimized to obtain the equilibrium configuration. The bent configurations obtained by this method are also compared.
The effect of inner relaxation on bending modulus is also investigated. 
The results obtained from the continuum formulation are compared with the molecular mechanics simulations where both the models use the same inter-atomic potential.
Energy density at various bending curvatures and the corresponding deformation patterns are presented in figure~\ref{fig:bending_curvature}\textcolor{blue}{a}. It shows an excellent match between the continuum model and molecular mechanics simulations for both of the following cases: (I) none of the atoms are allowed to move freely from the ideal cylindrical shape, (II) all of the atoms are allowed to move freely to attain equilibrium.

The bending modulus ($D$) is calculated by fitting the equation $E = \frac{1}{2}D\kappa^2$ to the energy--curvature ($E$--$\kappa$) data, plotted in figure~\ref{fig:bending_curvature}\textcolor{blue}{a}.
The bending modulus computed by  ~\cite{Jiang2013Nanotech} for MoS$_2$ is $D = 9.61eV$. However, this value is not for an  equilibrated  system -- none of the atoms were allowed to move freely from the ideal cylindrical shape that has the same thickness as the planar undeformed MoS$_2$. 
When these assumptions are maintained in the present continuum model, it yields a bending modulus of $D=9.6eV$, which matches well with \cite{Jiang2013Nanotech} (see the Case I of  figure~\ref{fig:bending_curvature}\textcolor{blue}{a}). However, not allowing the atoms to move freely would overestimate the bending energy. In the present continuum model, all atoms are allowed to freely move during the energy minimization, and the relative shifts are incorporated. The present model does not overestimate the bending energy and matches excellently with the molecular mechanics simulation that does not constraint the atoms as shown in  Case II of figure~\ref{fig:bending_curvature}\textcolor{blue}{a}. This  energy (Case II) yields the bending modulus as $D = 7.656eV$. This value of the bending modulus falls within the experimentally obtained range, $D = 6.62 - 13.24\;eV$~\citep{Bertolazzi2011ACSNano,cooper2013nonlinear}.
%

The equilibrium shapes obtained through the continuum model and the atomic positions obtained by the molecular simulation are compared for various curvatures in  figure~\ref{fig:bending_curvature}\textcolor{blue}{b}, demonstrating high accuracy of the continuum model. 

\subsection{Uniaxial Compression}
\label{sec:uniaxial_compression}
\begin{figure}[h!]
\centering
 \includegraphics[width=\textwidth]{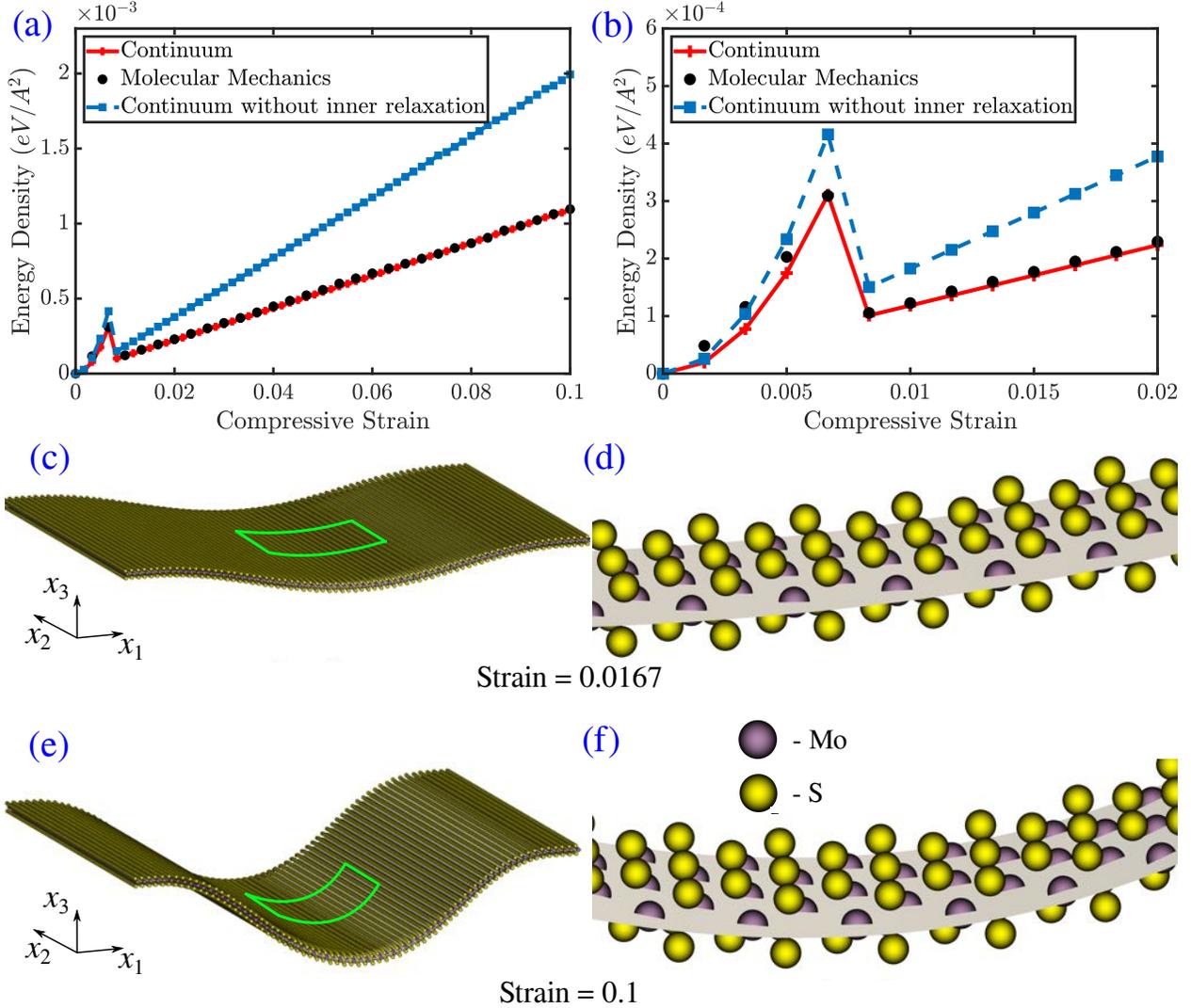}
\caption{Uniaxial compression test for monolayer MoS$_2$ sheet of dimension $200 \times 500$ \AA. (a) Energies under compression obtained by the present continuum model and molecular mechanics model. The energy under compression by the present continuum model without inner relaxation is also shown.
(b) The comparison of energetics close to buckling point. 
(c-d) Comparison of deformations obtained from the continuum and molecular mechanics simulations at 0.0167 strain.
(e-f) Similar comparison of deformations obtained from the continuum and molecular mechanics simulations at 0.1 strain. (d) and (f) show the magnified views of the deformations provided in (c) and (e) respectively.}
\label{fig:uniaxial_compression}
\end{figure}

To investigate the buckling behavior of the continuum membrane, a uniaxial compression test is performed. The deformed shapes and the energy are validated against the molecular mechanics  simulation as shown in figure~\ref{fig:uniaxial_compression}. 
A continuum membrane sample of size 200 \AA\, $\times$ 500 \AA\,  is subjected to uniaxial compression.
In the continuum model, in addition to compressive strain (along $x_1$ direction), periodic boundary conditions are applied along both $x_1$ and $x_2$--directions (see figure~\ref{fig:uniaxial_compression}\textcolor{blue}{(c,e)}).
This boundary condition is used in the continuum model to mimic the molecular mechanics simulation for validation.

\begin{figure}[t]
\centering
 \includegraphics[width=\textwidth]{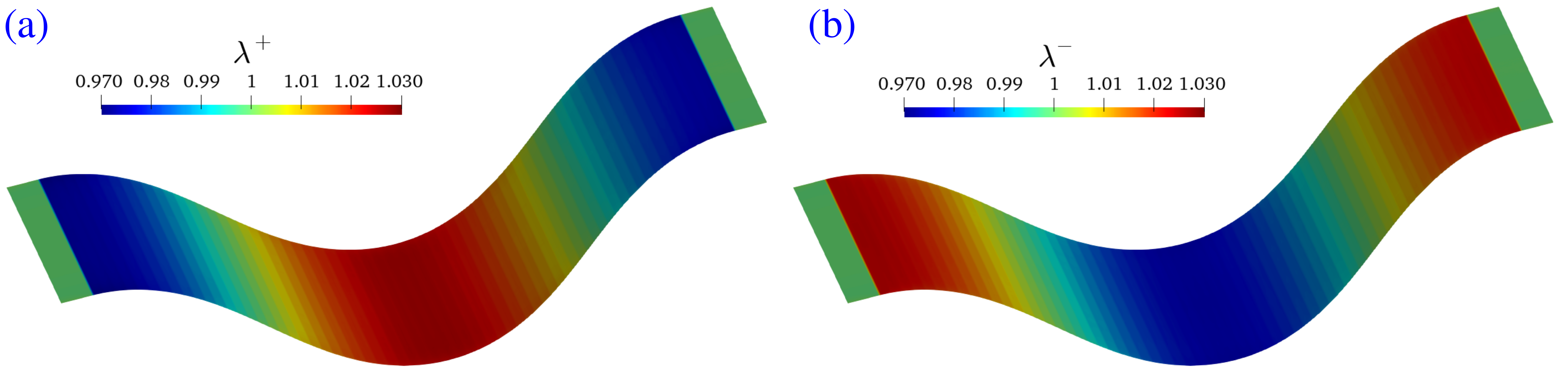}
\caption{Normal stretches: (a) above ($\lambda^+$) and  (b) below ($\lambda^-$) the middle surface at 10\% compression.}
\label{fig:uniaxial_compression_lambda}
\end{figure}
The same simulation is performed via molecular mechanics that uses the same inter-atomic potential and periodic boundary conditions. 
The energy minimization is performed to obtain the equilibrium configuration at each load increment. 
The comparison of energy and the deformation pattern is shown in figure~\ref{fig:uniaxial_compression}.
Under compression, the MoS$_2$ membranes tend to bend easily as their bending rigidity is much smaller compared to their in-plane rigidity. A similar deformation pattern for MoS$_2$ under compression is shown in the experimental work by  \cite{Castellanos-Gomez2013NanoLett}. 

The total energy of MoS$_2$ membranes as a function of compressive strain is plotted in figure~\ref{fig:uniaxial_compression}\textcolor{blue}{(a-b)}. 
Under small compression, the membrane gets compressed without buckling and the energy varies in a quadratic fashion. 
The buckling point is predicted very accurately by the continuum model.
The continuum model without the inner relaxation also predicts the buckling point correctly but significantly overestimates the total energy density.
After the buckling point, the energy grows linearly with compressive strain and the total energy is now dominated by the bending.
In the post-buckling region, the energy obtained by the continuum model with inner relaxation remarkably match with the molecular mechanics simulation. However, the continuum model without the inner relaxation incorrectly predicts the energy much higher than the molecular model.
This result highlights the need for inner relaxation.

The deformed configurations obtained from both the continuum and the molecular mechanics simulations at 0.0167 strain, just after buckling point, is shown in figure~\ref{fig:uniaxial_compression}\textcolor{blue}{(c-d)}.
A similar comparison at 0.1 strain is shown in figure~\ref{fig:uniaxial_compression}\textcolor{blue}{(e-f)}.
In figure~\ref{fig:uniaxial_compression}\textcolor{blue}{(c-f)}, the shaded surface denotes the middle surface predicted by the present continuum model, whereas the positions of the atoms obtained by the molecular mechanics simulation are shown by spheres of different colors.
The Mo-atoms coincided with the continuum surface ensuring remarkable accuracy by the present continuum model.

The change in thickness under compression are explored here by plotting the normal stretches in figure~\ref{fig:uniaxial_compression_lambda}. 
The portion of the membrane near to the boundaries has unit normal stretches  since they are kept fixed to impose the clamped  boundary condition.  
At 10\% strain, the membrane is buckled and its deformation is dominated by bending compared to the in-plane strains. In the middle portion of the membrane, the stretch above the middle surface ($\lambda^-$) is greater than 1, representing that the distance between the middle surface and the top surface has increased. Whereas, in this portion the normal stretch below the middle surface ($\lambda^-$) is less than 1, representing a reduction in distance between the middle surface and the bottom surface. These results are in accord with our intuition since in the middle of the sample, the top surface has lower radius of curvature than the middle surface; thus, the top surface is under compression. The top surface accommodates its extra length by moving away from the middle surface, thus showing a stretch greater than one. However, near the edges, the top surface has higher radius of curvature than the middle surface and hence shows stretch less than one.  The bottom surface has the opposite change in its radius of curvature than the top surface and  hence shows opposite change in  stretches than the top. This result justifies the incorporation of the two normal stretches in the present membrane formulation.

\subsection{Shear and Compression}
\label{sec:shear_compression}
\label{sec:shear_compression}
\begin{figure}[tbp]
\centering
 \includegraphics[width=\textwidth]{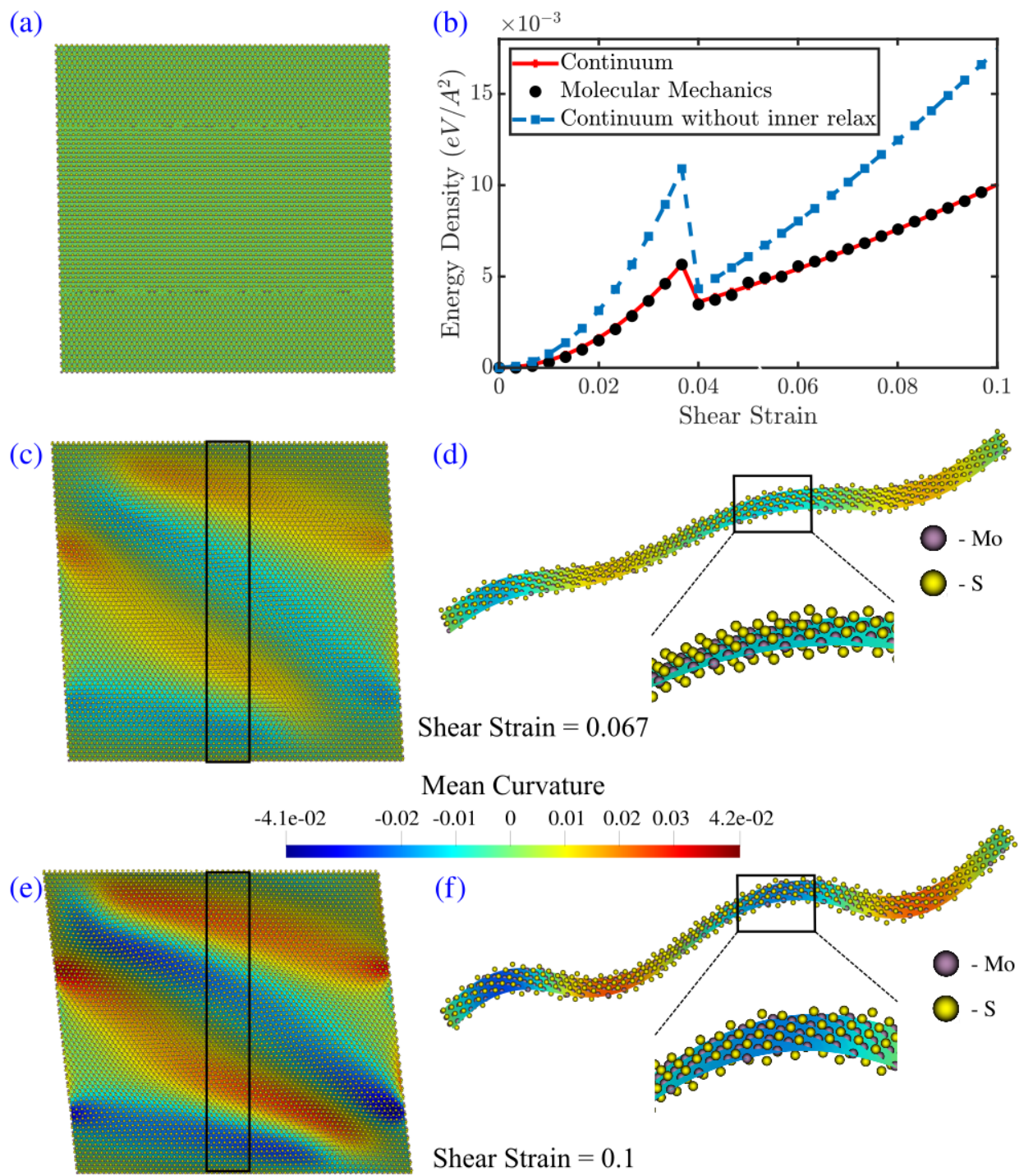}
\caption{Combined shear and compression test for a MoS$_2$ membrane of dimension $200 \times 200$ \AA \;by both the continuum model and the molecular mechanics model. The strain is the same in both shear and compression. 
(a) Solution of the continuum model is shown by the colored surface.
The atomic positions obtained by molecular mechanics are shown using the colored spheres. 
(b) The energies obtained by the continuum model with inner relaxation are compared against the molecular mechanics simulation. The energies by the continuum model without the inner relaxation is also shown.
(a,c,e) Top views of the deformed shapes of the MoS$_2$ membrane at strains 0.033, 0.067 and 0.1 respectively.
(d,f) Cross-sectional views and magnified views corresponding to figure (c) and (e) respectively. Colors on the surface obtained through the continuum model denote the mean curvature.
The atomic arrangement is obtained through the molecular mechanics simulation.
}
\label{fig:shear_compression}
\end{figure}

\begin{figure}[t]
\centering
 \includegraphics[width=\textwidth]{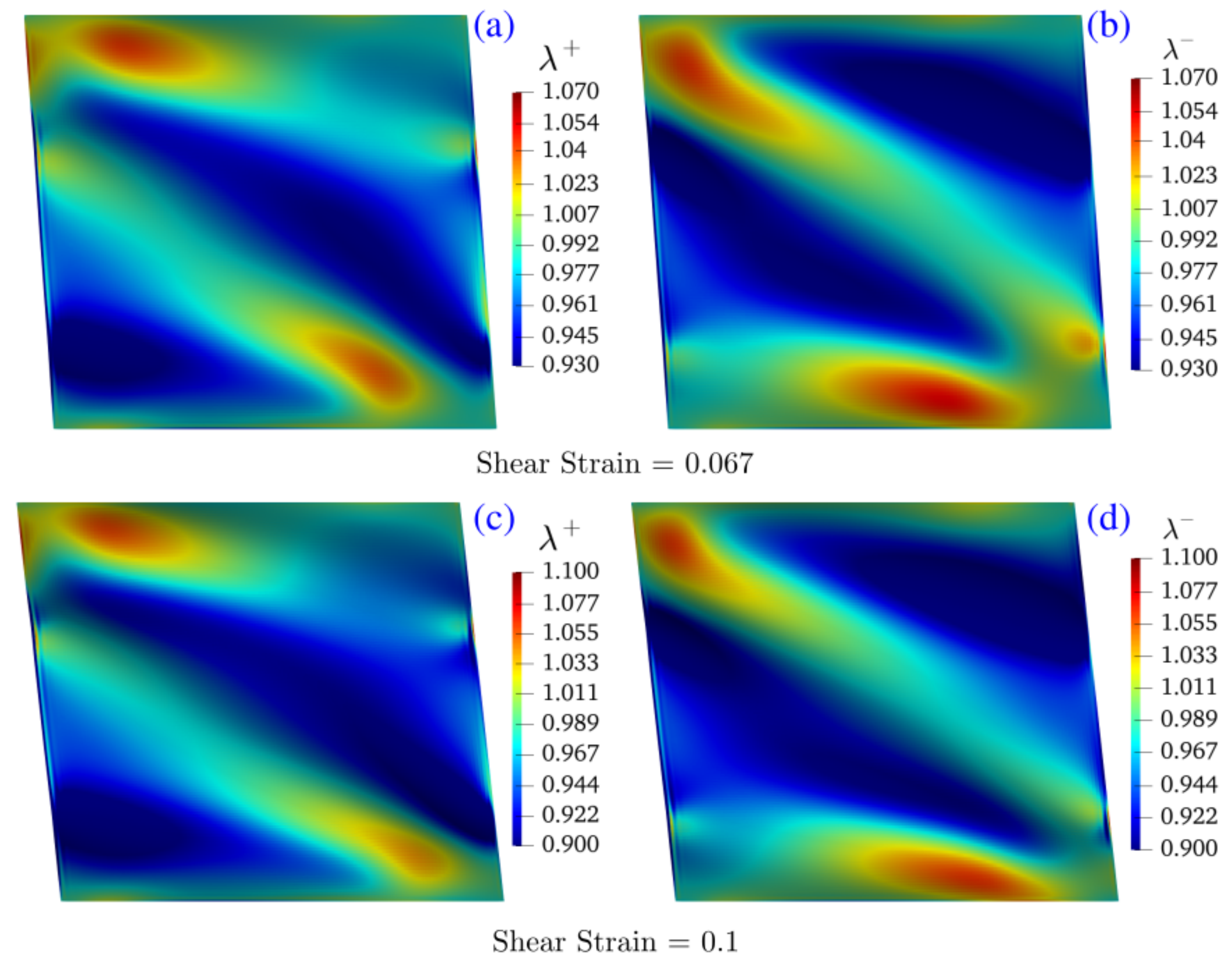}
\caption{Normal stretches ($\lambda^+$ and $\lambda^-$) on the deformed configurations at (a,b) 0.067 shear strain and (c,d) 0.1 shear strain.
}
\label{fig:shear_compression_lambda}
\end{figure}
The continuum model is put to the test to predict a more complicated post-buckling deformation by simultaneously applying shear and compression on a MoS$_2$ sample.
In this test, equal compression and shear strains are applied simultaneously on the top and bottom edges of MoS$_2$ membrane as shown in figure~\ref{fig:shear_compression}\textcolor{blue}{a}.
The magnitude of normal and tangential displacements at the boundaries are kept the same at each displacement increment.
In addition, a periodic boundary condition on the displacement normal to the undeformed membrane surface is also applied along both the directions.
The deformed shapes predicted by the continuum model are compared against that by the molecular mechanics simulation.
The boundary condition in the molecular mechanics simulation is kept the same as the continuum model
at each increment, and the atoms are allowed to obtain the equilibrium configuration through energy minimization. 

The deformed shapes and energies for combined shear and compression loading are shown in figure~\ref{fig:shear_compression}.
Under this boundary condition, the energy shows two different regions with respect to strain, before and after the buckling as shown in figure~\ref{fig:shear_compression}\textcolor{blue}{b}.
Before buckling, the membrane undergoes in-plane shear and compression without any out--of--plane deformation, as shown in figure~\ref{fig:shear_compression}\textcolor{blue}{a} for 0.033 shear strain and 0.033 compressive strain.
Before buckling, the energy increases in a quadratic fashion. 
The effect of inner relaxation is evident from the energy comparisons as shown in figure~\ref{fig:shear_compression}\textcolor{blue}{b}.
As expected, the energy predicted by the continuum model without the inner relaxation is higher than that with inner relaxation and the molecular mechanics simulation. 
The post-buckling deformations are shown in figure~\ref{fig:shear_compression}\textcolor{blue}{(c-f)} for combined shear and compressive strain of 0.067 and 0.1. The colormap of the continuum surface denotes the mean curvature. 
To show the post-buckling wrinkles, colormaps of the mean curvature are plotted on the deformed MoS$_2$ membrane in figure~\ref{fig:shear_compression}\textcolor{blue}{(c,e)}.
The undulations in the deformation are shown through magnified cross-sectional views in figure~\ref{fig:shear_compression}\textcolor{blue}{(d,f)}.
The magnified views in figure~\ref{fig:shear_compression}\textcolor{blue}{(d,f)} shows an excellent match between the present continuum model and the molecular mechanics model. 
The location of Mo-atoms predicted by the molecular mechanics simulation coincides with the deformed Mo-surface predicted by the continuum model.
This combined loading leads to a very complicated buckled deformation.
Despite this complexity in the deformation, the continuum model's prediction matches remarkably well with the molecular mechanics.

The thickness variation of the monolayer MoS$_2$ sheet under this combined loading is studied here. The normal stretches above ($\lambda^+$) and below ($\lambda^-$) the middle surface are shown in  Figure~\ref{fig:shear_compression_lambda}. The membrane shows significant normal stretch at the locations of high curvatures; the  maximum stretch goes beyond 10\%. 
This demonstrates the necessity to include the normal stretches in the present continuum membrane model to accurately capture the deformation.

The excellent match in both energy and deformation ensures a great accuracy by the proposed continuum formulation.
For such a small sample, the computational advantage of the proposed continuum model over molecular mechanics is not very significant as both the simulations takes a few seconds in a desktop computer.

\subsection{Simulation of Nano-indentation experiment}
\label{sec:indent}
\begin{figure}[b!]
\centering
\includegraphics[width=1\textwidth]{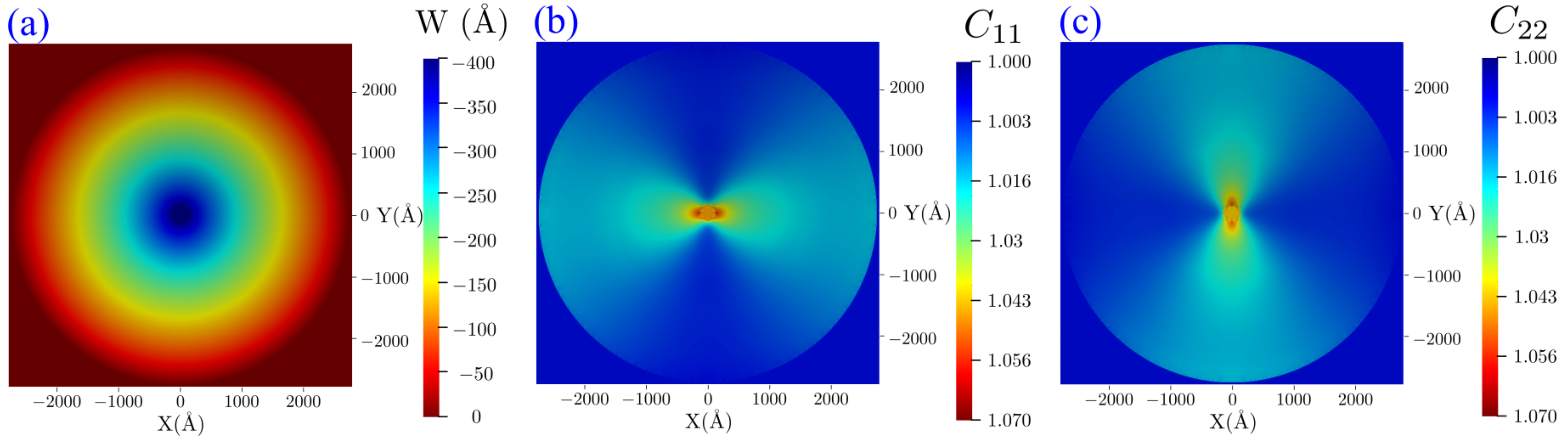}
\caption{Deformation patterns obtained for nano-indentation at 40nm. The colorplots show (a) the out-of-plane deformation, in-plane strains (b) $C_{11}$, and (c) $C_{22}$.} 
\label{fig:indent}
\end{figure}
We validate the continuum model against a nano-indentation experiment by simulating large-area samples.
Nano-indentation test has proven to be one of the most effective methods for material characterization of 2D materials. 
Nano-indentation experiments typically contain micron-size samples containing millions or billions of atoms.
In the absence of atomistic--based predictive models that can reach the length scale of experimental samples with a modest high performance computing facility, the following two routes are usually taken.
First, the small scale atomistic models containing a few hundreds of atoms \citep{hu2016molecular}.
Second, the phenomenological models whose parameters are obtained by fitting to experimental $F-\delta$ data~\citep{cooper2013nonlinear}.
The challenge with 2D materials is the dependency of material characteristics on the applied loads. 
Under small deformations, most of the 2D materials behave in an isotropic manner; however, for moderate to large deformations, they exhibit a nonlinear anisotropic behavior.
Due to this complexity in the material character, it is difficult to derive an accurate analytical relationship between the indenter displacement and the in-plane stresses \citep{cao2019mechanical}. 

In experiments, a sheet of 2D material is mounted on the substrate and indented by using an Atomic Force Microscopy (AFM) tip.
Depending on the absence or presence of substrate beneath the 2D material, the experiments can be classified in two categories \citep{cao2019mechanical}: (i) indent on free standing sheet, referred to as free standing indentation (FSI)
(ii) indent with substrate beneath the 2D material.
The material properties of 2D materials are obtained by fitting the experimental $F-\delta$ curve to the analytical expression. The analytical expression depends on the geometry of the sample and the boundary condition applied to it, as discussed in \cite{cao2019mechanical}. 
The experimental values of Young's modulus mentioned in Table~\ref{tab:elastic_constants} are obtained following this methodology.
\begin{figure}[b!]
\centering
\includegraphics[width=1\textwidth]{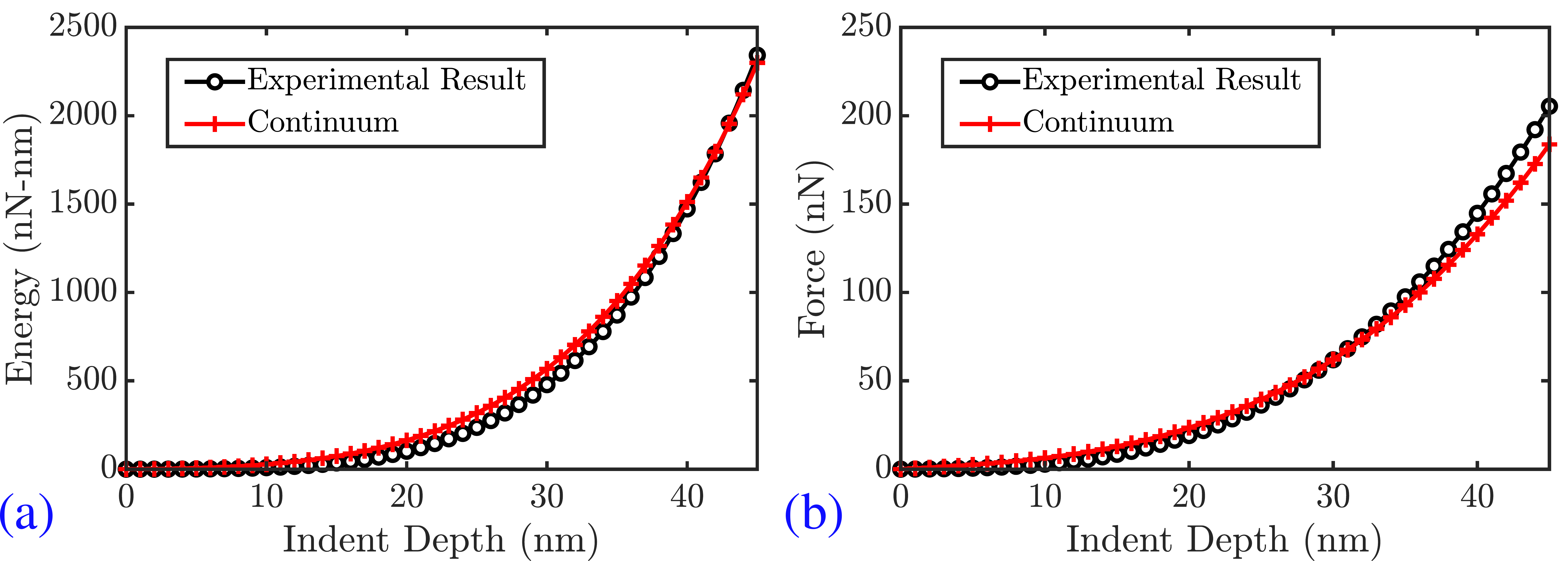}
\caption{Comparison of (a) energy and (b) force with respect to indentation depth obtained from the continuum simulation against experiment performed in  \cite{Bertolazzi2011ACSNano}.}
\label{fig:indent_energy}
\end{figure}


In the present work, the $F-\delta$ curve is obtained from the presented continuum model is compared against that by experiments performed in  \cite{Bertolazzi2011ACSNano}. 
In the experiment by~\cite{Bertolazzi2011ACSNano}, the MoS$_2$ monolayer is mounted on a substrate containing holes of diameter $550\pm 10$nm. The monolayer MoS$_2$ is indented by the atomic force microscopy (AFM) tip of radius 12 $\pm$ 2nm at the center of the substrate-hole.
To simulate the experiment by the continuum model, a sample of 560 nm $\times$ 560 nm is considered. 
The control points lying outside the hole radius are clamped and the rest of the control points are allowed to move freely under indentation.  
Figure~\ref{fig:indent}\textcolor{blue}{a} shows the deformation at 40 nm indentation depth.
The corresponding in-plane strains ($C_{11}$ and $C_{22}$) at 40 nm indenter depth are provided in figure~\ref{fig:indent}\textcolor{blue}{b} and figure~\ref{fig:indent}\textcolor{blue}{c}. 
It is evident from the figures that the membrane experiences maximum strains (about 7\% tension) under the indenter tip. As expected, it shows no compression anywhere in the membrane. We found that the mean curvature is almost zero everywhere except at the edge of the hole. Since the membrane is clamped at the edge, under tension, it does not slide over substrate and hence does not show any wrinkle. 

The comparison of total energy and force with respect to the indenter displacement is shown in figure~\ref{fig:indent_energy}.
Untill 38 nm indenter displacement, the comparison of $F-\delta$ is remarkable.
The difference in force and energy in the later part is because of the difference in Young's modulus predicted by two methods (continuum model and the experiment).
The difference in Young's modulus obtained by continuum and the experiment is mentioned in Table~\ref{tab:elastic_constants}.
In the experimental $F-\delta$ curve plotted here, the effect of pre-tension is eliminated as there exist no pre-tension in the continuum model.
The comparison of $F-\delta$ with the experiment can be considered as the validation of the present formulation and its numerical implementation. 

For this sample size, approximately 101,000,000 degrees of freedom are required to perform the purely atomistic simulation, whereas only 13,000,000 degrees of freedom are required for the present continuum model to obtain its response accurately.
This reduction in degrees of freedom without compromising the accuracy demonstrates the computational efficiency gained through the present continuum model. This enables the present continuum model to simulate larger scale samples while respecting the physics at the atomic length scale.

\section{Conclusion and Discussions}
\label{sec:conclusion}
A novel atomistic--based continuum membrane model for multi--atom--thick Transition metal Dichalcogenides (TMDs) is presented. The proposed crystal-elasticity formulation obtains the hyper-elastic potential of the material from the inter-atomic potential. 

TMDs have multiple atoms along their thickness and covalent bonds inclined to its  middle surface; thus, the \emph{crystal--elasticity}--based model for purely 2D membranes (e.g. Graphene) can not be directly applied to TMDs. This poses a key challenge for development of an efficient predictive model for TMDs. The present crystal-elasticity model overcomes this challenge by extending the purely 2D membrane model to incorporate the effect of thickness through two normal stretches. The covalent bonds inclined to the middle surface of TMDs are projected to the tangent and normal to the middle surface. The deformations of these tangential and normal components are computed using the \emph{exponential Cauchy-Born} and \emph{Cauchy-Born} rules, respectively, to compute the deformed bonds. Beyond this approximation, no other assumptions are used in the kinematics to incorporate the thickness of the membrane. The strain energy per unit area is represented in terms of the continuum strains.    
The strain energy of the continuum model depends on the strains  defined in the reference frame, hence it is material  frame-indifferent. The present model also incorporates the relative shifts between two simple lattices forming a complex lattice of TMD. In addition, a continuum energy is also computed for non-bonded interactions. Since the present continuum model builds on inter-atomic potentials, it is independent from any material modulus obtained from either the atomistic calculations or experiments. The continuum model is numerically implemented using a  smooth finite element framework based on B-spline basis, which provides a greater smoothness of the approximated fields than standard finite element discretization. This greater smoothness of the approximated fields is required for the present continuum model due to its dependence on curvature. 

Molecular mechanics simulations for small size TMDs are used as a reference for validation of the model  since the aim is to replicate the prediction of the discrete molecular models with the present continuum model and its numerical implementation.  The elastic material properties such as Young's modulus, shear modulus, and the bending modulus of an MoS$_2$ mono-layer obtained by the present model shows a good match with various \emph{ab initio} calculations, molecular mechanics simulations, and experiments reported in the literature.  
In addition, the present continuum model shows remarkable agreement with the molecular mechanics simulations for large post-buckled deformations of TMDs subjected to compression, shear, and their combinations. 
Both the complex deformed configurations and the equilibrium energies are compared to demonstrate the high accuracy of the present model. The present model also demonstrates that it can predict different normal stretches above and below the middle surface, which corroborates well with the deformation of the membrane. 
The effect of relative shift between two simple lattices (referred as inner relaxation) on the prediction of the model is found to be significant. 
The present model is also validated against nano-indentation experiments. It demonstrates that the present model can reach experimental length scales starting from inter-atomic potential while using a modest computational facility. 

The proposed model should offer significant computational efficiency over purely atomistic simulations of TMDs due to its continuum--finite element approach while offering accuracy similar  to purely atomistic simulations. 
A detailed study on the computational efficiency will be undertaken in our future work. 

The present results demonstrate the accuracy of the present continuum membrane model and validate it for a range of problems. However, further investigation is required to estimate the errors incurred due to the kinematic approximations and to identify the domain of applicability of the model.  \\

\noindent
\textbf{Acknowledgments:} 
The work is supported by NSF (CMMI MoMS) under grant number 1937983. We acknowledge Superior, a high-performance computing facility at MTU. This work used the Extreme Science and Engineering Discovery Environment (XSEDE), which is supported by the NSF grant number ACI-1548562. This work used the XSEDE Bridges at the Pittsburgh Supercomputing Center through allocation  MSS200004.

\appendix
\section{Optimization of equilibrium lattice parameters}
\label{app:opt}
\begin{figure}[htbp]
\centering
\includegraphics[width=\linewidth]{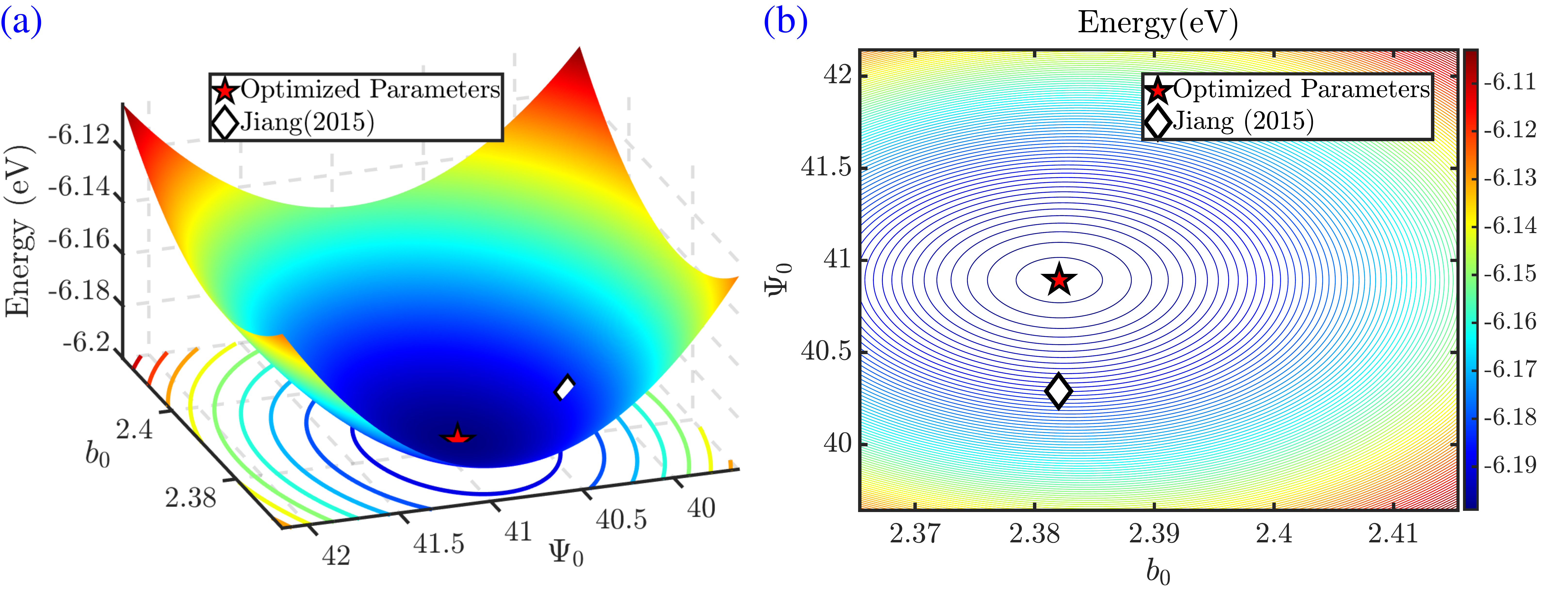}
\caption{Comparison of the lattice parameters reported in  from~\cite{Jiang2015a} and optimized by energy minimization in the present work are shown via (a) surface plot and (b) contour plot.}
\label{fig:opt_parameters}
\end{figure}

The unit cell of MoS$_2$ is shown in figure~\ref{fig:lattice_inner}\textcolor{blue}{b}. All the bond lengths and bond angles of the undeformed unit cell can be obtained only through two lattice variables $b_0$ and $\Psi_0$.
The thickness of the unit cell ($2h_0$) can be obtained using these two variables as
\begin{equation}
    2h_0 = 2b_0\sin\bigg(\frac{\Psi}{2}\bigg)
\end{equation}
Along with the parameters for the Stillinger-Weber potential, these two lattice variables are provided in~\cite{Jiang2015a}.
We found that these parameters do not correspond to minimum energy at equilibrium, which resulted in anomalous negative strain energy under compression. 
We optimized these parameters corresponding to the minimum energy. 
The comparison of values is shown in figure~\ref{fig:opt_parameters}.
The optimized parameters are used to perform the numerical validation of the present formulation.

\section{Properties of perpendicular projection operator}
\label{app:perpendicular_proj}
For any smooth and open surface $S$ defined in the basis $\mathscr{B}$, let $\bf{m}$ be the unit vector and $\mathscr{T}$ be the tangent space defined at any point on the surface. Assuming that the tangent space is defined in the basis $\mathscr{C} = \{\mathbf{g}_1,\mathbf{g}_2\}$, the perpendicular projection of $\mathbb{R}^3$ to $\mathscr{T}$ at that point can be expressed as

\begin{equation}
    \mathbb{P}_{\mathbf{m}} = \mathcal{I} - \mathbf{m} \otimes \mathbf{m}
    \label{eq:app_projection}
\end{equation}
where $\mathcal{I}$ is the identity operator which performs a linear transformation from $\mathbb{R}^3$ to $\mathbb{R}^3$. 
$\mathbf{g}_1$ and $\mathbf{g}_2$ represent the convected basis vectors.
The surface gradient $\nabla_{S_0}f$ of a scalar field $f$ at a point on the surface $S$ is defined as
\begin{equation}
   \nabla_{S_0}f = \mathbb{P}_m \nabla f 
\end{equation}
For a given vector field $\bf{v}$
defined on $\mathbb{R}^3$, the gradient and divergence with respect to the point can be denoted as $\nabla \bf{v}$ and Div $\bf{v} = \nabla \bf{v} \cdot \bf{I}$. The surface gradient of vector field $\bf{v}$ defined at a point, where $\mathbf{m}$ is the unit vector and $\mathscr{T}$ is the tangent space on the surface $S$, can be expressed by 
\begin{equation}
\nabla_S \bf{v} = (\nabla \bf{v}) \mathbb{P}_{\mathbf{m}} 
\end{equation}
Using the definition of gradient, we can obtain
\begin{equation}
\nabla \bf{v} = \nabla \bf{v}(\mathbb{P}_{\mathbf{m}} + \mathbf{m} \otimes \mathbf{m}) = \nabla_S \bf{v} + \frac{\partial \bf{v}}{\partial \mathbf{m}} \otimes \mathbf{m}
\end{equation}
where 
\begin{equation}
\frac{\partial \bf{v}}{\partial \mathbf{m}} = (\nabla \bf{v})\mathbf{m}
\end{equation}
\section{The Weingertan Map of a surface}
\label{app:surf}

The Weingarten map, $\mathcal{W}(\mathbf{a})$, at point $\mathbf{a}$ on the surface $S$ is defined by 
\begin{equation}
\mathcal{W}(\mathbf{a}) = -T_{\mathbf{a}}\mathscr{G}
\end{equation}
 where $\mathscr{G}$ is the Gauss map\footnote{
 \textit{Gauss Map:}\\
Consider a point $\mathbf{a}$ on $S$. A Gauss map, $\mathscr{G}$, maps a point $\mathbf{a} \in S$ to a point $\mathbf{n}_{\mathbf{a}}$ on the unit sphere, $S^2$,  centred at $(0,0,0)$. Here, $\mathbf{n}_{\mathbf{a}}$ is the unit normal at $\mathbf{a}$. The map can be expressed as
\begin{equation}
\mathscr{G}: \mathbf{a} \in S \to \mathbf{n}_{\mathbf{a}} \in S^2
\end{equation}
Moreover, the rate at which unit normal is varying can be obtained by taking the derivative of the Gauss map, such as
\begin{equation}
T_{\mathbf{a}}\mathscr{G} : T_{\mathbf{a}}S \to T_{\mathscr{G}(\mathbf{a})}S^2
\end{equation}
Since the unit normal at point $\mathbf{a}$ and $\mathscr{G}(\mathbf{a})$ is the same, the derivative of the Gauss map is a linear map from $T_{\mathbf{a}}S$ to itself.}.

The matrix representation of Weingarten map, $\mathcal{W}(\mathbf{a})$, with respect to the convected basis set $\{\mathbf{g}_1,\mathbf{g}_2\}$ can be expressed as
\begin{gather}
[\mathcal{W}(\mathbf{a})] = \begin{bmatrix} E & F \\ F & G \end{bmatrix}^{-1} \begin{bmatrix} L & M \\ M & N \end{bmatrix}.
\end{gather}
where the matrix $\begin{bmatrix} E & F \\ F & G \end{bmatrix}$ is the matrix representation of the First Fundamental Form and the matrix $\begin{bmatrix} L & M \\ M & N \end{bmatrix}$ is the matrix representation of the Second Fundamental Form of the surface $S$.
The gaussian curvature, $K$, and mean curvature, $H$, for a surface are defined as
\begin{equation}
K = \text{det}(\mathcal{W}), \; \; \; H = \text{trace}(\mathcal{W})
\end{equation} \\
For a point $\mathbf{a} \in S$, consider scalars $k_1$ and $k_2$ and a basis set $\{\mathbf{v}_1,\mathbf{v}_2\}$ defined on the tangent plane $T_{\mathbf{a}}S$, such that
\begin{equation}
\mathcal{W}(\mathbf{v}_1) = k_1 \mathbf{v}_1, \; \; \mathcal{W}(\mathbf{v}_2) = k_2 \mathbf{v}_2.
\end{equation}
The scalars $k_1$ and $k_2$ are the eigen values of the Weingarten map and are known as principal curvatures. Similarly, $\mathbf{v}_1$ and $\mathbf{v}_2$ are the principal vectors corresponding to $k_1$ and $k_2$. \\

\section{Principal directions and principal values of the curvature tensor}
\label{ap:princi_curv_ders}

Principal directions and principal values of the curvature tensor are obtained by solving the eigenvalue problem given in equation \ref{eq:eig}. See references \cite{Arroyo2004IJNME,Arroyo2002JMPS} for further details. The eigenvalues of the curvature tensor are also called the principal curvatures. Gaussian curvature, $G$, and mean curvature,$H$, at each point is expressed as
\begin{align}
   G & =\frac{\operatorname{det}[\widehat{\mathscr{K}}]}{\operatorname{det}[\widehat{\mathbf{C}}]}=\frac{\widehat{{K}}_{11} \widehat{{K}}_{22}-\widehat{{K}}_{12}^{2}}{\widehat{\text{C}}_{11} \widehat{\text{C}}_{22}-\widehat{\text{C}}_{12}^{2}} \\
   H & =\frac{1}{2} \operatorname{trace}\left([\widehat{\mathbf{C}}]^{-1}[\widehat{\mathscr{K}}]\right)=\frac{1}{2} \frac{\widehat{{K}}_{11} \widehat{\text{C}}_{22}-2 \widehat{{K}}_{12} \widehat{\text{C}}_{12}+\widehat{{K}}_{22} \widehat{\text{C}}_{11}}{\widehat{\text{C}}_{11} \widehat{\text{C}}_{22}-\widehat{\text{C}}_{12}^{2}}
\end{align}
Using these expressions, the principal curvatures can be obtained as
\begin{equation}
    k_{1,2} = H \pm \sqrt{H^2-G}
\end{equation}
Plugging the principal curvatures in the eigenvalue will result in eigenvectors or the principal directions, $\textbf{V}_1$ and $\textbf{V}_2$, corresponding to the principal curvatures. The derivatives of the principal curvatures with respect to $\widehat{\textbf{C}}$ and $\widehat{\mathscr{K}}$ can be obtained as
\begin{align}
    \frac{\partial k_{n}}{\partial \widehat{\mathscr{K}}} & =\mathbf{V}_{n} \otimes \mathbf{V}_{n} \nonumber \\ 
    \frac{\partial k_{n}}{\partial \widehat{\mathbf{C}}} & = -k_{n} \frac{\partial k_{n}}{\partial \widehat{\mathscr{K}}}
    \label{eq:dk_dCK}
\end{align}
Similarly, the derivatives of the principal directions with respect to $\widehat{\textbf{C}}$ and $\widehat{\mathscr{K}}$ can be obtained as
\begin{align}
    \frac{\partial \mathbf{V}_{n}}{\partial \widehat{\mathscr{K}}} & =\frac{1}{\left(k_{n}-k_{m}\right)} \mathbf{V}_{m} \otimes\left(\mathbf{V}_{n} \otimes_{\mathrm{symm}} \mathbf{V}_{m}\right) \nonumber \\
    \frac{\partial \mathbf{V}_{n}}{\partial \widehat{\mathbf{C}}} & =-\frac{1}{2} \mathbf{V}_{n} \otimes \mathbf{V}_{n} \otimes \mathbf{V}_{n}-k_{n} \frac{\partial \mathbf{V}_{n}}{\partial \widehat{\mathscr{K}}}
    \label{eq:dV_dCK}
\end{align}
where the $\otimes_{\mathrm{symm}}$ operation between two matrices, $\mathbf{A}$ and $\mathbf{B}$, is defined as
\begin{equation}
    \mathbf{A} \otimes_{\mathrm{symm}} \mathbf{B}=\frac{1}{2}(\mathbf{A} \otimes \mathbf{B}+\mathbf{B} \otimes \mathbf{A})
\end{equation}

\section{Derivatives of deformed lattice parameters with respect to strain measures}
\label{ap:bond_ders}
  Following references \cite{Arroyo2004IJNME,Arroyo2002JMPS} the derivative of the tangent lattice parameter, $\mathbf{w}$, with respect to $\widehat{\mathbf{C}}$ and $\widehat{\mathscr{K}}$ can be expressed as 
\begin{align}
    \frac{\partial \text{w}^{n}}{\partial \widehat{\mathbf{C}}} & =\widehat{\text{C}}_{A B} A^{A} \frac{\partial\left(V_{n}\right)^{B}}{\partial \widehat{\mathbf{C}}}+\mathbf{A} \otimes_{\mathrm{symm}} \mathbf{V}_{n} \\
    \frac{\partial w^{n}}{\partial \widehat{\mathscr{K}}} & =\widehat{\text{C}}_{A B} A^{A} \frac{\partial\left(V_{n}\right)^{B}}{\partial \widehat{\mathscr{K}}}
\end{align}
Then, the derivative of the bond vector with respect to the $\bullet = \widehat{\mathbf{C}} \: \text{or} \: \widehat{\mathscr{K}}$ can be expressed as
\begin{equation}
\frac{\partial[\mathbf{a}_t]}{\partial \bullet}=\left\{\begin{array}{c}
\mathscr{Q}_{1} \frac{\partial w^{1}}{\partial \bullet}+w^{1} \mathscr{Q}_{1}^{\prime}\left(w^{1} \frac{\partial k_{1}}{\partial \bullet}+k_{1} \frac{\partial w^{1}}{\partial \bullet}\right) \\
\mathscr{Q}_{2} \frac{\partial w^{2}}{\partial \bullet}+w^{2} \mathscr{Q}_{2}^{\prime}\left(w^{2} \frac{\partial k_{2}}{\partial \bullet}+k_{2} \frac{\partial w^{2}}{\partial \bullet}\right) \\
\frac{1}{2}\left[w^{1} \mathscr{Q}_{12}^{2}\left(w^{1} \frac{\partial k_{1}}{\partial \bullet}+2 k_{1} \frac{\partial w^{1}}{\partial \bullet}\right)+k_{1}\left(w^{1}\right)^{2} \mathscr{Q}_{12} \mathscr{Q}_{12}^{\prime}\left(w^{1} \frac{\partial k_{1}}{\partial \bullet}+k_{1} \frac{\partial w^{1}}{\partial \bullet}\right)+\cdots\right. \\
\left.w^{2} \mathscr{Q}_{22}^{2}\left(w^{2} \frac{\partial k_{2}}{\partial \bullet}+2 k_{2} \frac{\partial w^{2}}{\partial \bullet}\right)+k_{2}\left(w^{2}\right)^{2} \mathscr{Q}_{22} \mathscr{Q}_{22}^{\prime}\left(w^{2} \frac{\partial k_{2}}{\partial \bullet}+k_{2} \frac{\partial w^{2}}{\partial \bullet}\right)\right]
\end{array}\right\}
\end{equation}
and the derivative of the thickness component of the deformed bond with respect to the $\lambda^{\pm}$ can be expressed as
\begin{equation}
    \frac{\partial[\mathbf{a}_n^+]}{\partial \lambda^{+}} = \mathbf{A}^+_n\: \: ; 
        \frac{\partial[\mathbf{a}_n^-]}{\partial \lambda^{-}} = \mathbf{A}^-_n\: \: 
\end{equation}
whereas
\begin{equation}
    \frac{\partial[\mathbf{a}_n^+]}{\partial \lambda^{-}} = 0 \: \: ; 
        \frac{\partial[\mathbf{a}_n^-]}{\partial \lambda^{+}} = 0\: \: 
\end{equation}
Therefore, the derivatives of each bond and angle can be obtained as
\begin{align}
    \frac{\partial a_{i}}{\partial \circ} & =\frac{1}{a_{i}}\left(a_{i}\right)^{c} \frac{\partial\left(a_{i}\right)^{c}}{\partial \circ} \\
    \frac{\partial \theta_{i}}{\partial \circ} & =\frac{-1}{\sin \theta_{i} a_{j} a_{k}}\left\{\left(a_{j}\right)^{c} \frac{\partial\left(a_{k}\right)^{c}}{\partial \circ}+\left(a_{k}\right)^{c} \frac{\partial\left(a_{j}\right)^{c}}{\partial \circ}-\cos \theta_{i}\left[a_{j} \frac{\partial a_{k}}{\partial \circ}+a_{k} \frac{\partial a_{j}}{\partial \circ}\right]\right\}
\end{align}
Here, $\circ =  \widehat{\mathbf{C}} \: \text{or} \: \widehat{\mathscr{K}} \: \text{or} \: \lambda^+ \: \text{or} \: \lambda^-$

\section{Finite element approximation using B-splines}
\label{app:b_splines}
\begin{figure}[htbp]
   \centering
   \includegraphics[width=0.75\linewidth]{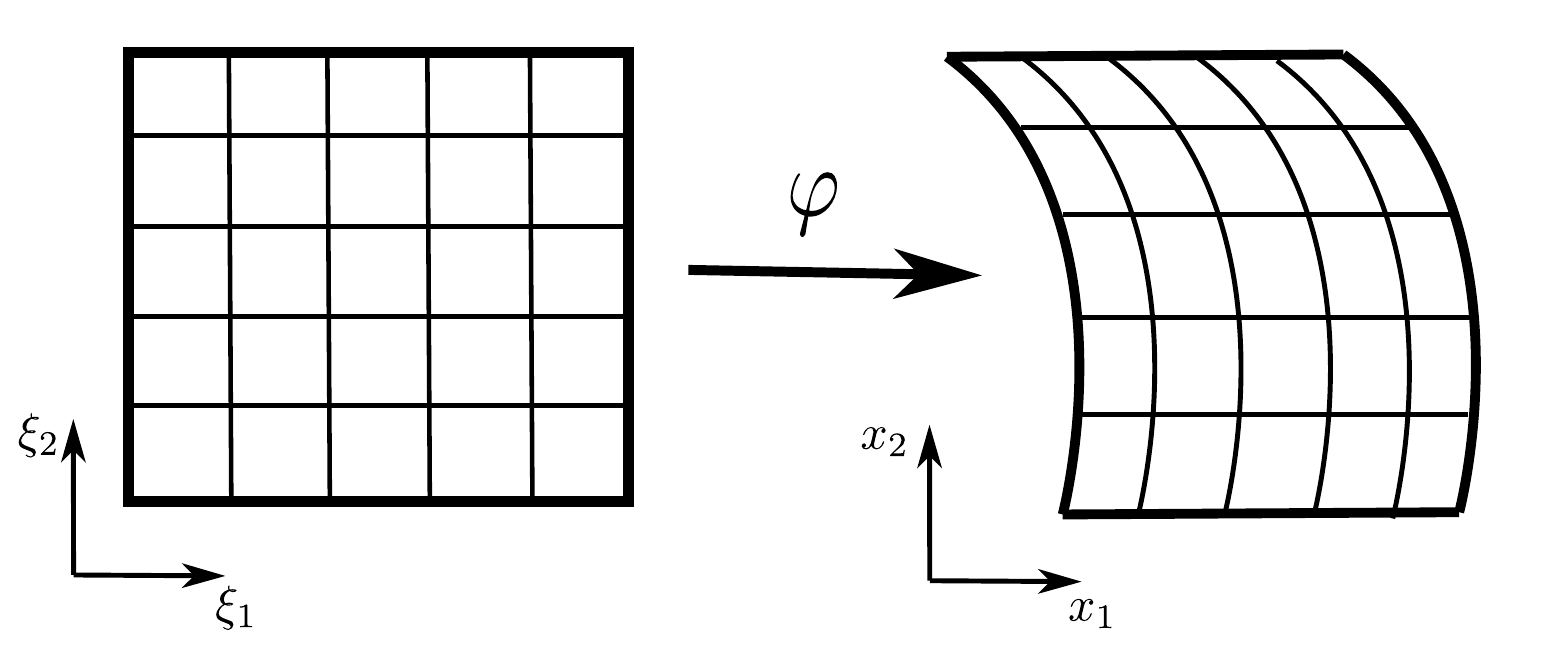}
   \caption{B-splines map between parametric domain and physical domain}
   \label{fig:spline_map}
\end{figure}
In the present work, B-spline basis functions are used to discretize the domain as the potential energy is a function of curvature which requires the basis function to have bounded second order derivatives. In B-spline formulation, a map between the parametric domain and the physical domain (either undeformed or deformed configuration) is defined, as shown in figure~\ref{fig:spline_map}. 
The map is defined based on the order of polynomial and the number of elements required to discretize the domain. 
Similar to the standard finite element methods, the order of polynomial, $p$, and the number of elements are chosen based on the desired accuracy.
Based on this, a knot vector is defined which contains non-descending breaking points in it. For example for a open uniform B-spline where, $\xi_{1,2} \in [0,1]$, the knot vector can be written as 
\begin{equation}
    \xi_{1,2} = [0,0,0,0.2,0.4,0.6,0.8,1,1,1] \: \text{for} \: p = 2
\end{equation}
In any open B-spline, the first and last knot points repeat themselves $p+1$ times. Open B-splines are used when interpolation of control points for first and last points on the boundaries of the physical domain is required. 
In the closed B-splines, the first and last control points do not interpolate to the boundaries of the physical domain.
The distinction between these two approximations starts from the knot vector itself.
In either open or closed B-splines, the number of elements is equal to the number of intervals between two distinct consecutive knot points. 
For example, in the above knot vector, there are 5 elements, one between $\xi=0$ and $\xi=0.2$, another between $\xi=0.2$ and $\xi=0.4$, and similarly, other elements can be obtained. Therefore, the knot vector is obtained based on the number of elements required in the discretization. The number of control points and basis functions to define the physical domain in each direction can be computed as
\begin{equation}
    n = m + p + 1
\end{equation}
Here, $n$ represents the number of basis function and control points, $m$ represents the size of the knot vector and $p$ is the order of polynomial. The basis functions are obtained as
    \begin{align}
    N_{i,0}  = & \begin{cases} 
     1 & \text{if} \: \xi_i \le \xi < \xi_{i+1} \\
     0 & \text{otherwise}
    \end{cases} \\
    N_{i,p} = & \frac{\xi-\xi_i}{\xi_{i+p} - \xi_i}N_{i,p-1}(\xi) + \frac{\xi_{i+p+1}-\xi}{\xi_{i+p+1}-\xi_{i+1}}N_{i+1,p-1}(\xi)
    \end{align}
$\xi$ in the equation represents the point in the parametric domain. 
In B-spline approximations, there are $p+1$ non-zero basis functions in each element. The map can then be expressed as
\begin{equation}
    \mathbf{\varphi}^e(\xi_1,\xi_2) = \sum_I \mathbf{Q}_I^e N_I(\xi_1,\xi_2)
\end{equation}
Here, $\xi_1$ and $\xi_2$ represent the co-ordinates in the parametric domain. $\mathbf{Q}_I^e$ represents the $I-th$ control point for element $e$. Control points are co-ordinates in the physical domain which define the shape. $N_I$ represents the $I$-th basis function to define that element. 


\newpage
\bibliographystyle{elsarticle-harv} 
\bibliography{MoS2_formulation}

\begin{thebibliography}{63}
\expandafter\ifx\csname natexlab\endcsname\relax\def\natexlab#1{#1}\fi
\providecommand{\url}[1]{\texttt{#1}}
\providecommand{\href}[2]{#2}
\providecommand{\path}[1]{#1}
\providecommand{\DOIprefix}{doi:}
\providecommand{\ArXivprefix}{arXiv:}
\providecommand{\URLprefix}{URL: }
\providecommand{\Pubmedprefix}{pmid:}
\providecommand{\doi}[1]{\href{http://dx.doi.org/#1}{\path{#1}}}
\providecommand{\Pubmed}[1]{\href{pmid:#1}{\path{#1}}}
\providecommand{\bibinfo}[2]{#2}
\ifx\xfnm\relax \def\xfnm[#1]{\unskip,\space#1}\fi
\bibitem[{Akinwande et~al.(2017)Akinwande, Brennan, Bunch, Egberts, Felts, Gao,
  Huang, Kim, Li, Li, Liechti, Lu, Park, Reed, Wang, Yakobson, Zhang, Zhang,
  Zhou and Zhu}]{Akinwande2017EML}
\bibinfo{author}{Akinwande, D.}, \bibinfo{author}{Brennan, C.J.},
  \bibinfo{author}{Bunch, J.S.}, \bibinfo{author}{Egberts, P.},
  \bibinfo{author}{Felts, J.R.}, \bibinfo{author}{Gao, H.},
  \bibinfo{author}{Huang, R.}, \bibinfo{author}{Kim, J.S.},
  \bibinfo{author}{Li, T.}, \bibinfo{author}{Li, Y.}, \bibinfo{author}{Liechti,
  K.M.}, \bibinfo{author}{Lu, N.}, \bibinfo{author}{Park, H.S.},
  \bibinfo{author}{Reed, E.J.}, \bibinfo{author}{Wang, P.},
  \bibinfo{author}{Yakobson, B.I.}, \bibinfo{author}{Zhang, T.},
  \bibinfo{author}{Zhang, Y.W.}, \bibinfo{author}{Zhou, Y.},
  \bibinfo{author}{Zhu, Y.}, \bibinfo{year}{2017}.
\newblock A review on mechanics and mechanical properties of 2D
  materials--Graphene and beyond.
\newblock \bibinfo{journal}{Extreme Mechanics Letters} \bibinfo{volume}{13},
  \bibinfo{pages}{42 -- 77}.
\bibitem[{Ansari et~al.(2016)Ansari, Shahnazari, Malakpour, Faghihnasiri and
  Sahmani}]{ansari2016dft}
\bibinfo{author}{Ansari, R.}, \bibinfo{author}{Shahnazari, A.},
  \bibinfo{author}{Malakpour, S.}, \bibinfo{author}{Faghihnasiri, M.},
  \bibinfo{author}{Sahmani, S.}, \bibinfo{year}{2016}.
\newblock {A DFT study on the elastic and plastic properties of MoS2 nanosheet
  subjected to external electric field}.
\newblock \bibinfo{journal}{Superlattices and Microstructures}
  \bibinfo{volume}{97}, \bibinfo{pages}{506--518}.
\bibitem[{Arroyo and Belytschko(2002)}]{Arroyo2002JMPS}
\bibinfo{author}{Arroyo, M.}, \bibinfo{author}{Belytschko, T.},
  \bibinfo{year}{2002}.
\newblock An atomistic-based finite deformation membrane for single layer
  crystalline films.
\newblock \bibinfo{journal}{Journal of the Mechanics and Physics of Solids}
  \bibinfo{volume}{50}, \bibinfo{pages}{1941--1977}.
\bibitem[{Arroyo and Belytschko(2003)}]{Arroyo2003MOM}
\bibinfo{author}{Arroyo, M.}, \bibinfo{author}{Belytschko, T.},
  \bibinfo{year}{2003}.
\newblock A finite deformation membrane based on inter-atomic potentials for
  the transverse mechanics of nanotubes.
\newblock \bibinfo{journal}{Mechanics of Materials} \bibinfo{volume}{35},
  \bibinfo{pages}{193--215}.
\bibitem[{Arroyo and Belytschko(2004)}]{Arroyo2004IJNME}
\bibinfo{author}{Arroyo, M.}, \bibinfo{author}{Belytschko, T.},
  \bibinfo{year}{2004}.
\newblock Finite element methods for the non-linear mechanics of crystalline
  sheets and nanotubes.
\newblock \bibinfo{journal}{International Journal for Numerical Methods in
  Engineering} \bibinfo{volume}{59}, \bibinfo{pages}{419--456}.
\bibitem[{Bertolazzi et~al.(2011)Bertolazzi, Brivio and
  Kis}]{Bertolazzi2011ACSNano}
\bibinfo{author}{Bertolazzi, S.}, \bibinfo{author}{Brivio, J.},
  \bibinfo{author}{Kis, A.}, \bibinfo{year}{2011}.
\newblock Stretching and Breaking of Ultrathin MoS2.
\newblock \bibinfo{journal}{ACS Nano} \bibinfo{volume}{5},
  \bibinfo{pages}{9703--9709}.
\bibitem[{Bhimanapati et~al.(2015)Bhimanapati, Lin, Meunier, Jung, Cha, Das,
  Xiao, Son, Strano, Cooper, Liang, Louie, Ringe, Zhou, Kim, Naik, Sumpter,
  Terrones, Xia, Wang, Zhu, Akinwande, Alem, Schuller, Schaak, Terrones and
  Robinson}]{Bhimanapati2015acsnano}
\bibinfo{author}{Bhimanapati, G.R.}, \bibinfo{author}{Lin, Z.},
  \bibinfo{author}{Meunier, V.}, \bibinfo{author}{Jung, Y.},
  \bibinfo{author}{Cha, J.}, \bibinfo{author}{Das, S.}, \bibinfo{author}{Xiao,
  D.}, \bibinfo{author}{Son, Y.}, \bibinfo{author}{Strano, M.S.},
  \bibinfo{author}{Cooper, V.R.}, \bibinfo{author}{Liang, L.},
  \bibinfo{author}{Louie, S.G.}, \bibinfo{author}{Ringe, E.},
  \bibinfo{author}{Zhou, W.}, \bibinfo{author}{Kim, S.S.},
  \bibinfo{author}{Naik, R.R.}, \bibinfo{author}{Sumpter, B.G.},
  \bibinfo{author}{Terrones, H.}, \bibinfo{author}{Xia, F.},
  \bibinfo{author}{Wang, Y.}, \bibinfo{author}{Zhu, J.},
  \bibinfo{author}{Akinwande, D.}, \bibinfo{author}{Alem, N.},
  \bibinfo{author}{Schuller, J.A.}, \bibinfo{author}{Schaak, R.E.},
  \bibinfo{author}{Terrones, M.}, \bibinfo{author}{Robinson, J.A.},
  \bibinfo{year}{2015}.
\newblock Recent Advances in Two-Dimensional Materials beyond Graphene.
\newblock \bibinfo{journal}{ACS Nano} \bibinfo{volume}{9},
  \bibinfo{pages}{11509--11539}.
\bibitem[{Cao and Gao(2019)}]{cao2019mechanical}
\bibinfo{author}{Cao, G.}, \bibinfo{author}{Gao, H.}, \bibinfo{year}{2019}.
\newblock Mechanical properties characterization of two-dimensional materials
  via nanoindentation experiments.
\newblock \bibinfo{journal}{Progress in Materials Science}
  \bibinfo{volume}{103}, \bibinfo{pages}{558--595}.
\bibitem[{Castellanos-Gomez et~al.(2012)Castellanos-Gomez, Poot, Steele,
  van~der Zant, Agraït and Rubio-Bollinger}]{Castellanos-Gomez2011AdvMat}
\bibinfo{author}{Castellanos-Gomez, A.}, \bibinfo{author}{Poot, M.},
  \bibinfo{author}{Steele, G.A.}, \bibinfo{author}{van~der Zant, H.S.J.},
  \bibinfo{author}{Agraït, N.}, \bibinfo{author}{Rubio-Bollinger, G.},
  \bibinfo{year}{2012}.
\newblock Elastic Properties of Freely Suspended MoS2 Nanosheets.
\newblock \bibinfo{journal}{Advanced Materials} \bibinfo{volume}{24},
  \bibinfo{pages}{772--775}.
\bibitem[{Castellanos-Gomez et~al.(2013)Castellanos-Gomez, RoldAn, Cappelluti,
  Buscema, Guinea, van~der Zant and Steele}]{Castellanos-Gomez2013NanoLett}
\bibinfo{author}{Castellanos-Gomez, A.}, \bibinfo{author}{RoldAn, R.},
  \bibinfo{author}{Cappelluti, E.}, \bibinfo{author}{Buscema, M.},
  \bibinfo{author}{Guinea, F.}, \bibinfo{author}{van~der Zant, H.S.J.},
  \bibinfo{author}{Steele, G.A.}, \bibinfo{year}{2013}.
\newblock Local Strain Engineering in Atomically Thin MoS2.
\newblock \bibinfo{journal}{Nano Letters} \bibinfo{volume}{13},
  \bibinfo{pages}{5361--5366}.
\bibitem[{Colas et~al.(2019)Colas, Serles, Saulot and
  Filleter}]{colas2019strength}
\bibinfo{author}{Colas, G.}, \bibinfo{author}{Serles, P.},
  \bibinfo{author}{Saulot, A.}, \bibinfo{author}{Filleter, T.},
  \bibinfo{year}{2019}.
\newblock Strength measurement and rupture mechanisms of a micron thick
  nanocrystalline MoS2 coating using AFM based micro-bending tests.
\newblock \bibinfo{journal}{Journal of the Mechanics and Physics of Solids}
  \bibinfo{volume}{128}, \bibinfo{pages}{151--161}.
\bibitem[{Conley et~al.(2013)Conley, Wang, Ziegler, Haglund, Pantelides and
  Bolotin}]{Conley2013NanoLett}
\bibinfo{author}{Conley, H.J.}, \bibinfo{author}{Wang, B.},
  \bibinfo{author}{Ziegler, J.I.}, \bibinfo{author}{Haglund, R.F.},
  \bibinfo{author}{Pantelides, S.T.}, \bibinfo{author}{Bolotin, K.I.},
  \bibinfo{year}{2013}.
\newblock Bandgap Engineering of Strained Monolayer and Bilayer MoS2.
\newblock \bibinfo{journal}{Nano Letters} \bibinfo{volume}{13},
  \bibinfo{pages}{3626--3630}.
\bibitem[{Cooper et~al.(2013)Cooper, Lee, Marianetti, Wei, Hone and
  Kysar}]{cooper2013nonlinear}
\bibinfo{author}{Cooper, R.C.}, \bibinfo{author}{Lee, C.},
  \bibinfo{author}{Marianetti, C.A.}, \bibinfo{author}{Wei, X.},
  \bibinfo{author}{Hone, J.}, \bibinfo{author}{Kysar, J.W.},
  \bibinfo{year}{2013}.
\newblock Nonlinear elastic behavior of two-dimensional molybdenum disulfide.
\newblock \bibinfo{journal}{Physical Review B} \bibinfo{volume}{87},
  \bibinfo{pages}{035423}.
\bibitem[{Cousins(1978)}]{cousins1978inner}
\bibinfo{author}{Cousins, C.}, \bibinfo{year}{1978}.
\newblock Inner elasticity.
\newblock \bibinfo{journal}{Journal of Physics C: Solid State Physics}
  \bibinfo{volume}{11}, \bibinfo{pages}{4867}.
\bibitem[{Cousins(2001)}]{cousinsPhD}
\bibinfo{author}{Cousins, C.}, \bibinfo{year}{2001}.
\newblock \bibinfo{title}{Inner elasticity and the higher-order elasticity of
  some diamond and graphite allotropes}.
\newblock Ph.D. thesis. University of Exeter.
\bibitem[{Dai et~al.(2020)Dai, Sanchez, Brennan and Lu}]{dai2020radial}
\bibinfo{author}{Dai, Z.}, \bibinfo{author}{Sanchez, D.A.},
  \bibinfo{author}{Brennan, C.J.}, \bibinfo{author}{Lu, N.},
  \bibinfo{year}{2020}.
\newblock Radial buckle delamination around 2D material tents.
\newblock \bibinfo{journal}{Journal of the Mechanics and Physics of Solids}
  \bibinfo{volume}{137}, \bibinfo{pages}{103843}.
\bibitem[{Do~Carmo(2016)}]{do2016differential}
\bibinfo{author}{Do~Carmo, M.P.}, \bibinfo{year}{2016}.
\newblock \bibinfo{title}{Differential geometry of curves and surfaces: revised
  and updated second edition}.
\newblock \bibinfo{publisher}{Courier Dover Publications}.
\bibitem[{Ghosh and Arroyo(2013)}]{GhoshJMPS2013}
\bibinfo{author}{Ghosh, S.}, \bibinfo{author}{Arroyo, M.},
  \bibinfo{year}{2013}.
\newblock An atomistic-based foliation model for multilayer graphene materials
  and nanotubes.
\newblock \bibinfo{journal}{Journal of the Mechanics and Physics of Solids}
  \bibinfo{volume}{61}, \bibinfo{pages}{235 -- 253}.
\bibitem[{Gilbert and Nocedal(1992)}]{gilbert1992global}
\bibinfo{author}{Gilbert, J.C.}, \bibinfo{author}{Nocedal, J.},
  \bibinfo{year}{1992}.
\newblock Global convergence properties of conjugate gradient methods for
  optimization.
\newblock \bibinfo{journal}{SIAM Journal on optimization} \bibinfo{volume}{2},
  \bibinfo{pages}{21--42}.
\bibitem[{Guo et~al.(2006)Guo, Wang and Zhang}]{Guo-IJSS}
\bibinfo{author}{Guo, X.}, \bibinfo{author}{Wang, J.}, \bibinfo{author}{Zhang,
  H.}, \bibinfo{year}{2006}.
\newblock Mechanical properties of single-walled carbon nanotubes based on
  higher order Cauchy--Born rule Mechanical properties of single-walled carbon
  nanotubes based on higher order Cauchy--Born rule.
\newblock \bibinfo{journal}{International Journal of Solids and Structures}
  \bibinfo{volume}{43}.
\bibitem[{Gupta and Vasudevan(2018)}]{gupta2018understanding}
\bibinfo{author}{Gupta, A.}, \bibinfo{author}{Vasudevan, S.},
  \bibinfo{year}{2018}.
\newblock Understanding surfactant stabilization of MoS2 nanosheets in aqueous
  dispersions from zeta potential measurements and molecular dynamics
  simulations.
\newblock \bibinfo{journal}{The Journal of Physical Chemistry C}
  \bibinfo{volume}{122}, \bibinfo{pages}{19243--19250}.
\bibitem[{He et~al.(2013)He, Poole, Mak and Shan}]{He2013NanoLett}
\bibinfo{author}{He, K.}, \bibinfo{author}{Poole, C.}, \bibinfo{author}{Mak,
  K.F.}, \bibinfo{author}{Shan, J.}, \bibinfo{year}{2013}.
\newblock Experimental Demonstration of Continuous Electronic Structure Tuning
  via Strain in Atomically Thin MoS2.
\newblock \bibinfo{journal}{Nano Letters} \bibinfo{volume}{13},
  \bibinfo{pages}{2931--2936}.
\bibitem[{Hill(1975)}]{hill1975elasticity}
\bibinfo{author}{Hill, R.}, \bibinfo{year}{1975}.
\newblock On the elasticity and stability of perfect crystals at finite strain,
  in: \bibinfo{booktitle}{Mathematical Proceedings of the Cambridge
  Philosophical Society}, \bibinfo{organization}{Cambridge University Press}.
  pp. \bibinfo{pages}{225--240}.
\bibitem[{Hollig(2003)}]{fe_bsplines}
\bibinfo{author}{Hollig, K.}, \bibinfo{year}{2003}.
\newblock \bibinfo{title}{Finite Element Methods with B-Splines}.
\newblock \bibinfo{publisher}{Society for Industrial and Applied Mathematics},
  \bibinfo{address}{USA}.
\bibitem[{Hu et~al.(2016)Hu, Li, Wang and Li}]{hu2016molecular}
\bibinfo{author}{Hu, J.}, \bibinfo{author}{Li, M.}, \bibinfo{author}{Wang, W.},
  \bibinfo{author}{Li, L.}, \bibinfo{year}{2016}.
\newblock Molecular Dynamics Simulations on Nanoindentation Experiment of
  Single-Layer MoS 2 Circular Nanosheets, in: \bibinfo{booktitle}{FZU-OPU-NTOU
  joint symposium on Advanced Mechanical Science \& Technology for Industrial
  Revolution 4.0}, \bibinfo{organization}{Springer}. pp.
  \bibinfo{pages}{333--339}.
\bibitem[{Jiang(2015)}]{Jiang2015a}
\bibinfo{author}{Jiang, J.W.}, \bibinfo{year}{2015}.
\newblock {Parametrization of Stillinger-Weber potential based on valence force
  field model: Application to single-layer MoS2 and black phosphorus}.
\newblock \bibinfo{journal}{Nanotechnology} \bibinfo{volume}{26}.
\bibitem[{Jiang and Park(2015)}]{Jiang2015d}
\bibinfo{author}{Jiang, J.W.}, \bibinfo{author}{Park, H.S.},
  \bibinfo{year}{2015}.
\newblock {A Gaussian treatment for the friction issue of Lennard-Jones
  potential in layered materials: Application to friction between graphene,
  MoS2, and black phosphorus}.
\newblock \bibinfo{journal}{Journal of Applied Physics} \bibinfo{volume}{117}.
\bibitem[{Jiang et~al.(2013a)Jiang, Park and Rabczuk}]{Jiang2013JAP}
\bibinfo{author}{Jiang, J.W.}, \bibinfo{author}{Park, H.S.},
  \bibinfo{author}{Rabczuk, T.}, \bibinfo{year}{2013}a.
\newblock Molecular dynamics simulations of single-layer molybdenum disulphide
  (MoS2): Stillinger-Weber parametrization, mechanical properties, and thermal
  conductivity.
\newblock \bibinfo{journal}{Journal of Applied Physics} \bibinfo{volume}{114},
  \bibinfo{pages}{064307}.
\bibitem[{Jiang et~al.(2013b)Jiang, Qi, Park and Rabczuk}]{Jiang2013Nanotech}
\bibinfo{author}{Jiang, J.W.}, \bibinfo{author}{Qi, Z.}, \bibinfo{author}{Park,
  H.S.}, \bibinfo{author}{Rabczuk, T.}, \bibinfo{year}{2013}b.
\newblock Elastic bending modulus of single-layer molybdenum disulfide (MoS 2
  ): finite thickness effect.
\newblock \bibinfo{journal}{Nanotechnology} \bibinfo{volume}{24},
  \bibinfo{pages}{435705}.
\bibitem[{Kandemir et~al.(2016)Kandemir, Yapicioglu, Kinaci and
  Sevik}]{Kandemir2016}
\bibinfo{author}{Kandemir, A.}, \bibinfo{author}{Yapicioglu, H.},
  \bibinfo{author}{Kinaci, A.}, \bibinfo{author}{Sevik, C.},
  \bibinfo{year}{2016}.
\newblock {Thermal transport properties of MoS$_2$ and MoSe$_2$ monolayers}.
\newblock \bibinfo{journal}{Nanotechnology} \bibinfo{volume}{27},
  \bibinfo{pages}{1--7}.
\bibitem[{Kolobov and Tominaga(2016)}]{kolobov2016two}
\bibinfo{author}{Kolobov, A.V.}, \bibinfo{author}{Tominaga, J.},
  \bibinfo{year}{2016}.
\newblock \bibinfo{title}{Two-dimensional transition-metal dichalcogenides}.
  volume \bibinfo{volume}{239}.
\newblock \bibinfo{publisher}{Springer}.
\bibitem[{Lennard-Jones(1931)}]{lennard1931cohesion}
\bibinfo{author}{Lennard-Jones, J.E.}, \bibinfo{year}{1931}.
\newblock Cohesion.
\newblock \bibinfo{journal}{Proceedings of the Physical Society (1926-1948)}
  \bibinfo{volume}{43}, \bibinfo{pages}{461}.
\bibitem[{Li(2012)}]{Li2012PRB}
\bibinfo{author}{Li, T.}, \bibinfo{year}{2012}.
\newblock Ideal strength and phonon instability in single-layer MoS${}_{2}$.
\newblock \bibinfo{journal}{Phys. Rev. B} \bibinfo{volume}{85},
  \bibinfo{pages}{235407}.
\bibitem[{Liang et~al.(2009)Liang, Phillpot and
  Sinnott}]{liang2009parametrization}
\bibinfo{author}{Liang, T.}, \bibinfo{author}{Phillpot, S.R.},
  \bibinfo{author}{Sinnott, S.B.}, \bibinfo{year}{2009}.
\newblock Parametrization of a reactive many-body potential for Mo--S systems.
\newblock \bibinfo{journal}{Physical Review B} \bibinfo{volume}{79},
  \bibinfo{pages}{245110}.
\bibitem[{Liu and Nocedal(1989)}]{liu1989limited}
\bibinfo{author}{Liu, D.C.}, \bibinfo{author}{Nocedal, J.},
  \bibinfo{year}{1989}.
\newblock On the limited memory BFGS method for large scale optimization.
\newblock \bibinfo{journal}{Mathematical programming} \bibinfo{volume}{45},
  \bibinfo{pages}{503--528}.
\bibitem[{Marsden and Hughes(1994)}]{marsden1994mathematical}
\bibinfo{author}{Marsden, J.E.}, \bibinfo{author}{Hughes, T.J.},
  \bibinfo{year}{1994}.
\newblock \bibinfo{title}{Mathematical foundations of elasticity}.
\newblock \bibinfo{publisher}{Courier Corporation}.
\bibitem[{Martin(1975)}]{martin1975many}
\bibinfo{author}{Martin, J.}, \bibinfo{year}{1975}.
\newblock Many-body forces in metals and the Brugger elastic constants.
\newblock \bibinfo{journal}{Journal of Physics C: Solid State Physics}
  \bibinfo{volume}{8}, \bibinfo{pages}{2837}.
\bibitem[{Milstein(1982)}]{milstein}
\bibinfo{author}{Milstein, F.}, \bibinfo{year}{1982}.
\newblock \bibinfo{title}{Mechanics of Solids}. \bibinfo{publisher}{Pergamon,
  Oxford}.
\newblock \bibinfo{note}{Edited by H. G. Hopkins and M. J. Sewell}.
\bibitem[{Morgan(1993)}]{morgan1993}
\bibinfo{author}{Morgan, F.}, \bibinfo{year}{1993}.
\newblock \bibinfo{title}{{Riemannian Geometry, a Beginner’s Guide}}.
\newblock \bibinfo{publisher}{Jones and Barlett Publishers},
  \bibinfo{address}{Boston, MA}.
\bibitem[{Nocedal(1980)}]{nocedal1980updating}
\bibinfo{author}{Nocedal, J.}, \bibinfo{year}{1980}.
\newblock Updating quasi-Newton matrices with limited storage.
\newblock \bibinfo{journal}{Mathematics of computation} \bibinfo{volume}{35},
  \bibinfo{pages}{773--782}.
\bibitem[{Ostadhossein et~al.(2017)Ostadhossein, Rahnamoun, Wang, Zhao, Zhang,
  Crespi and Van~Duin}]{ostadhossein2017reaxff}
\bibinfo{author}{Ostadhossein, A.}, \bibinfo{author}{Rahnamoun, A.},
  \bibinfo{author}{Wang, Y.}, \bibinfo{author}{Zhao, P.},
  \bibinfo{author}{Zhang, S.}, \bibinfo{author}{Crespi, V.H.},
  \bibinfo{author}{Van~Duin, A.C.}, \bibinfo{year}{2017}.
\newblock ReaxFF reactive force-field study of molybdenum disulfide (MoS2).
\newblock \bibinfo{journal}{The journal of physical chemistry letters}
  \bibinfo{volume}{8}, \bibinfo{pages}{631--640}.
\bibitem[{Park et~al.(2006)Park, Cho, Kim, Jun and Im}]{quasi-CNT}
\bibinfo{author}{Park, J.}, \bibinfo{author}{Cho, Y.}, \bibinfo{author}{Kim,
  S.}, \bibinfo{author}{Jun, S.}, \bibinfo{author}{Im, S.},
  \bibinfo{year}{2006}.
\newblock A quasicontinuum method for deformations of carbon nanotubes.
\newblock \bibinfo{journal}{Computer Modeling in Engineering \& Sciences}
  \bibinfo{volume}{11}, \bibinfo{pages}{61--72}.
\bibitem[{Peelaers and Van~de Walle(2014)}]{peelaers2014elastic}
\bibinfo{author}{Peelaers, H.}, \bibinfo{author}{Van~de Walle, C.},
  \bibinfo{year}{2014}.
\newblock Elastic constants and pressure-induced effects in MoS2.
\newblock \bibinfo{journal}{The Journal of Physical Chemistry C}
  \bibinfo{volume}{118}, \bibinfo{pages}{12073--12076}.
\bibitem[{Peng et~al.(2013)Peng, Liang, Ji and De}]{peng2013theoretical}
\bibinfo{author}{Peng, Q.}, \bibinfo{author}{Liang, C.}, \bibinfo{author}{Ji,
  W.}, \bibinfo{author}{De, S.}, \bibinfo{year}{2013}.
\newblock A theoretical analysis of the effect of the hydrogenation of graphene
  to graphane on its mechanical properties.
\newblock \bibinfo{journal}{Physical Chemistry Chemical Physics}
  \bibinfo{volume}{15}, \bibinfo{pages}{2003--2011}.
\bibitem[{Piegl and Tiller(1996)}]{PiegTill96}
\bibinfo{author}{Piegl, L.}, \bibinfo{author}{Tiller, W.},
  \bibinfo{year}{1996}.
\newblock \bibinfo{title}{{The NURBS Book}}.
\newblock \bibinfo{edition}{second} ed., \bibinfo{publisher}{Springer-Verlag},
  \bibinfo{address}{New York, NY, USA}.
\bibitem[{Pressley(2012)}]{diff_geom}
\bibinfo{author}{Pressley, A.}, \bibinfo{year}{2012}.
\newblock \bibinfo{title}{{Elementary Differential Geometry}}.
\newblock \bibinfo{publisher}{Springer}, \bibinfo{address}{London, United
  Kingdom}.
\bibitem[{Shenoy et~al.(1999)Shenoy, Miller, Tadmor, Rodney, Phillips and
  Ortiz}]{Shenoy1999JMPS}
\bibinfo{author}{Shenoy, V.}, \bibinfo{author}{Miller, R.},
  \bibinfo{author}{Tadmor, E.}, \bibinfo{author}{Rodney, D.},
  \bibinfo{author}{Phillips, R.}, \bibinfo{author}{Ortiz, M.},
  \bibinfo{year}{1999}.
\newblock An adaptive finite element approach to atomic-scale mechanics---the
  quasicontinuum method.
\newblock \bibinfo{journal}{Journal of the Mechanics and Physics of Solids}
  \bibinfo{volume}{47}, \bibinfo{pages}{611--642}.
\bibitem[{Smith et~al.(2000)Smith, Tadmor and Kaxiras}]{quasi4}
\bibinfo{author}{Smith, G.}, \bibinfo{author}{Tadmor, E.},
  \bibinfo{author}{Kaxiras, E.}, \bibinfo{year}{2000}.
\newblock Multiscale simulation of loading and electrical resistance in silicon
  nanoindentation.
\newblock \bibinfo{journal}{Physical Review Letters} \bibinfo{volume}{84},
  \bibinfo{pages}{1260--1263}.
\bibitem[{Splendiani et~al.(2010)Splendiani, Sun, Zhang, Li, Kim, Chim, Galli
  and Wang}]{Splendiani2010NanoLetters}
\bibinfo{author}{Splendiani, A.}, \bibinfo{author}{Sun, L.},
  \bibinfo{author}{Zhang, Y.}, \bibinfo{author}{Li, T.}, \bibinfo{author}{Kim,
  J.}, \bibinfo{author}{Chim, C.Y.}, \bibinfo{author}{Galli, G.},
  \bibinfo{author}{Wang, F.}, \bibinfo{year}{2010}.
\newblock Emerging Photoluminescence in Monolayer MoS2.
\newblock \bibinfo{journal}{Nano Letters} \bibinfo{volume}{10},
  \bibinfo{pages}{1271--1275}.
\bibitem[{Stillinger and Weber(1985)}]{PhysRevB.31.5262}
\bibinfo{author}{Stillinger, F.H.}, \bibinfo{author}{Weber, T.A.},
  \bibinfo{year}{1985}.
\newblock {Computer simulation of local order in condensed phases of silicon}.
\newblock \bibinfo{journal}{Phys. Rev. B} \bibinfo{volume}{31},
  \bibinfo{pages}{5262--5271}.
\bibitem[{Sun and Liew(2008)}]{liew-IJNME}
\bibinfo{author}{Sun, Y.}, \bibinfo{author}{Liew, K.}, \bibinfo{year}{2008}.
\newblock Application of the higher-order Cauchy--Born rule in mesh-free
  continuum and multiscale simulation of carbon nanotubes.
\newblock \bibinfo{journal}{International Journal for Numerical Methods in
  Engineering} \bibinfo{volume}{75}, \bibinfo{pages}{1238--1258}.
\bibitem[{Sundaram et~al.(2013)Sundaram, Engel, Lombardo, Krupke, Ferrari,
  Avouris and Steiner}]{Sundaram2013NanoLetter}
\bibinfo{author}{Sundaram, R.S.}, \bibinfo{author}{Engel, M.},
  \bibinfo{author}{Lombardo, A.}, \bibinfo{author}{Krupke, R.},
  \bibinfo{author}{Ferrari, A.C.}, \bibinfo{author}{Avouris, P.},
  \bibinfo{author}{Steiner, M.}, \bibinfo{year}{2013}.
\newblock Electroluminescence in Single Layer MoS2.
\newblock \bibinfo{journal}{Nano Letters} \bibinfo{volume}{13},
  \bibinfo{pages}{1416--1421}.
\bibitem[{Tadmor et~al.(1996)Tadmor, Ortiz and Phillips}]{Tadmor1996PhilMagA}
\bibinfo{author}{Tadmor, E.}, \bibinfo{author}{Ortiz, M.},
  \bibinfo{author}{Phillips, R.}, \bibinfo{year}{1996}.
\newblock Quasicontinuum analysis of defects in solids.
\newblock \bibinfo{journal}{Philosophical Magazine A} \bibinfo{volume}{73},
  \bibinfo{pages}{1529--1563}.
\bibitem[{Tadmor et~al.(1999)Tadmor, Smith, Bernstein and
  Kaxiras}]{Tadmor1999PRB}
\bibinfo{author}{Tadmor, E.}, \bibinfo{author}{Smith, G.},
  \bibinfo{author}{Bernstein, N.}, \bibinfo{author}{Kaxiras, E.},
  \bibinfo{year}{1999}.
\newblock Mixed finite element and atomistic formulation for complex crystals.
\newblock \bibinfo{journal}{Physical Review B} \bibinfo{volume}{59},
  \bibinfo{pages}{235--245}.
\bibitem[{Wen et~al.(2017)Wen, Shirodkar, Plech{\'{a}}{\v{c}}, Kaxiras, Elliott
  and Tadmor}]{Wen2017}
\bibinfo{author}{Wen, M.}, \bibinfo{author}{Shirodkar, S.N.},
  \bibinfo{author}{Plech{\'{a}}{\v{c}}, P.}, \bibinfo{author}{Kaxiras, E.},
  \bibinfo{author}{Elliott, R.S.}, \bibinfo{author}{Tadmor, E.B.},
  \bibinfo{year}{2017}.
\newblock {A force-matching Stillinger-Weber potential for MoS2:
  Parameterization and Fisher information theory based sensitivity analysis}.
\newblock \bibinfo{journal}{Journal of Applied Physics} \bibinfo{volume}{122}.
\bibitem[{Wright et~al.(1999)Wright, Nocedal et~al.}]{wright1999numerical}
\bibinfo{author}{Wright, S.}, \bibinfo{author}{Nocedal, J.}, et~al.,
  \bibinfo{year}{1999}.
\newblock Numerical optimization.
\newblock \bibinfo{journal}{Springer Science} \bibinfo{volume}{35},
  \bibinfo{pages}{7}.
\bibitem[{Wu et~al.(2008)Wu, Hwang and Huang}]{Huang-JMPS-2008}
\bibinfo{author}{Wu, J.}, \bibinfo{author}{Hwang, K.}, \bibinfo{author}{Huang,
  Y.}, \bibinfo{year}{2008}.
\newblock An atomistic-based finite-deformation shell theory for single-wall
  carbon nanotubes.
\newblock \bibinfo{journal}{Journal of the Mechanics and Physics of Solids}
  \bibinfo{volume}{56}, \bibinfo{pages}{279--292}.
\bibitem[{Xin and He(2012)}]{xin2012atomistic}
\bibinfo{author}{Xin, K.}, \bibinfo{author}{He, M.}, \bibinfo{year}{2012}.
\newblock Atomistic potential based cohesive modeling for surface separation.
\newblock \bibinfo{journal}{Blucher Mechanical Engineering Proceedings}
  \bibinfo{volume}{1}, \bibinfo{pages}{846--853}.
\bibitem[{Yang and E(2006)}]{weinan-prb}
\bibinfo{author}{Yang, J.}, \bibinfo{author}{E, W.}, \bibinfo{year}{2006}.
\newblock Generalized Cauchy-Born rules for elastic deformation of sheets,
  plates, and rods: Derivation of continuum models from atomistic models.
\newblock \bibinfo{journal}{Physical Review B} \bibinfo{volume}{74},
  \bibinfo{pages}{184110}.
\bibitem[{Yang et~al.(2018)Yang, Wang, Li, Gao, Chai and Yao}]{yang2018edge}
\bibinfo{author}{Yang, J.}, \bibinfo{author}{Wang, Y.}, \bibinfo{author}{Li,
  Y.}, \bibinfo{author}{Gao, H.}, \bibinfo{author}{Chai, Y.},
  \bibinfo{author}{Yao, H.}, \bibinfo{year}{2018}.
\newblock Edge orientations of mechanically exfoliated anisotropic
  two-dimensional materials.
\newblock \bibinfo{journal}{Journal of the Mechanics and Physics of Solids}
  \bibinfo{volume}{112}, \bibinfo{pages}{157--168}.
\bibitem[{Zhao et~al.(2019)Zhao, Wang, Katz, Mockensturm, Crespi and
  Zhang}]{zhao2019geometry}
\bibinfo{author}{Zhao, P.}, \bibinfo{author}{Wang, Y.}, \bibinfo{author}{Katz,
  B.}, \bibinfo{author}{Mockensturm, E.}, \bibinfo{author}{Crespi, V.},
  \bibinfo{author}{Zhang, S.}, \bibinfo{year}{2019}.
\newblock Geometry and chiral symmetry breaking of ripple junctions in 2D
  materials.
\newblock \bibinfo{journal}{Journal of the Mechanics and Physics of Solids}
  \bibinfo{volume}{131}, \bibinfo{pages}{337--343}.
\bibitem[{Zhao and Liu(2018)}]{zhao2018study}
\bibinfo{author}{Zhao, Z.Y.}, \bibinfo{author}{Liu, Q.L.},
  \bibinfo{year}{2018}.
\newblock Study of the layer-dependent properties of MoS 2 nanosheets with
  different crystal structures by DFT calculations.
\newblock \bibinfo{journal}{Catalysis Science \& Technology}
  \bibinfo{volume}{8}, \bibinfo{pages}{1867--1879}.
\bibitem[{Zhu et~al.(2013)Zhu, Wang, Liu, Marie, Qiao, Zhang, Wu, Fan, Tan,
  Amand and Urbaszek}]{Zhu2013PRB}
\bibinfo{author}{Zhu, C.R.}, \bibinfo{author}{Wang, G.}, \bibinfo{author}{Liu,
  B.L.}, \bibinfo{author}{Marie, X.}, \bibinfo{author}{Qiao, X.F.},
  \bibinfo{author}{Zhang, X.}, \bibinfo{author}{Wu, X.X.},
  \bibinfo{author}{Fan, H.}, \bibinfo{author}{Tan, P.H.},
  \bibinfo{author}{Amand, T.}, \bibinfo{author}{Urbaszek, B.},
  \bibinfo{year}{2013}.
\newblock Strain tuning of optical emission energy and polarization in
  monolayer and bilayer MoS${}_{2}$.
\newblock \bibinfo{journal}{Phys. Rev. B} \bibinfo{volume}{88},
  \bibinfo{pages}{121301}.

\end{thebibliography}





\end{document}